\documentclass[aps,prd,preprintnumbers,nofootinbib,floatfix,floats,groupedaddress]{revtex4} 

\usepackage{graphicx}
\usepackage{amsfonts}
\usepackage{amssymb}
\usepackage{amsbsy}
\usepackage{amsmath}
\usepackage{mathrsfs}
\usepackage{latexsym}
\usepackage{natbib}
\usepackage{bm}
\usepackage[utf8]{inputenc}
\usepackage{subfigure} 
\usepackage{color}
\usepackage[active]{srcltx}
\usepackage{wasysym}
\usepackage{mathbbol}
\usepackage{bigints}
\usepackage{hyperref}
\allowdisplaybreaks
\usepackage{comment}
\usepackage{empheq}
\usepackage{soul}
\usepackage[export]{adjustbox}
\usepackage{gensymb}

\DeclareFontFamily{OT1}{pzc}{}
\DeclareFontShape{OT1}{pzc}{m}{it}{<-> s * [1.10] pzcmi7t}{}
\DeclareMathAlphabet{\mathpzc}{OT1}{pzc}{m}{it}

\def\nn{\nonumber}

\def\l{\left}
\def\r{\right}
\def\DM{\mathrm{d}}

\DeclarePairedDelimiter\bra{\langle}{\rvert}
\DeclarePairedDelimiter\ket{\lvert}{\rangle}
\DeclarePairedDelimiterX\braket[2]{\langle}{\rangle}{#1 \delimsize\vert #2}

\newcommand{\gae}{\lower 3pt \hbox{$\,\, \buildrel {\scriptstyle >}\over {\scriptstyle
\sim}\,\,$}}
\newcommand{\lae}{\lower 2pt \hbox{$\, \buildrel {\scriptstyle <}\over {\scriptstyle
\sim}\,$}}

\reversemarginpar

\def\eq#1{{Eq.~(\ref{#1})}}

\def\fig#1{{Fig.~\ref{#1}}}

\def\erfc{\mathcal{E}\text{rfc}}
\def\erf{\mathcal{E}\text{rf}}

\begin{document}

\title{Universal role of curvature in vacuum entanglement}

 \author{Hari K}
 \email{harik@physics.iitm.ac.in}

 \author{Subhajit Barman}
 \email{subhajit.barman@physics.iitm.ac.in}

 \author{Dawood Kothawala}
 \email{dawood@iitm.ac.in}
 
 \affiliation{Centre for Strings, Gravitation and Cosmology, Department of Physics, Indian Institute of Technology Madras, Chennai 600 036, India}

\date{\today}
\begin{abstract}
\noindent
We highlight some universal features concerning the role of spacetime curvature in the entanglement induced between quantum probes coupled to a quantum field in a suitable vacuum state. The probes are initially causally disconnected and non-entangled. We explore the parameter space $\{{\omega}, d_0, \bm v_0\}$ spanned by the energy gap $\omega$ of the detectors, and the initial values of separation distance $d_0$ and relative velocity $\bm v_0$, both covariantly defined in arbitrary curved spacetime. We also obtain numerical results in de Sitter spacetimes and use these to explore strong curvature regime, while also corroborating our perturbative results in arbitrary curved spacetime. Our analysis shows that curvature can induce entanglement features in certain regions of the above parameter space, in a manner which facilitates using entanglement as a probe of spacetime curvature. 
\end{abstract}

\maketitle
\tableofcontents
\section{Introduction}
Entanglement -- one of the key features of quantum mechanics -- is also of fundamental importance in studying phenomena at the interface of gravity and quantum physics. Such an interface has conventionally been associated with the domain of quantum gravity, but of course it does not have to be. Interesting aspects associated with the study of low energy quantum systems in non-singular curved backgrounds can in fact provide significant insights into what one might expect in the extreme, quantum gravitational conditions. Work along these lines has gained considerable attention over the past decade, particularly in the light of universal role that gravity might play in quantum processes such as decoherence and entanglement \cite{pikovski2015universal}, and the question of whether this has any implication for the quantum nature of gravity \cite{bose-etal, marletto-vedral}. In this work, we will focus on identifying how entanglement between quantum probes, mediated through their coupling to a quantum field, is affected by spacetime curvature. A deeper question we would like to ask is whether a measure of entanglement can be used to re-construct the curvature of spacetime. This would be in the spirit of the broader formalism, developed by one of us and collaborators, which aims to describe spacetime in terms of non-local, bi-tensorial objects such as Synge world function and correlators; for recent reviews and relevant references, see \cite{qmetric-rev1, qmetric-rev2}. 

Unruh-de Witt (UdW) detectors provide the simplest and the most operational way to study such effects. It is well known that individual UdW detectors probe vacuum fluctuations of the quantum field which they are coupled to. When a pair of such detectors are employed, entanglement in the vacuum of the quantum field can be transferred to that between the two detectors. 
What makes this latter a more operationally and conceptually appropriate probe of the structure of quantum field vacuum is that the entanglement induced in detectors can be non-zero even for two inertial detectors coupled to a field in Minkowski vacuum \cite{Koga:2018the, Pozas-Kerstjens:2015gta}. (A single such detector would not respond at all.) 
Using two detectors as a probe of vacuum entanglement is an idea first introduced by Reznik \cite{Reznik:2002fz}, and has since been extensively studied in the literature. (It is also referred to as ``entanglement harvesting" in the literature \cite{Koga:2018the, Koga:2019fqh, Martin-Martinez:2015psa, Tjoa:2021roz}.) Many of these works have studied the nature of entanglement induced in the detectors for various scenarios, such as the effects of motion of the detectors \cite{Cong:2020nec, Zhang:2020xvo, Barman:2021kwg, Barman:2022xht, Hu:2015lda, Menezes:2017oeb}, the effects of presence of a thermal bath \cite{Brown:2013kia, Simidzija:2018ddw, Barman:2021bbw}, the dependence on spacetime dimensions \cite{Pozas-Kerstjens:2015gta}, and the effect of a passing gravitational wave \cite{Xu:2020pbj, Gray:2021dfk, Barman:2023aqk}. In a related but different context, radiative processes involving entangled detectors have also been extensively studied \cite{Ostapchuk:2011ud, Hu:2012jr, Menezes:2015uaa, Menezes:2015iva, Arias:2015moa, Menezes:2015veo, Liu:2018zod, Costa:2020aqa, Barman:2021oum, Barman:2022utm}.

However, the effect of background curvature in all its generality, explicitly in terms of the Riemann tensor, has not been studied so far. This is important since any entanglement dynamics involving the detector-field system will be affected by the background curvature, both explicitly, via additional terms in the vacuum two-point correlators, and implicitly, through the geometrical quantities appearing in the correlators pulled back onto the detector trajectories. One might expect to gain some insight by restricting to a special class of background spacetimes, taken as some exact solution of Einstein equations which is simple enough for the analysis to be tractable and yield analytical results \cite{Cliche:2010fi, Kukita:2017etu, Tjoa:2020eqh}. Evidently, such attempts are limited by the fact that the choice of a specific solution hides the manner in which Riemann curvature explicitly appears and affects the final results, thereby yielding only limited insights. We will comment further on these works vis-a-vis our results in the final section. In this work, we will set-up the problem directly in terms of Riemann tensor components, thereby highlighting explicitly their role in various measures of entanglement, and uncovering interesting universal features even from a perturbative analysis.

We will briefly describe here the key features of the setup conventionally used to examine the possibility of two uncorrelated detectors getting entangled over time and then quantify the measure of this entanglement. The simplest such probes of vacuum structure of a quantum field are the Unruh-DeWitt detectors, originally (and still widely) used to demonstrate and understand various aspects of the Unruh effect \cite{Unruh:1976db}. A study of entanglement involves two such detectors, 1 and 2, interacting with a massless background scalar field $\phi(x^{k})$. We will assume these detectors to be two-level systems with ground state $\ket{g}$, an excited state $\ket{e}$, with energy difference $\omega$ between them. The interaction action of this detector-field system is taken as
\begin{eqnarray}\label{eq:Int-action}
\mathcal{A}_{\rm int}= \int c_0 \, \chi(\tau) \, m_1 (\tau) \, \phi(x_1(\tau)) \,d \tau
+
\int c_0 \, \chi(s) \, m_2 (s) \, \phi(x_2(s)) \, d s~.
\end{eqnarray}
Here, $\chi$ is the switching function, and $m_{1,2}$ represent the monopole operators of the detectors. Let $\ket{\rm in}$ be the product state in the asymptotic past, $\ket{{\rm in}}=\ket{0}\ket{g}_1\ket{g}_2$. Here, $\ket{0}$ is the vacuum state of the field. In the asymptotic future, the state will evolve into, $\ket{{\rm out}}=T\big\{e^{i \mathcal{A}_{\rm{int}}} \ket{{\rm in}}\big\}$, with $T$ being the time ordering operator.
The reduced density matrix corresponding to the detector system is obtained by tracing over all the final field states, i.e., $\rho_{12}={\rm Tr}_{\phi}\, \ket{{\rm out}}\bra{{\rm out}}$. Introducing $c_0$ as a book-keeping variable for the interaction strength, one can express the reduced density matrix in the bases $\left\{\; \ket{g}_1 \ket{g}_2,\; \ket{e}_1 \ket{g}_2,\; \ket{g}_1 \ket{e}_2,\; \ket{e}_1 \ket{e}_2 \;\right\}$ (ref. \cite{Koga:2018the}) as
\begin{eqnarray}\label{eq:density-matrix}
\rho_{12}=
\begin{bmatrix}
1-c_0^2\l(\mathcal{P}_{1}+\mathcal{P}_{2} \r) & 0 & 0 & c_0^2 \mathscr{E}^{*}\\
0 & c_0^2\mathcal{P}_{1} & c_0^2\mathcal{P}_{12} & 0\\
0 & c_0^2\mathcal{P}_{12} & c_0^2\mathcal{P}_{2} & 0\\
c_0^2 \mathscr{E} & 0 & 0 & 0\\
\end{bmatrix} + \mathcal{O}(c_0^4)~.
\end{eqnarray}
For understanding entanglement, the quantities we will need are $\mathcal{P}_{1}, \mathcal{P}_{2}$ and $\mathscr{E}$ \cite{Koga:2018the}, and these are given by 
\begin{subequations}\label{eq:Pj-E-general}
\begin{eqnarray}
\mathcal{P}_{1} &=& \left \lvert \bra{e} m_{1}(0)\ket{g}_1 \right \rvert^2 \, \mathcal{I}_{1}
\hspace{1cm} ({\rm similarly \; for \; 2})
\\ 
\mathscr{E} &=&  \bra{e} m_{2}(0) \ket{g}_2 \;  \bra{e} m_{1}(0)\ket{g}_1 \, \mathcal{I}_{\varepsilon}
\end{eqnarray}
\end{subequations}
where
\begin{subequations}\label{eq:Ij-Ie-general}
\begin{eqnarray}\label{eq:Ij-general}
\mathcal{I}_{1} &=& 
\int_{-\infty}^{\infty} d\tau^\prime \int_{-\infty}^{\infty}d\tau \;
\chi(\tau) \chi(\tau^{\prime}) 
e^{i \omega (\tau - \tau^{\prime})}
G_{W}(x_{1}(\tau^{\prime}), x_{1}(\tau))
\\ 
i \, \mathcal{I}_{\varepsilon} &=& \int_{-\infty}^{\infty} ds \int_{-\infty}^{\infty}d\tau \; \chi(\tau) \chi(s) 
e^{i \omega (\tau + s)}
G_{F}(x_{2}(s), x_{1}(\tau))~.
\label{eq:Ie-general}
\end{eqnarray}
\end{subequations}
Here, $G_{W}(x^{\prime},x)$ is the Wightmann function, and $G_F(x_2,x_1)$ is the Feynman propagator. The quantities $\mathcal{P}_{1,2}$ are local to each detector, and correspond to individual detector responses, whereas $\mathscr{E}$ represents the non-local correlation between the detectors.

Now according to \cite{Peres:1996dw}, a state is entangled if the partial transposition of the relevant density matrix has at least one negative eigenvalue. With the reduced detector density matrix from \eq{eq:density-matrix}, this requirement leads to the condition:
\begin{eqnarray}\label{eq:cond-entanglement-1}
    \mathcal{P}_{1} \mathcal{P}_{2} < |\mathscr{E}|^2~
\end{eqnarray}
which can be rewritten as
\begin{equation}\label{eq:cond-entanglement-2}
\mathcal{I}_{1} \mathcal{I}_{2} < |\mathcal{I}_{\varepsilon}|^2
\end{equation}
where, $\mathcal{I}_{1}$ and $\mathcal{I}_{2}$ are the individual detector responses. The above condition can be used to analyze the possibility of two Unruh-DeWitt detectors coupled to a free scalar field getting entangled. It is worth highlighting that the non-local term (RHS) above depends on the Feynman propagator, while the individual responses (LHS) depend on the Wightman function.

Several measures quantify the entanglement induced between the detectors by providing a bound on this entanglement in the system. For example, entanglement negativity $\mathcal{N}(\rho_{12})$ is defined as the sum of all negative eigenvalues of the partial transpose of the density matrix $\rho_{12}$. One can also define the logarithmic negativity $\log_{2}{(\mathcal{N}(\rho_{12})+1)}$, see \cite{Koga:2018the}, which provides an upper bound on the induced entanglement. Another very relevant measure of entanglement is concurrence $\mathcal{C}(\rho_{12})$ \cite{Koga:2018the}. With the reduced density matrix from \eq{eq:density-matrix}, we have the expressions for the negativity and the concurrence (to lowest order in perturbation, up to $c_0^2$ terms), respectively as 
\begin{subequations}\label{eq:negativity-concurrence}
    \begin{eqnarray}\label{eq:negativity-general}
    \mathcal{N}(\rho_{12}) &=& c^2_{0}\,\Big[\sqrt{(\mathcal{P}_{1}-\mathcal{P}_{2})^2+4\,|\mathscr{E}|^2}-\mathcal{P}_{1}-\mathcal{P}_{2}\Big]/2 
    \\
    ~\mathcal{C}(\rho_{12}) &=& 2\,c^2_{0}\, \Big[|\mathscr{E}|-\sqrt{\mathcal{P}_{1}\,\mathcal{P}_{2}}\Big]~.
    \label{eq:concurrence-general}
\end{eqnarray}
\end{subequations}
The expression of the monopole moment operator (for each detector) is given by 
\begin{eqnarray}\label{eq:monopole-operator}
    m(0) = \ket{e}\bra{g} + \ket{g}\bra{e}~.
\end{eqnarray}
Then one can observe that the expectation values of these operators as used in \eq{eq:Pj-E-general} is $\bra{e}m(0)\ket{g} =1$. Utilizing this observation, in the special scenario of $\mathcal{I}_{1}=\mathcal{I}_{2}=\mathcal{I}$, one can express the negativity and the concurrence as 
$\mathcal{N}(\rho_{12}) = \mathcal{C}(\rho_{12})/2 = \big[|\mathcal{I}_{\varepsilon}|-\mathcal{I}\big]c^2_{0}$, taking terms upto $\mathcal{O}(c_0^2)$. This is the case that we shall encounter in our system. Therefore, it is convenient to define negativity specifically as $\mathcal{N}^{(2)} = |\mathcal{I}_{\varepsilon}|-\mathcal{I}$ to denote a measure of the entanglement in our system.

The rest of the manuscript is organised as follows:
In Sec. \ref{sec:GeodesicInterval-GreensFn}, we provide a description for some of the quantities in curved spacetime, like Green's functions in terms of  the geodesic distance, that are essential in realizing the phenomenon of entanglement. 
In the subsequent Sec. \ref{sec:Negativity-GenCurved}, we study the nature of negativity in a generalized curved spacetime with a relatively small curvature compared to the distance between the two detectors, i.e., the curvature length scale is larger than the initial separation between the detectors. Following in Sec. \ref{sec:Negativity-de-Sitter}, we consider the maximally symmetric de Sitter background and investigate the characteristics of negativity. In Sec. \ref{sec:Entanglement-Probe-Curvature}, we describe how our results quantifying the effect of curvature on entanglement can be utilized as a tool to probe curvature of arbitrary spacetime, in the spirit of the comment made at the end of the first paragraph of this section. We discuss our observed outcomes regarding negativity in a de Sitter background and elucidate their physical interpretations in Sec. \ref{sec:Observations-discussion}. 
Finally, we summarize with a concluding discussion in Sec. \ref{sec:Concluding-remarks}.

\section{Geometric aspects of quantum probes coupled to a quantum field in curved spacetime}\label{sec:GeodesicInterval-GreensFn}

As elucidated in the previous section, the criteria for entanglement from \eq{eq:cond-entanglement-2} and its quantification depends on two specific kinds of terms in the expression for negativity: the local terms involving $\mathcal{I}_{1,2}$ that denote individual detector transition probabilities, and the non-local term $\mathcal{I}_{\varepsilon}$ that signifies the correlation between the two detectors. To obtain explicit expressions of these quantities, an understanding of different types of Green's functions is necessary. These Green's functions will again contain the information about the detector trajectories and the background in them. In this section, we intend to provide general expressions of Green's functions in curved background. In particular, we shall start by providing a general expression for the geodesic interval between two geodesic detectors and delineate the necessary Green's functions in terms of this geodesic distance. In this section, we shall also discuss the curved background's causal structure and compare it to the Minkowski spacetime. As we will see, it has a close correspondence with the nature of Green's functions.

\subsection{Geodesic interval between two detectors in curved spacetime} \label{sec:geodesic-interval}

\begin{figure*}[!htb]%
    \includegraphics[width=0.5\textwidth]{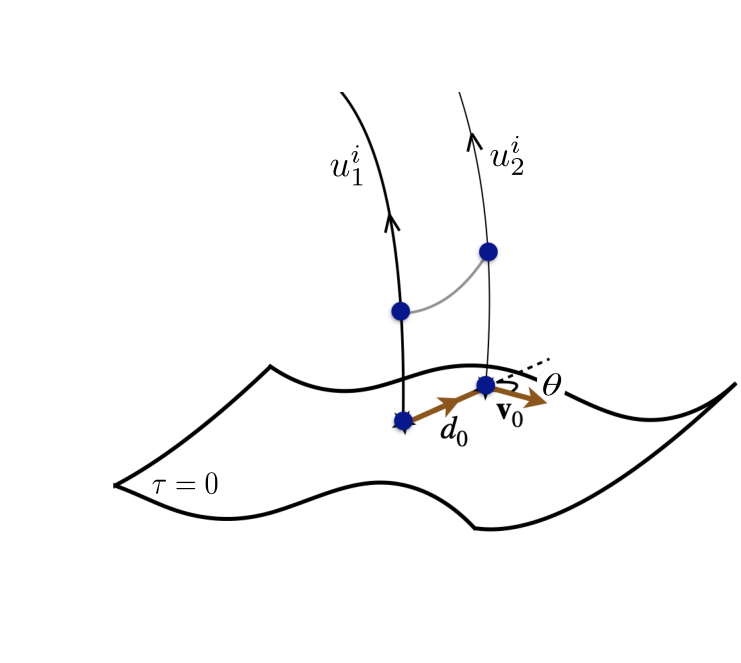}
    \includegraphics[width=0.38\textwidth]{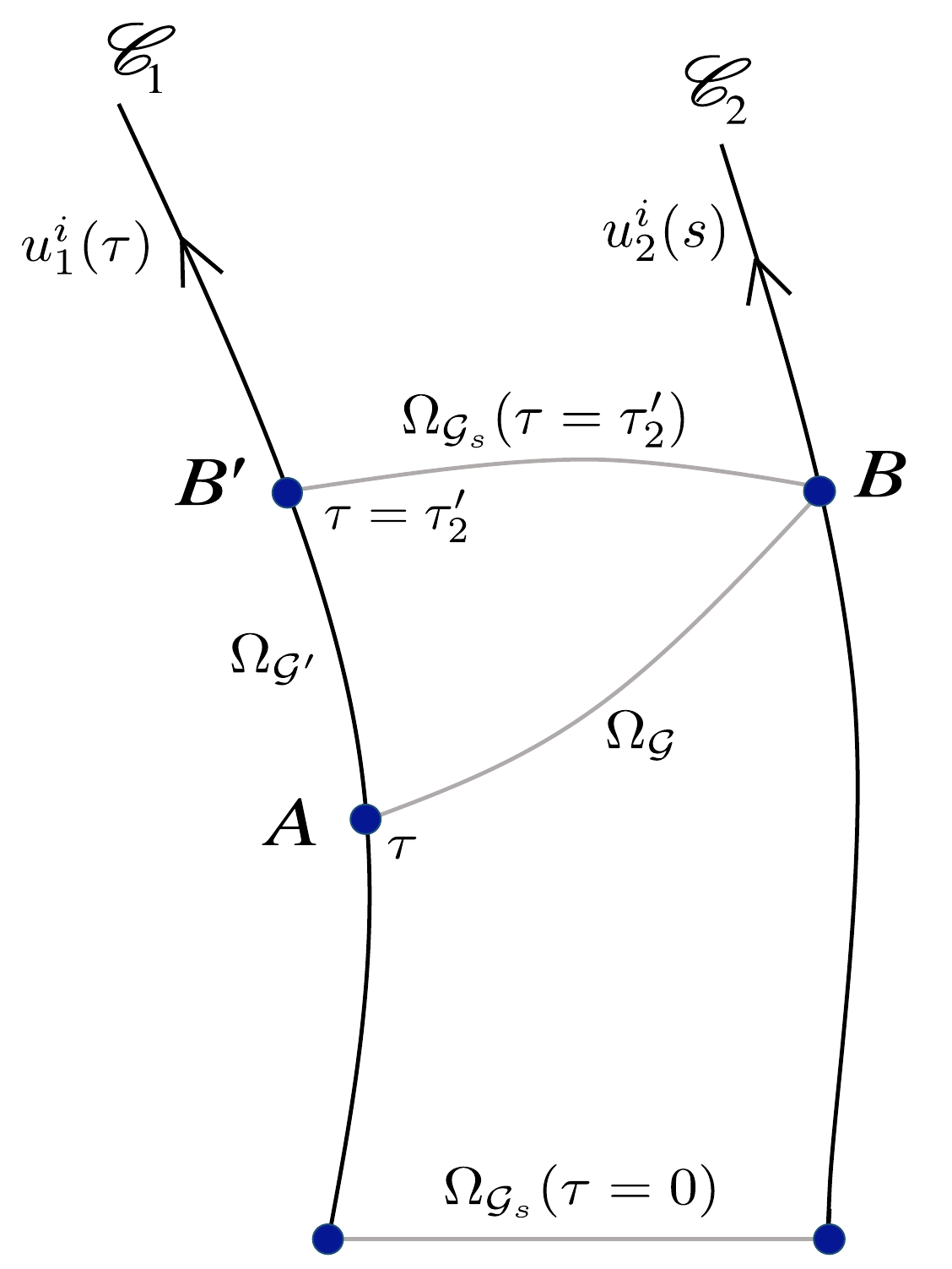}%
    \caption{Schematic diagrams for two point-like detectors in geodesic trajectories in general curved spacetime. In left we depict the two detectors as denoted by $u_{1}^{i}$ and $u_{2}^{i}$. On the $\tau=0$ hypersurface, the shortest distance between these detectors is $d_{0}$. Note the angle between this distance and the velocity vector of the second detector is $\theta$. On the right, we depict the same scenario, which now corresponds to the construction of the Fermi coordinates for the two detectors with respect to the trajectory of detector $A$. Here, $\Omega_{AB}\equiv \Omega_{\mathcal{G}}$, $\Omega_{AB^{\prime}}\equiv \Omega_{\mathcal{G}^{\prime}}$, and $\Omega_{B^{\prime}B}\equiv \Omega_{\mathcal{G}s}$. Also note that $\Omega_{\mathcal{G}s}(\tau=0)=d_{0}$ denotes the separation between the two detectors at the $\tau_{1}=0$ hypersurface.}
    \label{fig:setup}
\end{figure*}

The geometric set-up and kinematical parameters associated with two detectors (timelike curves) is described in Fig. \ref{fig:setup}. To obtain the geodesic interval between the two detectors, we construct Fermi Normal Coordinates(FNC) along the curve $\mathscr{C}_1$ and identify the point on curve $\mathscr{C}_2$ by sending out space-like geodesics orthogonal to curve $\mathscr{C}_1$. 
Let the first detector follows the curve $\mathscr{C}_1$ with proper time $\tau$ and be at $A$ and the second detector follows the curve $\mathscr{C}_2$ with proper time $s$ be at $B$. Let $\tau_{2}^{\prime}$ be the time component of the FNC on $\mathscr{C}_1$ of the detector $B$ at $B^{\prime}$. Now, the geodesic interval between $A$ and $B$ can be expressed as a series expansion in terms of the spacelike geodesic interval between $B$ and $B^{\prime}$, and the geodesic interval between $A$ and $B^{\prime}$. The point $O$ with $\tau=0$ provides the Cauchy data. Employing a generalized variant of an expansion scheme proposed by Synge, we can obtain the geodesic interval between arbitrary events $A$ and $B$ on the two probe trajectories using coincidence limits of covariant derivatives of Synge's world function. This rather long calculation is described in Appendix \ref{app:sigma2}. We introduce the unit vector $\xi^{\alpha}$ (representing direction cosines) defined by ${X}^\alpha(0):=d_{0}\, \xi^{\alpha}$ where, $\xi^{\alpha}$ (the directional cosines) satisfy $\eta_{\alpha \beta}\xi^{\alpha} \xi^{\beta}=1$, and the parallel and perpendicular components of initial $v_0^\alpha$ w.r.t. $\xi^\alpha$ by $v_0^\alpha = v_{||} \, \xi^{\alpha}+ v_{\perp} \, n^\alpha$, with $n^\alpha$ an arbitrary unit vector in the 2-plane orthogonal to $\bm \xi$ (and, of course, $\bm u_1$). (Note that $v_{||}=v_0 \cos\theta$ and $v_{\perp}=v_0 \sin\theta$.)

%
%

The final expression in terms of these quantities can be written in a compact form
\begin{eqnarray}\label{eq:geodesic-dist-gen}
\nonumber \\
-\sigma^2_{\mathcal{G}} \;=\; \Delta \tau_{\mathcal{G}}^2 \;=\; \l(\tau_{2}^{\prime}-\tau\r)^2 \mathpzc{f}(\tau_{2}^{\prime})-\mathpzc{g}(\tau_{2}^{\prime})+ \mathcal{O}(\nabla R,R^2)~
\\
\nonumber
\end{eqnarray}
where the auxiliary functions $\mathpzc{f}(\tau_{2}^{\prime})$ and $\mathpzc{g}(\tau_{2}^{\prime})$ are given by 
\begin{subequations}\label{eq:exp-f&g}
\begin{eqnarray}\label{eq:exp-f}
    \mathpzc{f}(\tau_{2}^{\prime}) &=& 1+\frac{1}{3}\l[ R_{\xi 0\xi 0} d_{0}^2 + 2 d_{0} \l( R_{\xi 0\xi 0} v_{||} + R_{\xi 0n 0} v_{\perp} \r)  \tau_{2}^{\prime} + \l( R_{\xi 0\xi 0} v_{||}^2 + 2 R_{\xi 0n 0} v_{||} v_{\perp} + R_{n0n0} v_{\perp}^2 \r) {\tau_{2}^{\prime}}^2 \r] \hspace{1cm}
    \\
    \mathpzc{g}(\tau_{2}^{\prime}) &=& \l( d_{0}^2 + v_0^2 {\tau_{2}^{\prime}}^2+ 2d_{0} v_{||} \tau_{2}^{\prime} \r) + d_{0}^2\,\l( \frac{2}{3} R_{n\xi 0 \xi}v_{\perp} - R_{\xi 0\xi 0}   + \frac{1}{3} R_{n\xi n\xi} v_{\perp}^2 \r) {\tau_{2}^{\prime}}^2 \nn \\
    && -\frac{4}{3} d_{0} \l( R_{\xi 0\xi 0} v_{||} + R_{\xi 0n 0}  v_{\perp} \r) {\tau_{2}^{\prime}}^3 - \frac{1}{3} \l( R_{\xi 0\xi 0} v_{||}^2 + 2 R_{n0\xi 0} v_{||} v_{\perp} + R_{n0n0} v_{\perp}^2 \r) {\tau_{2}^{\prime}}^4
    \label{eq:exp-g}
\end{eqnarray}
\end{subequations}  

To improve readability, we have introduced the following compact notation: $R_{\xi 0 \xi 0}:=R_{\alpha i \beta j} \xi^{\alpha} \xi^{\beta} u^{i} u^{j}, R_{\xi 0 n 0}:=R_{\alpha i \beta j} \xi^{\alpha} n^{\beta} u^{i} u^{j}, R_{n\xi 0 \xi}:=R_{\alpha \mu i \nu} n^{\alpha} \xi^{\mu} \xi^{\nu} u^{i}$ and so on for the other Riemann tensor components. 

It is easy to recover the following results from the above expression:

\textit{\underline{Flat spacetime}}:

\begin{equation}
    \Delta \tau_{\mathcal{G}}^2 = \l(\tau_{2}^{\prime}-\tau\r)^2 - 
    \Biggl[ \l( d_0 + v_\parallel \tau_{2}^{\prime}  \r)^2 + v_\perp^2 {\tau_{2}^{\prime}}^2 \Biggl]
\end{equation}

\textit{\underline{Initially static detectors; $v_0=0$}}:

When $v_0=0$, the above equation simplifies to
\begin{equation}
    \Delta \tau_{\mathcal{G}}^2 = \l(\tau_{2}^{\prime}-\tau\r)^2 
    \Biggl[1+ \frac{1}{3} R_{\xi 0\xi 0} d_{0}^2 \Biggl]
    - 
    d_{0}^2 \Biggl[1 + R_{\xi 0\xi 0} {\tau_{2}^{\prime}}^2 \Biggl]
\end{equation}

\vspace{.5cm}

To compute the non-local term in entanglement measures, we also need to map the proper time $s$ of trajectory $\mathscr{C}_2$ to the Fermi time $\tau_{2}^{\prime}$ based on $\mathscr{C}_1$'s trajectory. This is best done using the Fermi metric, and the derivation is given in Appendix \ref{app:sigma2}. The final result is: 
\begin{eqnarray}\label{eq:gamma-bar-tbp}
    s &=&  \frac{\tau_{2}^{\prime}}{\gamma_0} + \frac{\gamma_0}{2}  \tau_{2}^{\prime} \l[ \mathpzc{h}_1(v_{0},d_{0}) + \tau_{2}^{\prime} \l\{ d_0\,\mathpzc{h}_2(v_{0},d_{0}) + \frac{\tau_{2}^{\prime}}{3} \mathpzc{h}_3(v_{0},d_{0})\r\}\r] + \mathcal{O}(R^2,\nabla R)~.
\end{eqnarray}
where, $\gamma_0=1/\sqrt{1-v_0^2}$ and $\mathpzc{h}_1, \mathpzc{h}_2, \mathpzc{h}_3$ are given in Appendix \ref{app:sigma2}. When the initial velocity of the second observer measured with respect to the first one's FNC vanishes, we have $v_{0}=0$ and also $v_{||}=0=v_{\perp}$. In that scenario, the relation between $s$ and $\tau_{2}^{\prime}$ boils down to $s = \tau_{2}^{\prime}(1 + R_{\xi 0 \xi 0}\, d_0^2/2)$. We also mention that to obtain the expression of \eq{eq:gamma-bar-tbp}, we have resorted to expanding $\gamma$ from (\ref{eq:gamma-tbp}) perturbatively, assuming small curvature terms. Let us also consider $s = \tau_{2}^{\prime}/\bar{\gamma}$, where the expression of $\bar{\gamma}$ is evident from the previous equation.

\subsection{Green's function in Hadamard representation}

The Hadamard representation of the Feynman propagator\cite{hadamard2003lectures, Decanini:2005gt} for KG equation of massless scalar field in 4 dimensions is given by,
\begin{equation}\label{eq:FeynmanP-hadamard}
G_{F}\l(x,x^{\prime}\r)=\frac{i}{4 \pi^2}\l[\frac{\Delta^{1/2}}{\sigma^2(x,x^{\prime})+i\epsilon} + V_n(x,x^{\prime}) \ln\l(\sigma^2(x,x^{\prime})+i\epsilon\r) + W_n(x,x^{\prime}) \r]~.
\end{equation}
Here, the functions, $V_n(x,x^{\prime})$ and $W_n(x,x^{\prime})$ are expressed as series, $V_n(x,x^{\prime})=\sum_{n=0}^{\infty}v_n(x,x^{\prime})\sigma^{2n}(x,x^{\prime})$, $W_n(x,x^{\prime})=\sum_{n=0}^{\infty}w_n(x,x^{\prime})\sigma^{2n}(x,x^{\prime})$, and $-(1/2)\sigma^2(x,x^\prime)\equiv \Omega(x,x^\prime)$ is half the square of geodesic distance also known as Synge's world function\cite{synge1960relativity}. 

Similarly, for Wightman function we use the form\cite{Kay:1988mu},
\begin{eqnarray}\label{eq:Wightman-haramard}
    G^{+}\l(x,x^{\prime}\r)=\frac{1}{4 \pi^2}\l[ \frac{\Delta^{1/2}}{\sigma^2_{\epsilon}(x,x^{\prime})} + V_n(x,x^{\prime}) \ln\l(\sigma^2_{\epsilon}(x,x^{\prime})\r) + W_n(x,x^{\prime}) \r]~.
\end{eqnarray}
Here, $\sigma^2_{\epsilon}:=\sigma^2(x,x^{\prime})+i\epsilon\l[T(x)-T(x^{\prime})\r]+\epsilon^2$, with $T(x)$ being any global time function that increases towards the future and other parameters are similar to that defined for Feynman propagator.

The Hadamard coefficient, $V_n(x,x^{\prime})$ obeys certain recursion relation with specific boundary condition and can be solved, whereas the boundary condition $W_0(x,x^\prime)$ for $W_n(x,x^{\prime})$ is not restrained by the recursion relations. But the Hadamard coefficients, $W_n(x,x^{\prime})$ can be determined once $W_0(x,x^{\prime})$ is specified.
For our evaluation of the integral, we are interested in the leading $1/\sigma^2$ term since this term determines the pole structure of the Feynman propagator, and $\Delta^{1/2}$ will not contribute to the poles. We also approximate, $\Delta^{1/2}\approx 1$. Hence, the geodesic interval between the two detectors is required to evaluate the Feynman propagator.

\underline{\emph{The van Vleck determinant}}:

For both single detector probability and non-local term in the density matrix, the Hadamard form of the Wightmann function and the Feynman propagator have van Vleck determinant in the expression. The van Vleck determinant can be expanded in terms of geodesic distance as in Ref. \cite{Christensen:1976vb},
\begin{equation}\label{eq:van-Vleck-det}
    \Delta^{1/2}=1 + \frac{1}{12} R_{a b}\sigma^a \sigma^b -\frac{1}{24} \nabla_{c}R_{ab} \sigma^a \sigma^b \sigma^c + \mathcal{O}(\sigma^4)
\end{equation}

Note that $\sigma^a(x,x^\prime)=-(\Delta \tau_{\rm geod}) u^a$, where $\Delta \tau$ is the geodesic interval and $u^a$ is the tangent vector of the geodesic. When this expansion is used in the Hadamard form,
\begin{align}\label{eq:vV-det-expansion}
    \frac{\Delta^{1/2}}{4 \pi^2 \sigma^2_{\epsilon}} \approx & \frac{1}{4 \pi^2 \sigma^2_{\epsilon}} + \frac{(1/12) R_{ab}\sigma^a\sigma^b}{4 \pi^2 \sigma^2_{\epsilon}} + \mathcal{O}(\sigma^3) \nn \\
    \approx & \frac{1}{4 \pi^2 \sigma^2_{\epsilon}} + \frac{(1/12) R_{u u}\sigma^2}{4 \pi^2 \sigma^2_{\epsilon}} + \mathcal{O}(\sigma^3) \nn
\end{align}

Hence the contribution from the curvature through the van Vleck determinant will be only an additive constant. The additional corrections from derivatives of curvature and higher-order terms in the curvature are ignored.

\subsection{Causal structure and the nature of entanglement induced between the probes}
\label{subsec:NpNm-forms}

The non-local entangling term from \eq{eq:Ie-general} will depend on different parameters such as $d_0,v_0,\theta,$ and different curvature components. To compare the contribution of genuine entanglement and entanglement through field-mediated communication channel, we express the non-local term in the form 
\begin{eqnarray}
    \mathcal{I}_{\varepsilon} = -\mathcal{I}_{\varepsilon}^{+} - \mathcal{I}_{\varepsilon}^{-}~,
\end{eqnarray}
where $\mathcal{I}_{\varepsilon}^{+}$ and $\mathcal{I}_{\varepsilon}^{-}$ depend exclusively on the expectation values of the field anti-commutator and field commutators. The part $\mathcal{I}_{\varepsilon}^{-}$ depending on the field commutator, is considered in the literature to signify the entanglement due to the causal connection. Whereas $\mathcal{I}_{\varepsilon}^{+}$ corresponds to the contribution that signifies entanglement in the true sense. In particular, these quantities can be explicitly expressed in terms of the Wightman functions $G_{W}(\bar{u},\bar{v})$ as 
\begin{subequations}\label{eq:Ipme-general}
\begin{eqnarray}
    \mathcal{I}_{\varepsilon}^{+} &=& \frac{1}{2}\int_{-\infty}^{\infty}d\bar{u}\int_{-\infty}^{\infty}d\bar{v}\,e^{i\,\omega\,v}\chi_{A}(\bar{u},\bar{v})\,\chi_{B}(\bar{u},\bar{v})\,\text{Re}[G_{W}(\bar{u},\bar{v})]~,\\
    \mathcal{I}_{\varepsilon}^{-} &=& \frac{1}{2}\int_{-\infty}^{\infty}d\bar{u}\int_{-\infty}^{\infty}d\bar{v}\,e^{i\,\omega\,v}\chi_{A}(\bar{u},\bar{v})\,\chi_{B}(\bar{u},\bar{v})\,\l[1+2\,\Theta(\bar{u})\r]\,\text{Im}[G_{W}(\bar{u},\bar{v})]~.
\label{Ip-and-Im}
\end{eqnarray}
\end{subequations}
Here, $\Theta(\bar{u})$ denotes the Heaviside step function, which arrives from the decomposition of the Feynman propagator in terms of time-ordered Wightman functions, e.g., $i\,G_{F}(x,x')=\Theta(t-t')G_{W}(x,x')+\Theta(t'-t)G_{W}(x',x)$. In the special case, when under the transformation of $\bar{v}\to-\bar{v}$ the quantities $\chi_{A}(\bar{u},\bar{v})$, $\chi_{B}(\bar{u},\bar{v})$, and $G_{W}(\bar{u},\bar{v})$ remain unchanged, the above two quantities $\mathcal{I}_{\varepsilon}^{\pm}$ are simply given by 
\begin{eqnarray}\label{eq:Ipme-simple}
    \mathcal{I}_{\varepsilon}^{+} &=& \text{Re}[\mathcal{I}_{\varepsilon}]~,~~\textup{and}~~
    \mathcal{I}_{\varepsilon}^{-} = \text{Im}[\mathcal{I}_{\varepsilon}]~.
\end{eqnarray}
The Wightman function in Minkowski spacetime with the Gaussian switching satisfies this condition. Furthermore, we shall see that the Wightman function is de Sitter background also satisfies this condition.

On the other hand, in situations when at least one of the quantities $\chi_{A}(\bar{u},\bar{v})$, $\chi_{B}(\bar{u},\bar{v})$, or $G_{W}(\bar{u},\bar{v})$ is not invariant under the transformation of $\bar{v}\to-\bar{v}$, one cannot simply use the expressions from \eq{eq:Ipme-simple} for the evaluation of $\mathcal{I}_{\varepsilon}^{\pm}$. In this scenario, one should use the general expressions from \eq{eq:Ipme-general}. We should mention that the Green's functions in a generalized curved background with the geodesic interval from \eq{eq:geodesic-dist-gen} do not remain invariant under the aforementioned transformation. Therefore, when investigating the entanglement in a generalized curved spacetime, one has to use the expressions from \eq{eq:Ipme-general} to understand the natures of $\mathcal{I}_{\varepsilon}^{\pm}$.

\begin{figure*}[!h]
\centering
\includegraphics[width=5cm]{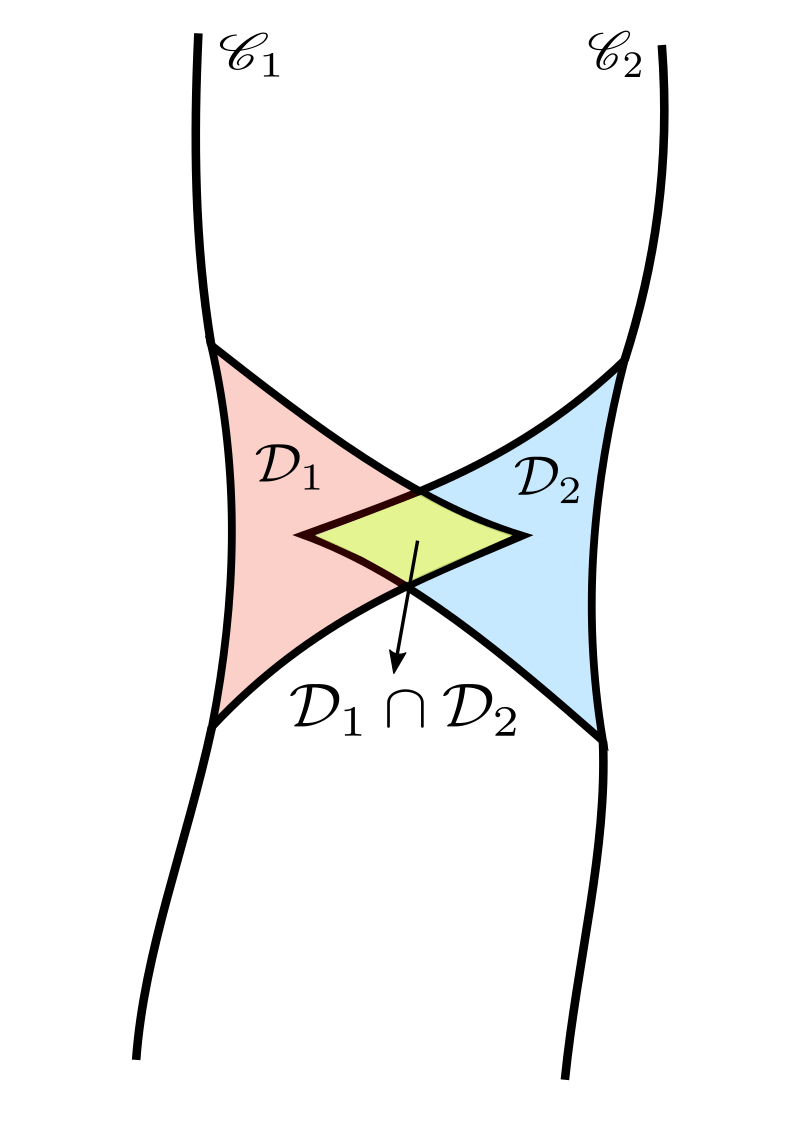}
\hskip 20pt
\includegraphics[width=5cm]{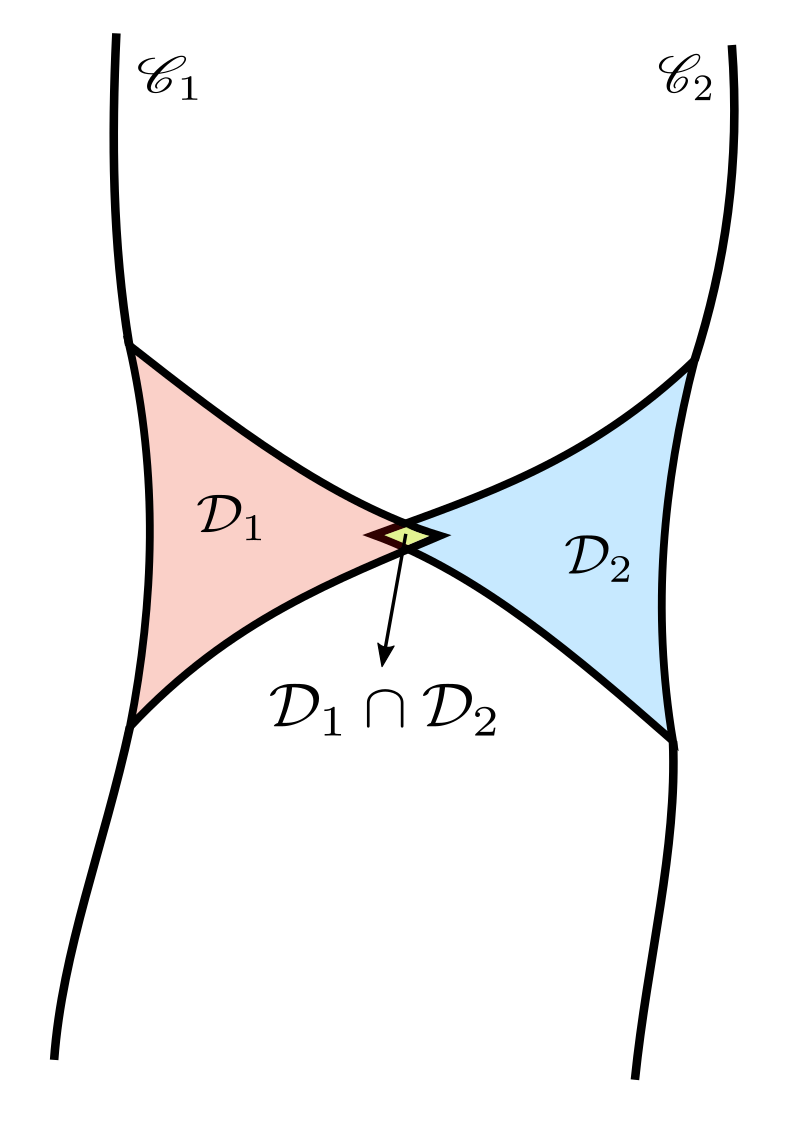}
\hskip 20pt
\includegraphics[width=5cm]{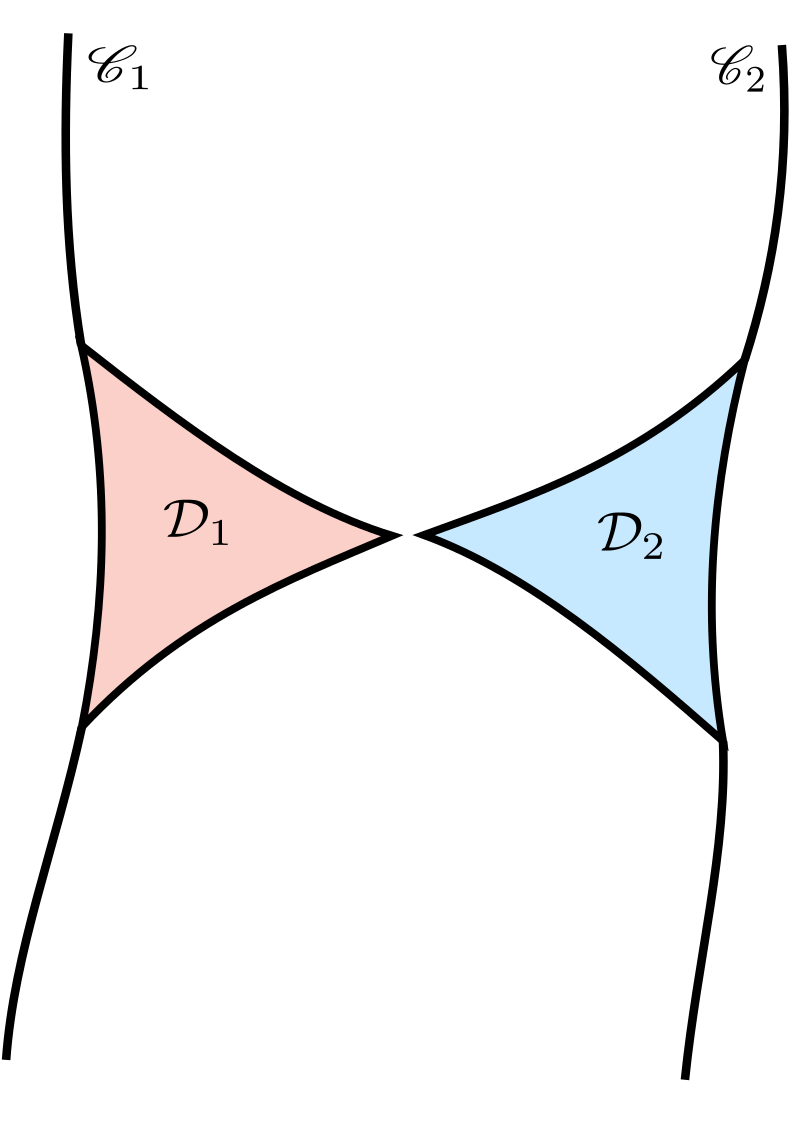}
\caption{The overlap of the causal diamond formed between the switch on and switch off time for each detector determines the contribution from genuine entanglement and field-mediated communication. We will choose our parameter values for the detector setup such that this overlap is minimal but the value of negativity is large enough to see the effects due to curvature.} 
\label{fig:causal-diamond}
\end{figure*}

\begin{figure*}[!h]
\centering
\includegraphics[height=6.8cm]{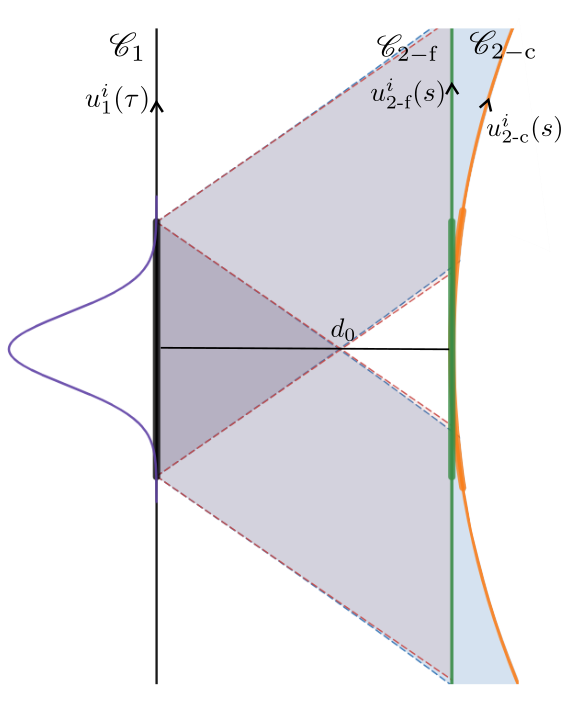}
\hskip 60pt
\includegraphics[height=6.8cm]{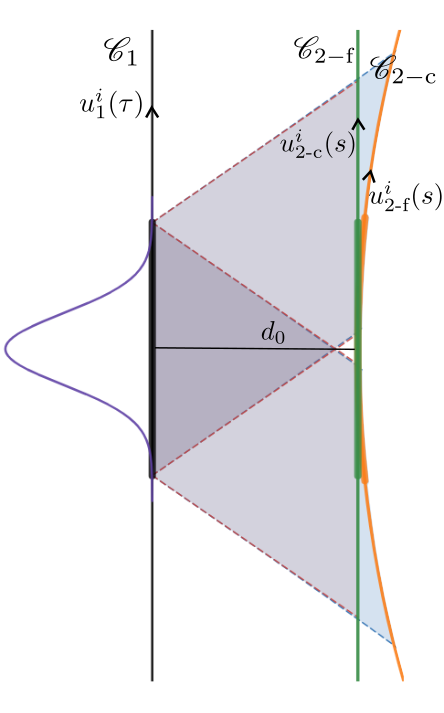}
\caption{The above figure represents schematic diagrams for the causal structure in a general curved background with two different detector trajectories. We have used Fermi normal coordinates to represent the curves and light cones to compare it with Minkowski. As the distance and time progresses, the curved spacetime light cones deviate from the Minkowski light cones. We also observe that the two detector trajectories move away from each other due to curvature. With finite switching, the first detector $A$ can access a lesser portion of the lightcone on the trajectory of detector $B$ as it moves away.} \label{fig:Geodesic-trajectory-distance}
\end{figure*}

Furthermore, as discussed in \cite{Martin-Martinez:2015psa, Tjoa:2021roz} one can define suitable quantities denoting the spacelike and communication-based entanglement as 
\begin{eqnarray}\label{eq:Npm-general}
 \mathcal{N}^{\pm} =|\mathcal{I}_{\varepsilon}^{\pm}|-\mathcal{I}~.
\end{eqnarray}
Here also, the superscript $'+'$ or $'-'$ respectively correspond to the spacelike and communication channel-mediated entanglement. In the later part of our present analysis, we shall observe how the light-cone diagrams from Fig. \ref{fig:causal-diamond} and \ref{fig:Geodesic-trajectory-distance} are closely related and give a physical interpretation to the quantitative values of these specific components $\mathcal{N}^{\pm}$.
%

The commutator and anti-commutator can be used to characterize the entanglement through field-mediated communication channels and the entanglement through non-local correlations. The contribution from these correlators will depend on the causal structure of the spacetime and the given detector configuration. The overlap of the causal diamonds $\mathcal{D}_{1, 2}$ determined by the switching functions (see Figs. \ref{fig:causal-diamond}, \ref{fig:Geodesic-trajectory-distance}) can be used to estimate these contributions. Any non-zero intersection of the causal diamonds would indicate possible entanglement due to the communication channel will be present. Hence, if the detectors are placed far apart to ensure there is little to no overlap between causal diamonds, the genuine entanglement can be obtained.
%

\section{Mathematical details for computing the entanglement}

\subsection{General curved spacetime}\label{sec:Negativity-GenCurved}

As discussed previously, the formulation to understand the effects of tidal curvature on entanglement constrains the interaction time between the detectors and the background field. Then it will be suitable to consider a specific switching function $\chi(\tau)$, through which we can implement this constraint. In particular, introducing a nontrivial switching function will complicate the computations, but it will provide physically more relevant and consistent results with the mathematical formulations. In this regard, we consider the Gaussian switching $\chi(\tau)=e^{-\tau^2/\Bar{T}^2}$, and shall evaluate the local $\mathcal{I}$ and non-local $\mathcal{I}_{\varepsilon}$ terms in entanglement.

\vspace{.5cm}

\textbf{\textit{The local and non-local terms in negativity:}}

\vspace{.5cm}

\underline{\textit{Evaluation of $\mathcal{I}$}}: 

Let us first evaluate the local terms $\mathcal{I}$ in the concurrence, which corresponds to the individual detector transition probabilities. We take the expression of this integral from Eq. (\ref{eq:Ij-general}), and shall be using the Gaussian switching functions in its evaluation. One can notice that the integral looks simplified with a change of variables to $u=\tau'-\tau$ and $\tau$. In particular, with this change of variables, this integral looks like
\begin{eqnarray}\label{eq:Ij-Gaussian-1}
   \mathcal{I} = -\frac{1}{4 \pi ^2}\int_{-\infty}^{\infty}d\tau\,\int_{-\infty}^{\infty}du\, \frac{ e^{-i \omega\, u}}{(u-i \epsilon )^2 } \,e^{-[\tau ^2+(\tau +u)^2]/\Bar{T}^2} ~.
\end{eqnarray}
Here the $\tau$ integral is easily doable and results in 
\begin{eqnarray}\label{eq:Ij-Gaussian-2}
   \mathcal{I} = -\frac{1}{4 \pi ^2}\sqrt{\frac{\pi }{2}} \Bar{T} \int_{-\infty}^{\infty}du\, \frac{ e^{- \left[ i \omega\,u +  u^2/(2\Bar{T}^2)\right]}}{(u-i \epsilon )^2} ~.
\end{eqnarray}
To analytically perform this last integral, we specifically Fourier transform the factor $e^{-  u^2/(2\Bar{T}^2)}=\Bar{T}/(\sqrt{2 \pi })\int_{-\infty}^{\infty}d\kappa\, e^{i \kappa  u-\kappa ^2 \Bar{T}^2/2}$. Then we shall get a factor of $e^{i (\kappa-{\omega})  u}$ in the numerator and $(u-i \epsilon )^2$ in the denominator. The integral will have a pole of order two at $u=i \epsilon$. This integral can be performed through contour integration, with non-vanishing contributions when the contour is taken in the upper half complex plane. When we take the contour in the upper half complex plane, we must satisfy $\kappa-{\omega}>0$ for the damping of other integration. This leads to the expression:
\begin{eqnarray}\label{eq:Ij-Gaussian-3}
   \mathcal{I} &=& -\frac{\Bar{T}^2}{8 \pi ^2} \int_{{\omega}}^{\infty}d\kappa\, e^{-\kappa ^2 \Bar{T}^2/2}\,(i (\kappa-{\omega}) e^{{\omega} \epsilon -\kappa  \epsilon })2\pi i ~\nonumber\\
   &\simeq& \frac{\Bar{T}^2}{4 \pi} \int_{{\omega}}^{\infty}d\kappa\, e^{-\kappa^2 \Bar{T}^2/2}\,(\kappa-{\omega}) ~.
\end{eqnarray}
After carrying out this integration over $\kappa$, we get 
\begin{eqnarray}\label{eq:Ij-Gaussian-4}
   \mathcal{I} &=& \frac{1}{4 \pi}\bigg[ e^{-{\omega}^2 \Bar{T}^2/2} -\sqrt{\frac{\pi}{2}}{\omega}\,\Bar{T}\, \erfc\left(\frac{{\omega}\, \Bar{T}}{\sqrt{2}}\right)\bigg] ~.
\end{eqnarray}
Here one can notice that with the use of the identity $\erfc (z)= \Gamma(1/2,z^2)/\sqrt{\pi}$, this expression matches exactly with the one provided in \cite{LSriramkumar_1996}. Thus we have obtained that the integrals $\mathcal{I}$ are non-vanishing with finite switching, and they depend on the form of the window function. This expression will be used in the expression \eq{eq:negativity-general} along with the non-local terms $\mathcal{I}_{\varepsilon}$ to investigate the entanglement measure negativity. It should be noted that in the infinite interaction time limit, i.e., when $\bar{T}\to \infty$, the entire $\mathcal{I}$ from Eq. (\ref{eq:Ij-Gaussian-4}) vanishes (as $\Bar{T} \erfc\left({\omega}\, \Bar{T}/\sqrt{2}\right)\sim e^{-{\omega}^2 \Bar{T}^2/2}$ when $\bar{T}\to \infty$), providing the usual eternal interaction result for the inertial detectors.

\underline{\textit{Evaluation of the non-local term $\mathcal{I}_{\varepsilon}$}}: 

Now let us focus on evaluating the non-local entangling term $\mathcal{I}_{\varepsilon}$ using the Gaussian switching function $\chi(\tau)=e^{-\tau^2/\Bar{T}^2}$. In this regard, the Hadamard form of the Feynman propagator from Eq. (\ref{eq:FeynmanP-hadamard}) with the expression of the geodesic distance (\ref{eq:geodesic-dist-gen}), is used in the integral of Eq. (\ref{eq:Ie-general}). We should mention that here providing a final analytical expression does not seem possible. However, we have tried to obtain as much analytical results as possible and then numerically evaluate the final form of the integral $\mathcal{I}_{\varepsilon}$.
For the simplicity of calculation, we consider two identical detectors, i.e., ${\omega}^{1}={\omega}^{2}={\omega}$, and then the concerned integral can be represented as
\begin{eqnarray}\label{eq:Ie-Gaussian-Gen-1}
\mathcal{I}_{\varepsilon}=\frac{1}{4\pi^2}\int_{-\infty}^{\infty} \frac{d \tau_{2}^{\prime}}{\gamma} \int_{-\infty}^{\infty}d\tau  \frac{e^{i {\omega}\l(\tau+\tau_{2}^{\prime}/\bar{\gamma}\r)}}{\l[\sigma_{\mathcal{G}}^2(\tau_{2}^{\prime}-\tau,\tau_{2}^{\prime})+i\epsilon\r]} e^{-\tau^2/\Bar{T}^2} e^{-{\tau_{2}^{\prime}}^2/(\bar{\gamma}^2\Bar{T}^2)}~.
\end{eqnarray}
We take the geodesic distance between the detectors $\sigma_{\mathcal{G}}^2(\tau_{2}^{\prime}-\tau,\tau_{2}^{\prime})$ from Eq. (\ref{eq:geodesic-dist-gen}), which was expressed in a form, $\sigma_{\mathcal{G}}^2=-(\tau_{2}^{\prime}-\tau)^2 \mathpzc{f}(\tau_{2}^{\prime})+\mathpzc{g}(\tau_{2}^{\prime})$. We consider a change of variables $\tau_{2}^{\prime}-\tau=\bar{u}$ and $\tau_{2}^{\prime}=\bar{\tau}$, similar to the one done in the previous case for evaluating $\mathcal{I}_{j}$. To perform the integration over $\bar{u}$ we express the factor $e^{-(\bar{u}-\bar{\tau})^2/\Bar{T}^2}$ in the integral in terms of Fourier transform, i.e., $e^{-(\bar{u}-\bar{\tau})^2/\Bar{T}^2} = \Bar{T}/(2\sqrt{ \pi })\int_{-\infty}^{\infty}d\kappa\, e^{i \kappa  (\bar{u}-\bar{\tau})-\kappa ^2 \Bar{T}^2/4}$. The resulting integral looks like
\begin{eqnarray}\label{eq:Ie-Gaussian-Gen-2}
    \mathcal{I}_{\varepsilon} &=& \frac{\Bar{T}}{8\pi^2\sqrt{\pi}}\int_{-\infty}^{\infty} \frac{d \bar{\tau}\,e^{-\bar{\tau}^2/(\bar{\gamma} ^2 \Bar{T}^2})}{\gamma} \int_{-\infty}^{\infty}d\kappa\,e^{-\kappa ^2 \Bar{T}^2/4}e^{i \left\{{\omega} \left(1+1/\bar{\gamma} \right)-\kappa\right\} \bar{\tau}} \,\int_{-\infty}^{\infty}d\bar{u}\,\frac{e^{i (\kappa-{\omega})  \bar{u}}}{ \mathpzc{f}(\bar{\tau})\, \bar{u}^2-\mathpzc{g}(\bar{\tau})-i \epsilon }~.
\end{eqnarray}
This integral has poles at $\bar{u} = \pm \sqrt{\mathpzc{g}(\bar{\tau})+i \epsilon }/\sqrt{\mathpzc{f}(\bar{\tau})}$ of order one, which is respectively in the upper and lower half complex plane depending on the `+' or `-' sign. When the $\kappa>{\omega}$, one should consider a contour in the upper half complex plane and the pole $\bar{u} = \sqrt{\mathpzc{g}(\bar{\tau})+i \epsilon }/\sqrt{\mathpzc{f}(\bar{\tau})}$ would contribute to the non-vanishing integral. Let us denote this pole as $\sqrt{\mathpzc{g}(\bar{\tau})+i \epsilon }/\sqrt{\mathpzc{f}(\bar{\tau})}=\bar{u}_{0}(\bar{\tau})$. Whereas, when $\kappa<{\omega}$, one should consider a contour in the lower half complex plane and the pole {$\bar{u} =-\bar{u}_{0}(\bar{\tau})$} contributes to the non-zero integral. Therefore we will have the expression of the previous integral as
\begin{eqnarray}\label{eq:Ie-Gaussian-Gen-3}
    \mathcal{I}_{\varepsilon} &=& \frac{2\pi\,i\,\Bar{T}}{8\pi^2\sqrt{\pi}}\int_{-\infty}^{\infty} \frac{d \bar{\tau}\,e^{-\bar{\tau}^2/(\bar{\gamma} ^2 \Bar{T}^2})}{\gamma} e^{i {\omega} \left(1+1/\bar{\gamma} \right) \bar{\tau}} \Bigg[\int_{{\omega}}^{\infty}d\kappa\,e^{-\kappa^2 \Bar{T}^2/4} e^{-i \,\kappa \bar{\tau}} \,\frac{e^{i (\kappa -{\omega}) \bar{u}_{0}(\bar{\tau})}}{2\, \mathpzc{f}(\bar{\tau})\, \bar{u}_{0}(\bar{\tau})}~\nonumber\\
    ~&&~~~~~~~~~~~~~~~~~ +\, \int_{-\infty}^{{\omega}}d\kappa\,e^{-\kappa^2 \Bar{T}^2/4} e^{-i \,\kappa \bar{\tau}} \, \frac{e^{-i (\kappa -{\omega}) \bar{u}_{0}(\bar{\tau})}}{2\, \mathpzc{f}(\bar{\tau}) \,\bar{u}_{0}(\bar{\tau})}\Bigg]\, .
\end{eqnarray}
After carrying out these integrations over $\kappa$, the entire non-local term $\mathcal{I}_{\varepsilon}$ looks like 
\begin{eqnarray}\label{eq:Ie-Gaussian-Gen-4}
    \mathcal{I}_{\varepsilon} &=& \frac{i\,\Bar{T}}{4\pi\sqrt{\pi}}\int_{-\infty}^{\infty} \frac{d \bar{\tau}\,e^{-\bar{\tau}^2/(\bar{\gamma} ^2 \Bar{T}^2})}{\gamma} \frac{e^{i {\omega} \left(1+1/\bar{\gamma} \right) \bar{\tau}}}{\bar{T}\, \mathpzc{f}(\bar{\tau}) \,\bar{u}_{0}(\bar{\tau})} \Bigg[i \pi ^{3/2} \exp \left(-\frac{ \bar{u}_{0}(\bar{\tau}) \left(2 \bar{\tau}-i \bar{T}^2 {\omega}\right)+\bar{\tau}^2 +\bar{u}_{0}^2(\bar{\tau}) }{\bar{T}^2}\right) \nonumber\\
    ~&\times& ~ \left\{\erf\left(\frac{2 i \bar{u}_{0}(\bar{\tau})+2 i \bar{\tau}+\bar{T}^2 {\omega}}{2 \bar{T}}\right)+\erfc\left(\frac{-2 i \bar{u}_{0}(\bar{\tau})+2 i \bar{\tau}+\bar{T}^2 {\omega}}{2 \bar{T}}\right) \exp \left(\frac{2\, \bar{u}_{0}(\bar{\tau}) \left(2 \bar{\tau}-i \bar{T}^2 {\omega}\right)}{\bar{T}^2}\right)+1\right\}\Bigg]\, .
\end{eqnarray}
We could not find an analytical result for this integral. However, one can always do a numerical analysis, and here also, we have numerically evaluated this integral.

\underline{\emph{Initially static detectors $(v_0=0)$}} :

Let us first consider the case when the initial coordinate velocity of the detector $B$ in the frame of detector $A$ is zero. In flat spacetime, this corresponds to two static detectors separated by a distance, $d_0$. Since the detectors are static, the separation distance will not change for a flat spacetime configuration. But when these detectors are placed in curved spacetime, the curvature of the spacetime will produce deviation between the initially static detectors. For $v_0=0$, the only curvature tensor component in the expression for geodesic distance (\ref{eq:geodesic-dist-gen}) will be $R_{\xi 0 \xi 0}$. 

We define $R_{\xi 0 \xi 0}:=k_1$ and to make the approximations valid, we are choosing the parameter range in the order, $\Bar{T} \sim d_0,\, \Bar{T} \ll \ell_R$ where $\ell_R$ is the curvature length scale corresponding to the tidal curvature. Scaling all the length scales associated with the problem by $\bar T$ leads to the domain of validity of our calculations for $d_{0}/\bar{T} \sim 1,\, \ell_R/\bar{T} \gg 1$. Furthermore, for $v_0=0$ we have the expressions of the auxiliary functions from Eq. (\ref{eq:exp-f&g}) as $\mathpzc{f}_{0}(\tau_{2}^{\prime}) = 1+ k_{1} d_{0}^2/3 $, and $\mathpzc{g}_{0}(\tau_{2}^{\prime}) = d_{0}^2(1 - k_{1}\, {\tau_{2}^{\prime}}^2)$. Whereas we have from Eqs. (\ref{eq:gamma-tbp}) and (\ref{eq:gamma-bar-tbp}) the expressions $\gamma = (1 + k_{1}\, d_{0}^2 )^{-1/2}$ and $\bar{\gamma} = (1 + k_{1}\, d_{0}^2\,\tau_{2}^{\prime}/2 )^{-1}$.

For initially static detectors, $\mathcal{I}_{\varepsilon}$ depends on the energy gap of the detectors, initial separation, and curvature length scale. In \fig{fig:Ij-Ie-v0e0} we have plotted the local terms $\mathcal{I}$ and the non-local entangling term $|\mathcal{I}_{\varepsilon}|$ as a function of the dimensionless detector transition energy $({\omega}\,\bar{T})$ in the case of $v_{0} = 0$. One should note that the local terms $\mathcal{I}$ from \eq{eq:Ij-Gaussian-4} remain the same in both flat and curved spacetimes and even in the scenario of $v_{0} \neq 0$. The mentioned figure asserts that both the local and the non-local terms decrease with increasing detector transition energy. We also observe that the non-local term is $\sim 10^2$ times larger than the local terms. Therefore, the characteristics of negativity, which is expressed as $\mathcal{N}^{(2)} = |\mathcal{I}_{\varepsilon}|-\mathcal{I}$, will also be dictated by the non-local entangling term.

On the other hand, in \fig{fig:DN-vs-diff-para-v0-Gaussian}, we have plotted the difference in negativity $\Delta\mathcal{N}^{(2)}$ between the curved and flat spacetimes as functions of the dimensionless detector transition energy $({\omega}\,\bar{T})$, the initial separation $d_{0}/\bar{T}$, and the curvature length scale $\ell_R/\bar{T}$. This difference decreases as the curvature length scale increases, i.e., as the curvature itself decreases. The difference $\Delta\mathcal{N}^{(2)}$ also decreases with increasing initial separation $d_{0}/\bar{T}$ between the detectors. Whereas, in Fig. \ref{fig:DN-vs-diff-para-v0-Gaussian-negativeL} we have plotted the difference $\Delta\mathcal{N}^{(2)}$ for opposite sign of curvature than Fig. \ref{fig:DN-vs-diff-para-v0-Gaussian}.

\underline{\emph{General geodesic detectors $(v_0\neq0)$}} :

Let us now talk about the scenario when the initial coordinate velocity of the detector $B$ in the frame of detector $A$ is not zero, i.e., $v_{0}\neq 0$. In that scenario, there are six distinct curvature terms that affect the non-local entangling term $\mathcal{I}_{\varepsilon}$. For instance, let us consider $R_{\xi 0\xi 0}=k_{1}$, $R_{\xi 0n 0}=k_{2}$, $R_{n 0 n 0}=k_{3}$, $R_{\xi 0\xi n}=k_{4}$, $R_{\xi n\xi n}=k_{5}$, and $R_{\xi nn 0}=k_{6}$. Then the quantities $\mathpzc{f}(\tau_{2}^{\prime})$ and $\mathpzc{g}(\tau_{2}^{\prime})$ will be given by
\begin{subequations}\label{eq:}
\begin{eqnarray}\label{}
    \mathpzc{f}(\tau_{2}^{\prime}) &=& 1+\frac{1}{3}\l[ k_{1} d_{0}^2 + 2 d_{0} \l( k_{1} v_{||} + k_{2} v_{\perp} \r)  \tau_{2}^{\prime} + \l( k_{1} v_{||}^2 + 2 k_{2} v_{||} v_{\perp} + k_{3} v_{\perp}^2 \r) {\tau_{2}^{\prime}}^2 \r]~,\\
    \mathpzc{g}(\tau_{2}^{\prime}) &=& \l( d_{0}^2 + v_0^2 {\tau_{2}^{\prime}}^2+ 2d_{0} v_{||} \tau_{2}^{\prime} \r) + d_{0}^2\,\l( \frac{2}{3} k_{4}v_{\perp} - k_{1}   + \frac{1}{3} k_{5} v_{\perp}^2 \r) {\tau_{2}^{\prime}}^2 \nn \\
    && -\frac{4}{3} d_{0} \l( k_{1} v_{||} + k_{2}  v_{\perp} \r) {\tau_{2}^{\prime}}^3 - \frac{1}{3} \l( k_{1} v_{||}^2 + 2 k_{2} v_{||} v_{\perp} + k_{3} v_{\perp}^2 \r) {\tau_{2}^{\prime}}^4~.\label{}
\end{eqnarray}
\end{subequations}  
While the auxiliary functions $\mathpzc{h}_1(v_{0},d_{0}),\, \mathpzc{h}_2(v_{0},d_{0})$ and $ \mathpzc{h}_3(v_{0},d_{0})$ necessary to define $\gamma$ and $\bar{\gamma}$ will be
\begin{subequations}\label{eq:gamma-tbp-hj-2}
\begin{eqnarray}
    \mathpzc{h}_1(v_{0},d_{0}) &:=& k_{1} d_{0}^2 + \frac{2}{3} k_{4} d_{0}^2 v_{\perp} \\
    \mathpzc{h}_2(v_{0},d_{0}) &:=& k_{1} v_{||}  + k_{2} v_{\perp} + \frac{2}{3} v_{\perp}\l( k_{4} v_{||} + k_{6} v_{\perp} \r)  \\
    \mathpzc{h}_3(v_{0},d_{0}) &:=& k_{1} v_{||}^2 + 2 k_{2} v_{||} v_{\perp} + k_{3} v_{\perp}^2~.
\end{eqnarray}
\end{subequations}
We use these expressions to evaluate the entangling term $\mathcal{I}_{\varepsilon}$. The behavior of this term for inertial probes in flat spacetime with the Gaussian switching function is similar to that reported in Ref. \cite{Koga:2018the}. This is expected as we have considered the curvature terms to be perturbatively small. In \fig{fig:DN-vs-diff-para-vne0-Gaussian}, we have plotted the difference in negativity $\Delta\mathcal{N}^{(2)}$ between the curved and flat spacetime in this case with respect to different system parameters. On the other hand, in Figs. \ref{fig:DN-vperp} and \ref{fig:DIE-vs-diff-para-v0-Gauss-Rabcd} we have presented the difference $\Delta\mathcal{N}^{(2)}$ fixing different components of the Riemann tensor and also varying some of these components.

\subsection{de Sitter background}\label{sec:Negativity-de-Sitter}

We
continue our investigation on entanglement in a background that is a maximally symmetric solution of Einstein's equations with a positive cosmological constant, i.e., in the de Sitter spacetime. The motivation behind considering the de Sitter spacetime is twofold. First, the de Sitter era signifies an accelerated expansion of the universe among the general homogeneous and isotropic descriptions given by the Friedmann–Lema\^{i}tre–Robertson–Walker (FLRW) metric. Our universe, as observed, is in accelerated expansion in its current state, and the study of the de Sitter background becomes relevant from this point of view. Second, obtaining an exact expression of the geodesic interval between two geodesic observers is straightforward in de Sitter spacetime due to its maximal symmetry.

In de Sitter spacetime, the line element \cite{Deser:1997ri} in standard coordinates is expressed as
\begin{eqnarray}\label{eq:DS-metric}
    ds^2 &=&- dt^2+R^2 \cosh^2{(t/R)}\Big[d\chi^2+\sin^2{(\chi)}\big\{d\theta^2+\sin^2{(\theta)}\,d\phi^2\big\}\Big]~.
\end{eqnarray}
For static observers in this coordinates of de Sitter spacetime, the coordinate time $t$ acts as a proper time.
We utilize the embedding coordinates for the de Sitter spacetime to perceive a geodesic distance between two static detectors. 
We use these embedding coordinates along with the considerations of constant $\theta$ and constant $\phi$ plane to obtain the geodesic distance. In the de Sitter background, with the identification of $\Delta \chi=d_{0}/R$, this geodesic distance is obtained to be
\begin{eqnarray}\label{eq:DS-geodesic-distance}
    \sigma^{2}_{DS} = \Lambda^{-1}~\cos^{-2}\Big\{-\sinh(\tau\sqrt{\Lambda})\sinh(s\sqrt{\Lambda})+\cosh(\tau\sqrt{\Lambda})\cosh(s\sqrt{\Lambda})\cos(d_{0}\sqrt{\Lambda})\Big\}~,
\end{eqnarray}
where $\Lambda=1/R^2$ and the Ricci scalar in this background is $\mathcal{R}=12\,\Lambda$. In a de Sitter background, considering a massless scalar field with conformal coupling, one can obtain the Green's function, see \cite{Garbrecht:2004du}, as,
\begin{eqnarray}\label{eq:DS-GreenFn}
    G(x,x^{\prime}) &=& -\frac{i}{4\pi^2} \, \frac{\Lambda}{4\sin^2{(\sqrt{\Lambda}\,\sigma_{DS}/2)}}~.
\end{eqnarray}
For the coincidence limit, Green's function reduces to Hadamard form and with $d_0 \to 0$ in \eq{eq:DS-geodesic-distance}, the Wightman function that characterizes the single detector transition can be obtained with appropriate $i\epsilon$ prescription.

\vspace{.5cm}

\textbf{The local and non-local terms in negativity}

\vspace{.5cm}

We use the expression of $\sigma^{2}_{DS}$ from Eq. (\ref{eq:DS-geodesic-distance}) to get the expression of the Wightman function (\ref{eq:DS-GreenFn}). One can notice that for a single detector $t_{A}=t_{B}$ and $d_{0}=0$. In that scenario, with the assumption $\tau-\tau'=u$ and $\tau+\tau'=v$, we can express the local individual detector transition energy denoting terms $\mathcal{I}$ in the de Sitter background as
\begin{eqnarray}\label{eq:DS-Ij}
    \mathcal{I} = -\frac{\Lambda}{32\pi^2}\int_{-\infty}^{\infty} dv\int_{-\infty}^{\infty} du\,\frac{e^{-i\,\omega\,u} e^{-\frac{u^2+v^2}{2\bar{T}^2}}}{\sinh^2{\big\{(\sqrt{\Lambda}\,u/2)-i\,\epsilon\big\}}}~.
\end{eqnarray}
Here one can perform the $v$ integration analytically, which results in $\int_{-\infty}^{\infty} dv\, e^{-v^2/(2\bar{T}^2)} = \sqrt{2\pi}\,\bar{T}$. To perform the $u$ integration, we have used numerical methods. In Fig. \ref{fig:DS-Ij-1}, we have plotted these local terms denoting the individual detector transition probabilities. We elaborately discuss the characteristics of these local terms in a later section of our manuscript.

On the other hand, we use the relation $\sin^2\big\{{(\cos^{-1}{z})/2}\big\}=(1-z)/2$ to get a simplified expression of the Feynman propagator from Eq. (\ref{eq:DS-GreenFn}). Using the expression of the Feynman propagator, and with a change of variables to $\bar{u}=s-\tau$ and $\bar{v}=s+\tau$, we get the non-local entangling term as,
\begin{eqnarray}\label{eq:DS-Ie}
    \mathcal{I}_{\varepsilon} &=& \frac{\Lambda}{8\pi^2}\int_{-\infty}^{\infty} d\bar{v}\int_{-\infty}^{\infty} d\bar{u}\,\frac{e^{i\,\omega\,\bar{v}} e^{-\frac{\bar{u}^2+\bar{v}^2}{2\bar{T}^2}}}{2-\alpha_{+}\,\cosh{(\sqrt{\Lambda}\,\bar{u})}-\alpha_{-}\,\cosh{(\sqrt{\Lambda}\,\bar{v})}+i\,\epsilon\,\Lambda}~,
\end{eqnarray}
where, we have the expressions of $\alpha_{\pm} = \cos{(d_{0}\,\sqrt{\Lambda})} \pm 1$~. Here, we evaluate both the $u$ and $v$ integrals numerically.

In Fig. \ref{fig:DS-DIe-vs-dE-1} we plot the negativity in a de Sitter background, which is the subtraction of the modulus of this non-local entangling term and the local terms. In a later section, we discuss elaborately the characteristics of these plots. Subsequently, in Fig. \ref{fig:DS-DIe-vs-dE-2} we have plotted the difference in negativity between the de Sitter and the Minkowski case for a very small curvature length scale. The discussion on its characteristics and its similarity with the general curved case is discussed in the same later section. We also plot the different components of negativity and understand their characteristics through the plots from Figs. \ref{fig:DS-DIe-vs-dE-3} and \ref{fig:DS-DIe-vs-dE-4}.

\section{Entanglement as a probe of curvature}\label{sec:Entanglement-Probe-Curvature}

In this part, we shall discuss our observations on entanglement in a general curved spacetime, for which we have used the perturbative technique as described in Sec. \ref{sec:Negativity-GenCurved}. 

\begin{figure}
\centering
\includegraphics[width=8.0cm]{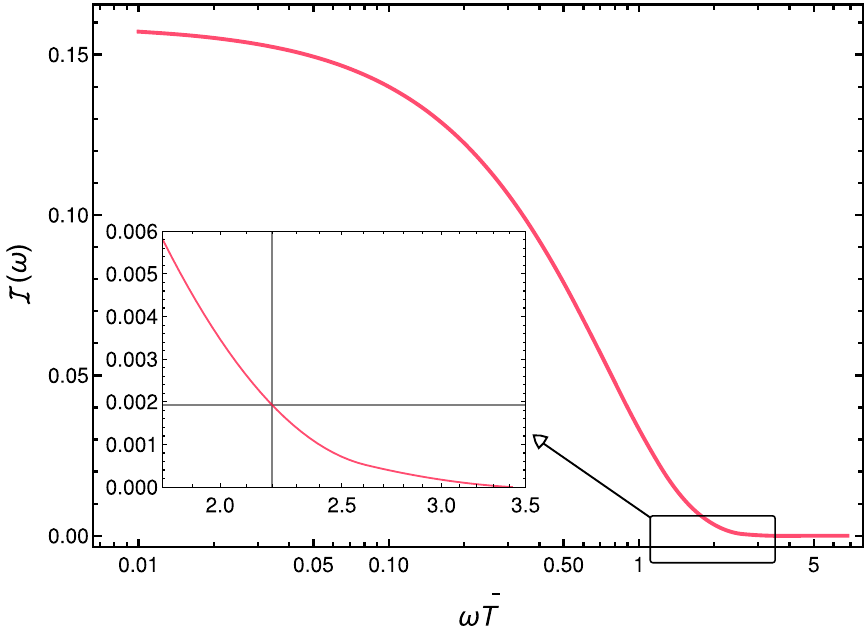}
\hskip 10pt
\includegraphics[width=8.0cm]{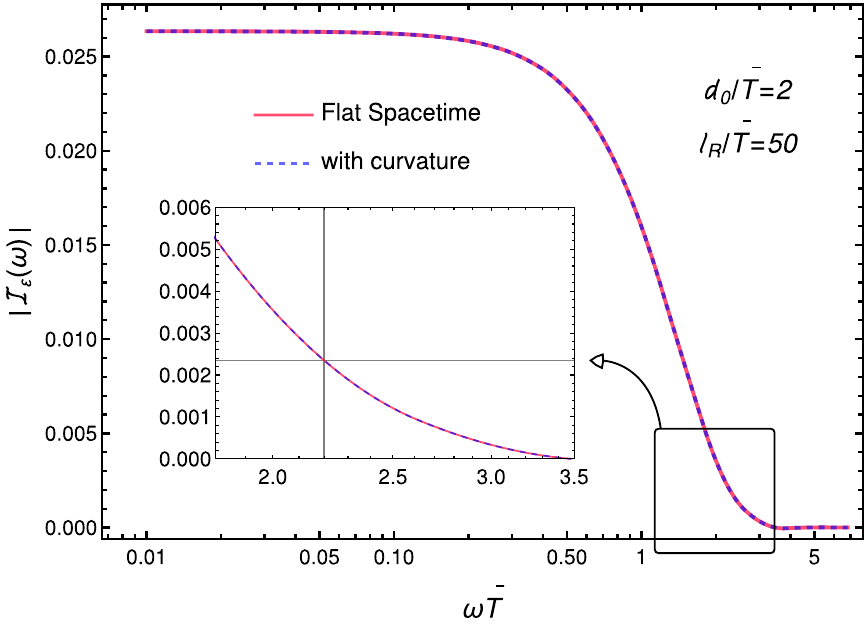}
\caption{The local term $\mathcal{I}$ and the non-local term $|\mathcal{I}_{\varepsilon}|$ (in $v_{0}=0$ scenario) in the negativity are plotted as functions of the dimensionless detector transition energy $\bar{T}\,{\omega}$ for the Gaussian switching function $\chi(\tau)=e^{-\tau^2/\Bar{T}^2}$. The local terms $\mathcal{I}$ signify individual detector transition probabilities and are identical in the flat and the curved backgrounds, at least for the geodesic detectors. From the right figure, we observe that the non-local term $\mathcal{I}_{\varepsilon}$ is indistinguishable in the curved background compared to the flat spacetime. However, there is a perceivable difference between them that will be clear in our subsequent analysis. The epilogue in the figure signifies the appropriate value for the energy we have to choose, such that $|\mathcal{I}_{\varepsilon}|>I$ and hence leads to entanglement between detectors.}
    \label{fig:Ij-Ie-v0e0}
\end{figure}

\vspace{0.5cm}
\underline{\textit{Initially static detectors, $v_0=0$}}:
\vspace{0.5cm}

\begin{figure}
\centering
\includegraphics[width=8.0cm]{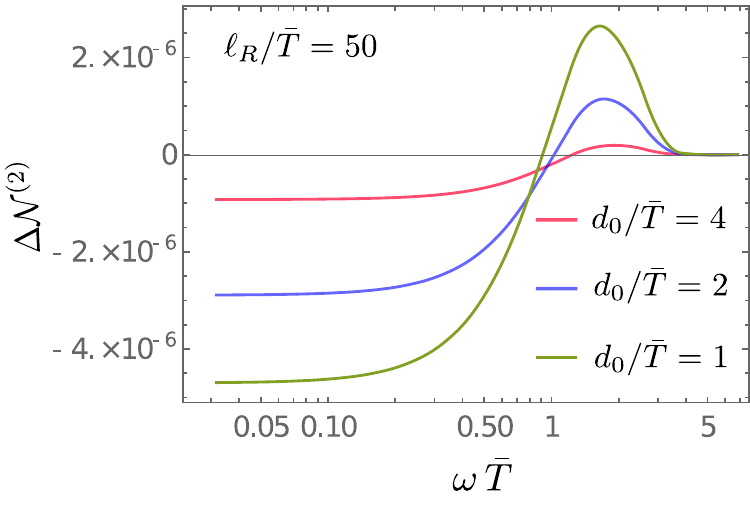}
\hskip 20pt
\includegraphics[width=8.0cm]{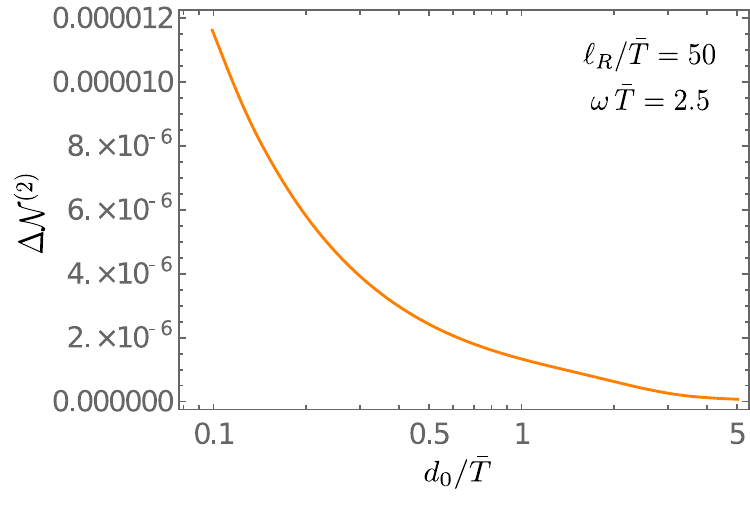}
\vskip 20pt
\includegraphics[width=8.0cm]{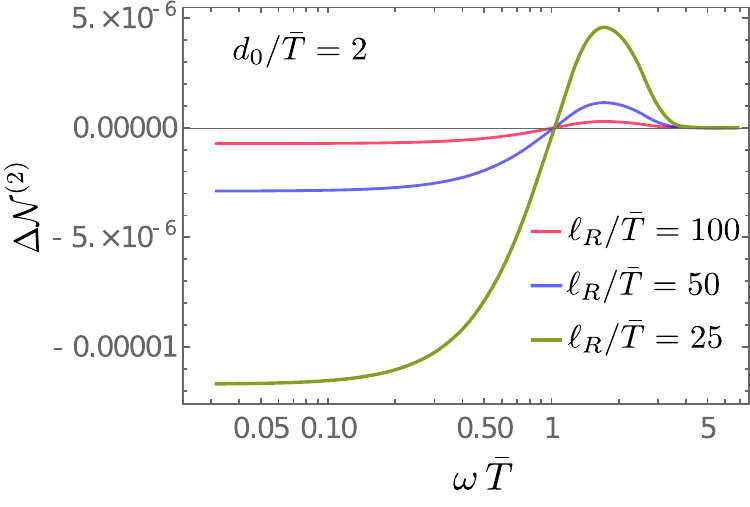}
\hskip 20pt 
\includegraphics[width=7.8cm]{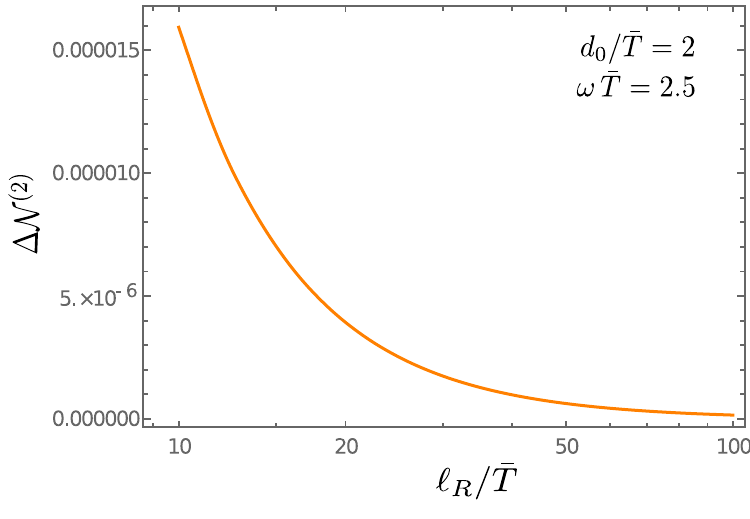}
\caption{The difference in the negativity $\Delta\mathcal{N}^{(2)} =\big[\mathcal{N}^{(2)}\big]_{\text{Curv}}-\big[\mathcal{N}^{(2)}\big]_{\text{Flat}}$ is plotted as functions of the dimensionless detector transition energy ${\omega}\,\bar{T}$ in the two left figures. In the upper left figure, different curves correspond to different initial separations $d_{0}/\bar{T}$, and fixed curvature length scale $\ell_{R}/\bar{T}=50$. While in the lower left plot, different curves correspond to different curvature length scales $\ell_{R}/\bar{T}$, and fixed initial separation $d_{0}/ \bar{T} =2$. In the upper right figure, we have plotted $\Delta\mathcal{N}^{(2)}$ as a function of the initial separation $d_{0}/\bar{T}$. In this plot, the curvature length scale is fixed at $\ell_{R}/\bar{T}=50$. While in the lower right subfigure, we have plotted $\Delta\mathcal{N}^{(2)}$ as a function of the dimensionless curvature length scale, and the separation between the detectors is fixed at $d_{0}/ \bar{T} =2$. This plot signifies that the correction due to the curvature becomes smaller as the curvature length scale increases, i.e., as the curvature decreases. In all of these plots the initial velocity is $v_{0}=0$.} \label{fig:DN-vs-diff-para-v0-Gaussian}
\end{figure}

\begin{figure}
\centering
\includegraphics[width=8.0cm]{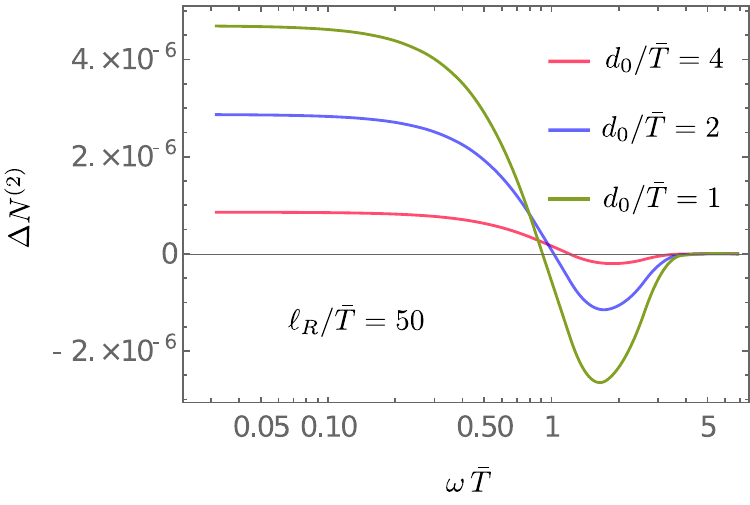}
\hskip 20pt
\includegraphics[width=8.0cm]{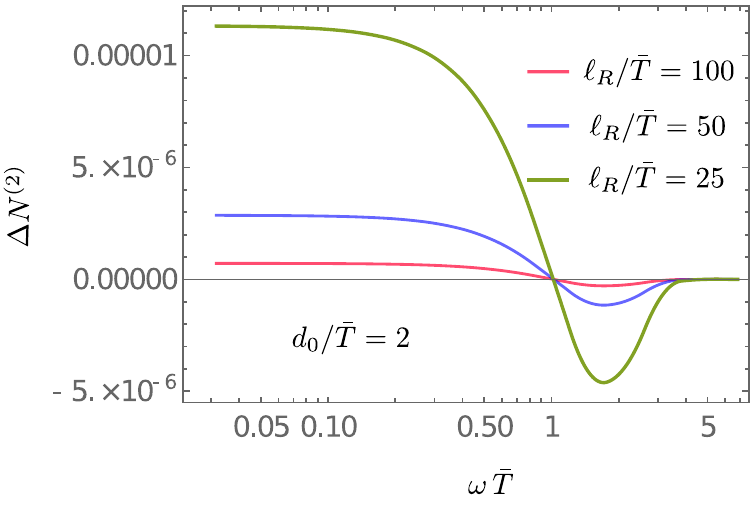}
\caption{ The difference in the negativity $\Delta\mathcal{N}^{(2)} =\big[\mathcal{N}^{(2)}\big]_{\text{Curv}}-\big[\mathcal{N}^{(2)}\big]_{\text{Flat}}$ is plotted as functions of the dimensionless detector transition energy ${\omega}\,\bar{T}$ for negative curvature. One can observe that these plots are flipped upside down compared to the ones from Fig. \ref{fig:DN-vs-diff-para-v0-Gaussian}, i.e., the curved background now reports more entanglement compared to the flat background, unlike the previous case. However, certain characteristics remain the same, like as the initial separation between the detectors increases and the absolute value of curvature decreases, the difference in negativity $\Delta\mathcal{N}^{(2)}$ from the flat background decreases.} 
\label{fig:DN-vs-diff-para-v0-Gaussian-negativeL}
\end{figure}

In Fig. \ref{fig:Ij-Ie-v0e0}, we have plotted the local individual detector transition probability denoting term $\mathcal{I}$ and the modulus of the non-local entangling term $|\mathcal{I}_{\varepsilon}|$. As we are dealing with a nearly flat spacetime background, the individual detector transition probabilities $\mathcal{I}$ from \eq{eq:Ij-Gaussian-4} are the same as Minkowski. 
We will see that, in de Sitter background, this quantity depends on the curvature, but for a very small curvature, like the one considered here, it does not change and remains the same as the one obtained in Fig. \ref{fig:Ij-Ie-v0e0}. 
From the right plot of Fig. \ref{fig:Ij-Ie-v0e0}, it seems that $|\mathcal{I}_{\varepsilon}|$ from the curved and Minkowski backgrounds overlap. However, there is a subtle difference in these curves, which will be prominent in the low transition energy regime, see Fig. \ref{fig:DN-vs-diff-para-v0-Gaussian}.

In Fig. \ref{fig:DN-vs-diff-para-v0-Gaussian}, we have plotted the difference in negativity $\Delta\mathcal{N}^{(2)}$ between the curved and flat backgrounds as a function of the dimensionless detector energy gap for initial static detectors. In the top row, we have depicted the variation of $\Delta\mathcal{N}^{(2)}$ with respect to different initial detector separations. Whereas, in the lower plots, the variation of $\Delta\mathcal{N}^{(2)}$ is shown for different curvature. These lower plots signify that as the curvature length scale increases, i.e., the curvature decreases, the negativity profile from the curved background tends to resemble the flat spacetime negativity, which is expected and consistent in our analysis.

From the top row of Fig. \ref{fig:DN-vs-diff-para-v0-Gaussian}, we observe that the negativity becomes more Minkowski-like as the initial separation $d_{0}$ between the detectors increases. To elaborate on this, for a smaller initial separation between the detectors, the curvature effects in the entanglement become prominent. At first glance, this outcome may seem a bit counter-intuitive, as some of the curvature corrections in our perturbative analysis, see \eq{eq:exp-f&g} and \eq{eq:gamma-tbp-hj}, arrive with multiplicative $d_{0}$ terms. Thus curvature corrections should increase with increasing $d_{0}$, which is not observed from our plots. However, with a deeper investigation, one can perceive that this is not unphysical. The reason can be found in the fact that the correlation between the detectors degrades with increasing distance. If this degradation is more than the increasing curvature effects, then the negativity profile from the curved background tends to become Minkowski-like as the separation increases.
Further, for very small initial separations, the entanglement due to field-mediated communication channels will dominate and if the detectors are placed farther apart, the total negativity itself will decrease. Even though the genuine entanglement starts to dominate after, $d_0\ge 2 c\,\bar{T}$, the net negativity is small.

With an increasing distance between inertial detectors, the decay of correlation is reported in \cite{Koga:2018the} for eternal switching, i.e., $\chi(\tau)=1$. Moreover, when one considers a finite switching like the Gaussian switching considered in our present work, this decay of correlation with increasing detector separation becomes rapid. To support this argument, one can look into the expressions of the non-local terms $\mathcal{I}^{M}_{\varepsilon} ({\omega})$ and $\mathcal{I}^{C}_{\varepsilon} ({\omega})$ from Appendix \ref{Appn:perturbative-Ie} (Eqs. \ref{eq:Appn-IeM-2} and \ref{eq:Appn-IeC-2}), which respectively have $e^{-d_{0}^2/\Bar{T}^2}$ and $e^{-d_{0}^2/(2\Bar{T}^2)}$ terms multiplied in them.
To further elucidate this, we refer the reader to Fig. \ref{fig:Geodesic-trajectory-distance}, where one can observe that as two geodesic trajectories move away from each other, the detectors moving on these curves interact less with each other with finite switching. Therefore, with increasing initial separation between the detectors, one should get diminished negativity and negativity like near flat spacetime.

Furthermore, in Fig. \ref{fig:DN-vs-diff-para-v0-Gaussian-negativeL}, we have plotted the difference in negativity $\Delta\mathcal{N}^{(2)}$ for two initial static detectors for opposite sign of curvature than Fig. \ref{fig:DN-vs-diff-para-v0-Gaussian}. This plot shows that the entanglement from the curved spacetime is now greater than the flat background. Therefore, if in the previous case (Fig. \ref{fig:DN-vs-diff-para-v0-Gaussian}) the curvature moved the geodesic trajectories apart, in the present scenario (Fig. \ref{fig:DN-vs-diff-para-v0-Gaussian-negativeL}) it can be considered to get closer. Moreover, we observed that the overall behavior of negativity to the variation of initial separation and curvature remains the same. As the initial separation between the detectors increases and the absolute value of curvature decreases, the difference in negativity $\Delta\mathcal{N}^{(2)}$ between the curved and flat backgrounds decreases.

\vspace{0.5cm}
\underline{\textit{General geodesic detectors, $v_0\ne0$}}:
\vspace{0.5cm}
\begin{figure}[!h]
   \centering
    \includegraphics[width=5.25cm]{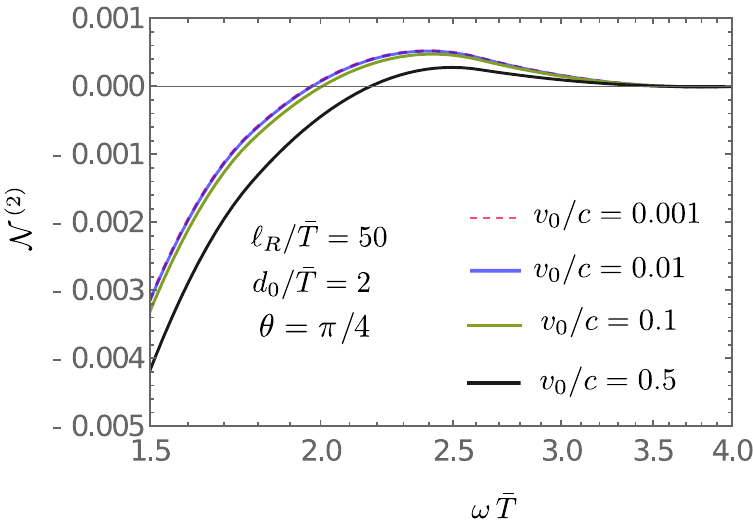}
    \hskip 10pt
    \includegraphics[width=5.5cm]{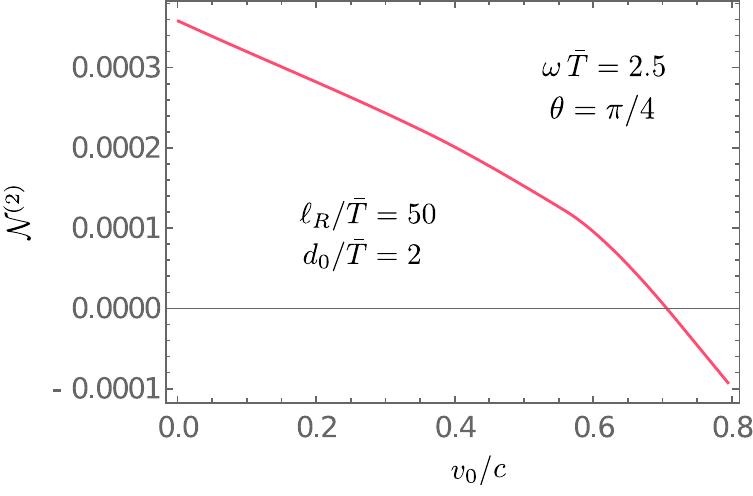}
    \hskip 10pt
    \includegraphics[width=5.5cm]{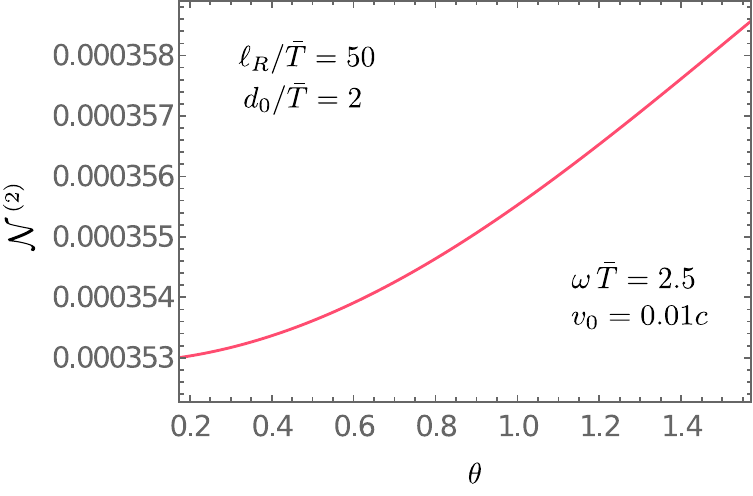}
    \vskip 10 pt
    \includegraphics[width=5.25cm]{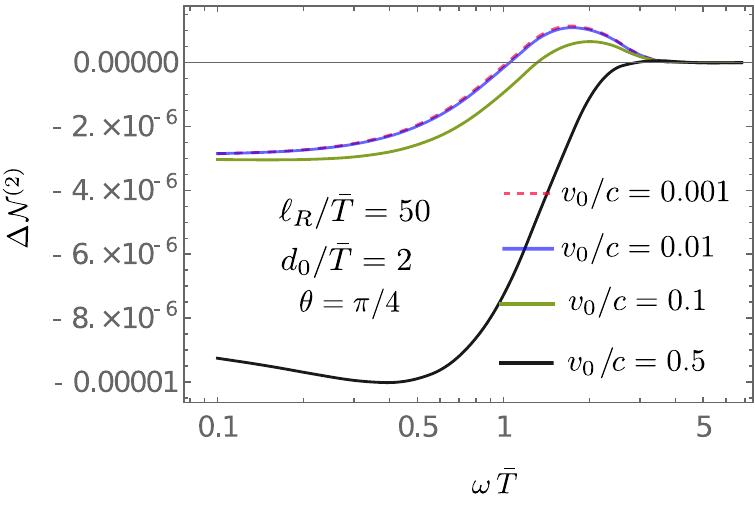}
    \hskip 10pt
    \includegraphics[width=5.5cm]{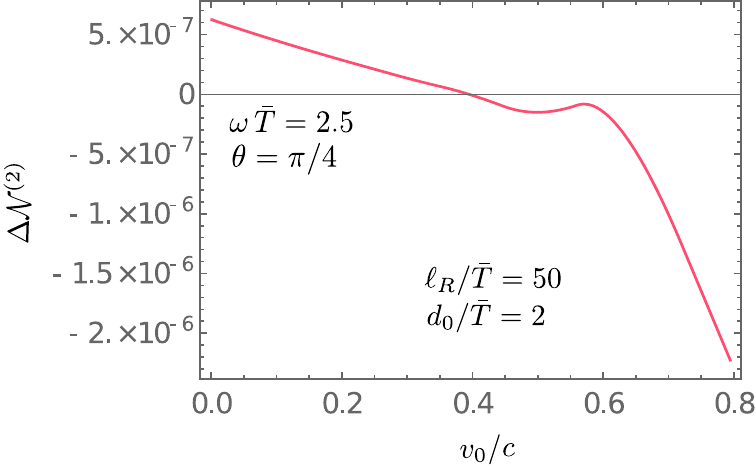}
    \hskip 10pt
    \includegraphics[width=5.5cm]{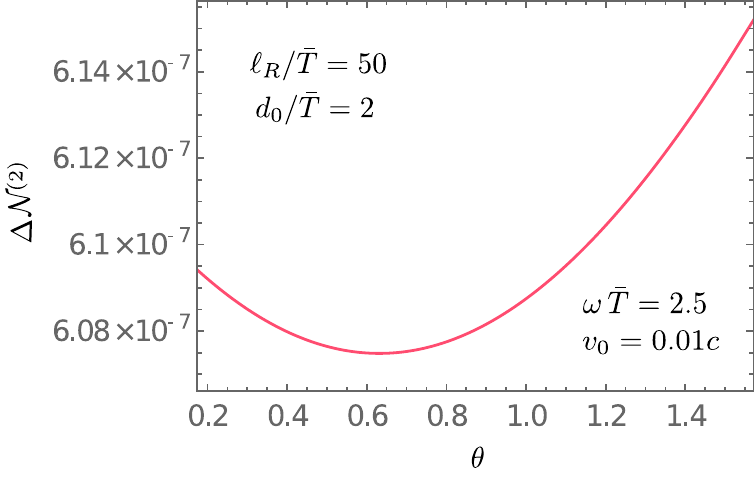}
    \caption{\textbf{Top row:} Behavior of $\mathcal{N}^{(2)}$ as functions of ${\omega}\, \bar{T}$, initial detector velocity $v_{0}/c$ and the angle $\theta$ between the initial separation and the velocity vector. In the extreme left plot, we have considered different initial velocities of the second detector. The angle between the initial velocity and the initial separation is taken as $\theta=\pi/4$, and the curvature length scale for all the components of the Riemann tensor is $\ell_R/\bar{T}=50$, and the initial separation is $d_0/\bar{T}=2$. In the right of both the top and bottom rows, the initial velocity is taken as $v_0=0.01 c$ for the second detector, and the energy gap as $\omega\,\bar{T} =2.5$.
    \newline \textbf{Bottom row:} Behavior of $\Delta\mathcal{N}^{(2)}=\big[\mathcal{N}^{(2)}\big]_{\text{Curv}}-\big[\mathcal{N}^{(2)}\big]_{\text{Flat}}$ as functions of ${\omega}\, \bar{T}$, initial detector velocity $v_{0}/c$ and the angle $\theta$ between the initial separation and the velocity vector. We have chosen the same parameter values as the top row.}\label{fig:DN-vs-diff-para-vne0-Gaussian}
\end{figure}

\begin{figure}[]
   \centering
    \includegraphics[width=16cm]{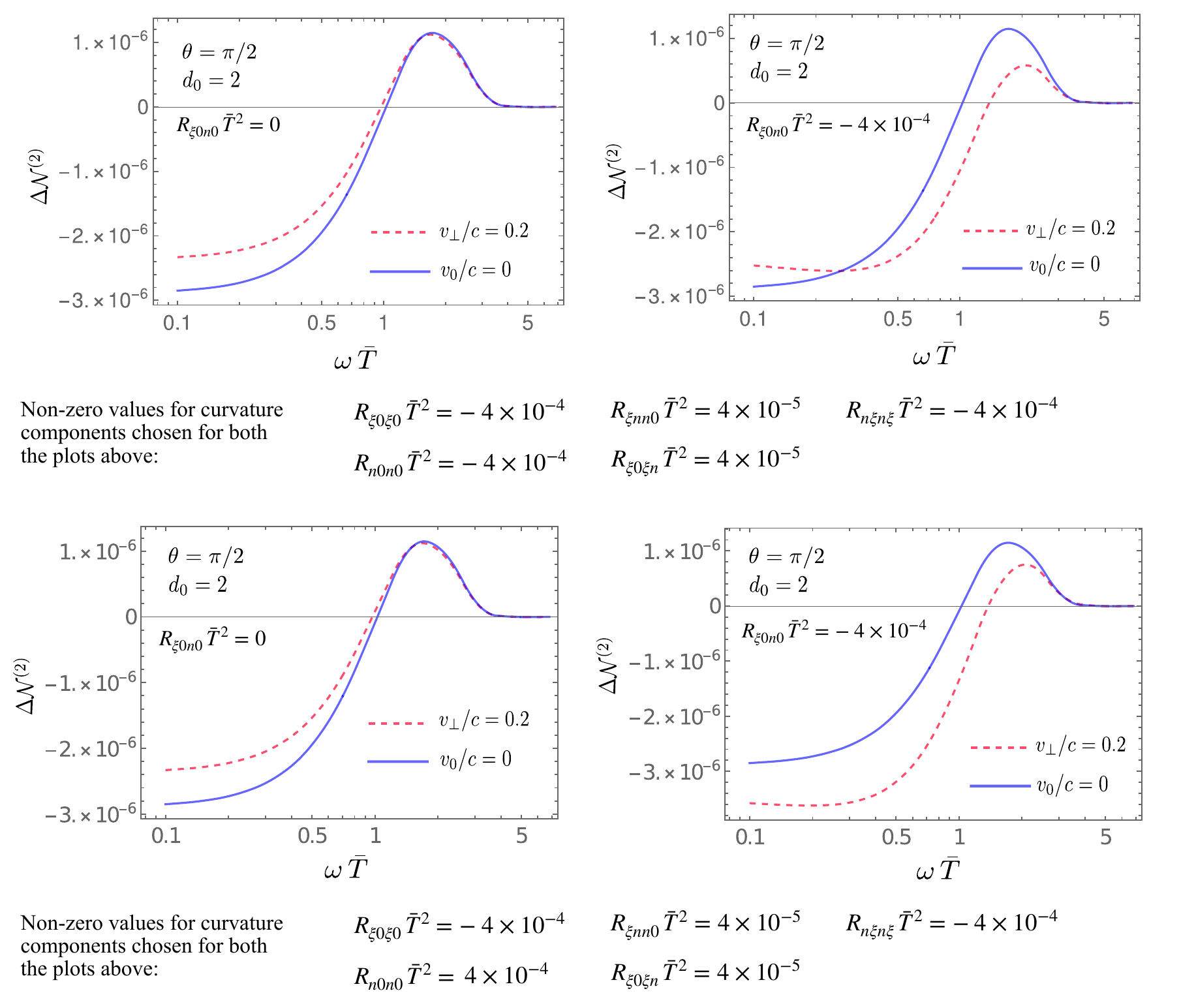}
    \caption{In both the above plots, we have considered the initial configuration such that the relative velocity of the detectors is perpendicular to their relative separation; that is, $\theta=\pi/2$. This facilitates isolating a combination of Riemann tensor components. When $v_0=0$, the only Riemann component that contributes is $R_{\xi 0 \xi 0}$. For $\theta=\pi/2$, if we choose $\bm \xi$ and $\bm n$ along the eigen-directions of the electric part of Riemann, the difference from the initially static $(v_0=0)$ case arises solely due to magnetic parts of the Riemann tensor. This is depicted in the left plot. In the right plot, all the electric parts of the Riemann tensor are considered in the same order of magnitude.}
\label{fig:DN-vperp}
\end{figure}

\begin{figure*}
\centering
\includegraphics[width=8.5cm]{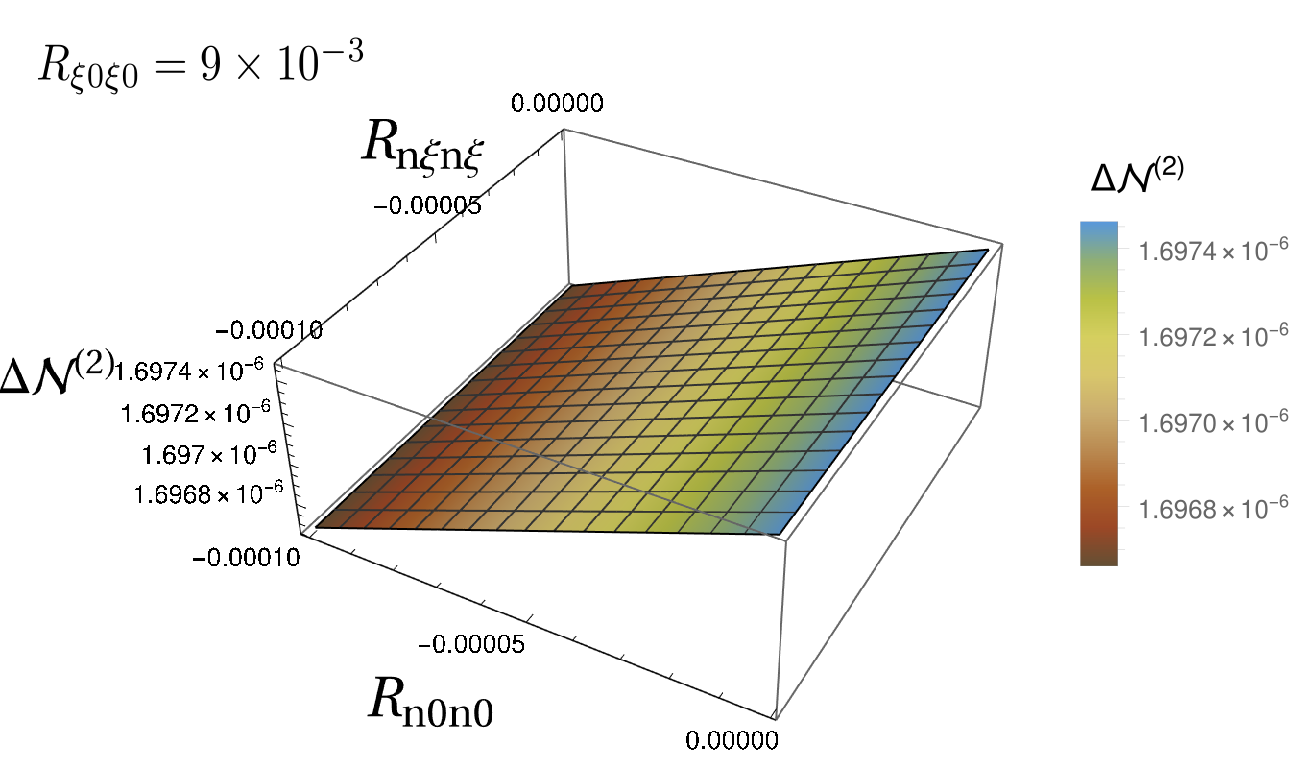}
\hskip 10pt
\includegraphics[width=8.5cm]{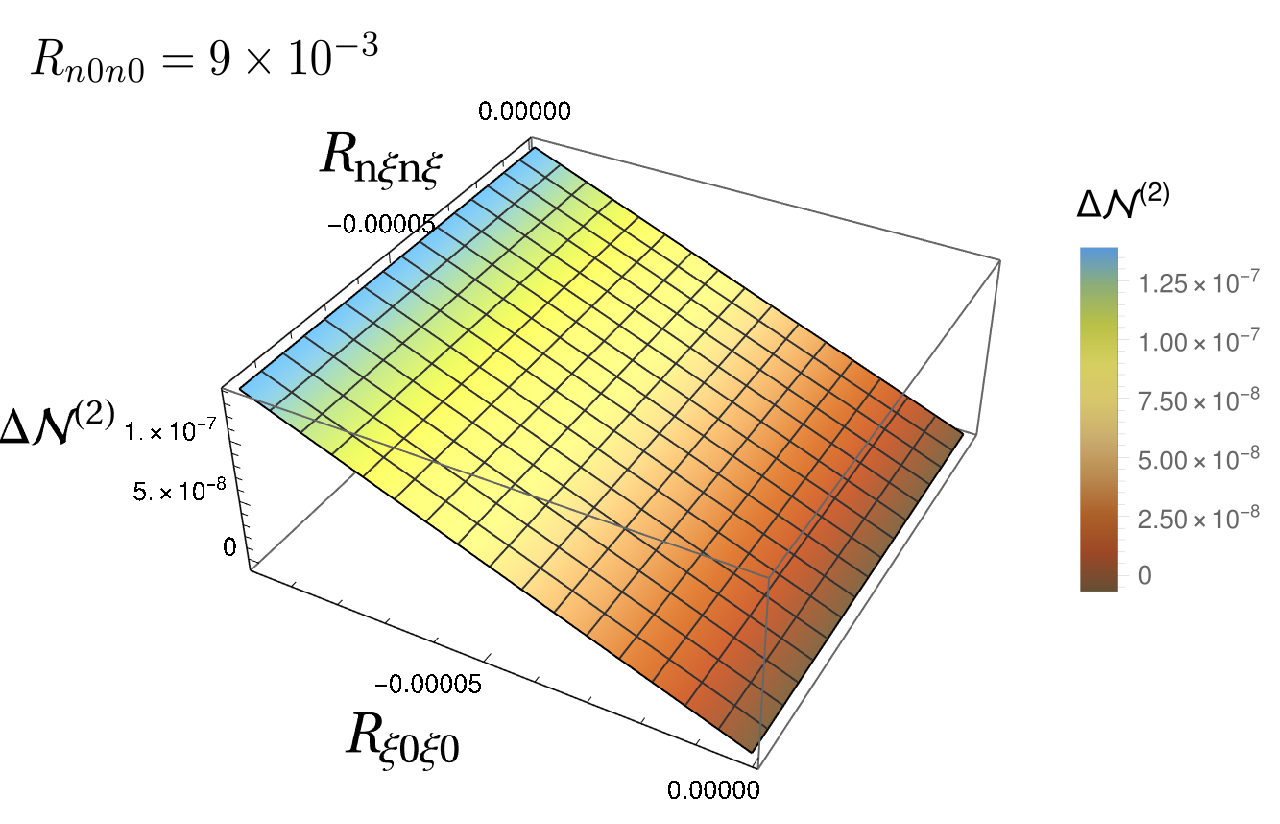}
\vskip 10pt
\includegraphics[width=8.5cm]{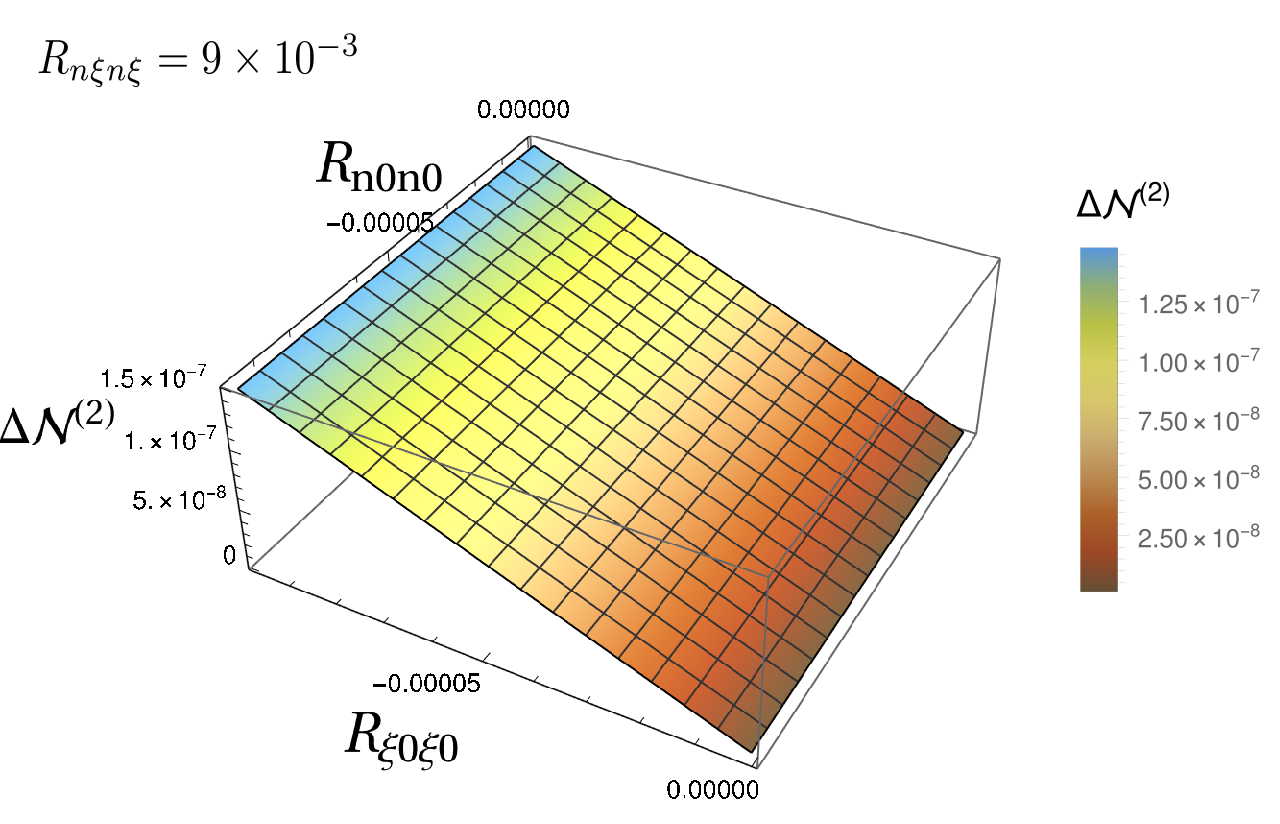}
\caption{The difference in negativity between the curved and flat spacetimes $\Delta\mathcal{N}^{(2)}$ is plotted as functions of different electric components of the curvature tensor, i.e., with respect to $R_{\xi 0 \xi 0}$, $R_{n 0 n 0}$, and $R_{n \xi n \xi}$. The initial separation is chosen as $d_0/\bar{T}=2$ and the energy gap as $\omega\,\bar{T}=2.5$. In the upper left plot, $R_{\xi 0 \xi 0}$ is kept fixed, while in the upper right plot, $R_{n 0 n 0}$ is fixed. On the other hand, in the lower plot, $R_{n \xi n \xi}$ is kept fixed. The upper two plots indicate no visible variation of $\Delta\mathcal{N}^{(2)}$ with respect to the particular curvature term $R_{n \xi n \xi}$. While there is a finite visible change in $\Delta\mathcal{N}^{(2)}$ when $R_{\xi 0 \xi 0}$ or $R_{n 0 n 0}$ is varied. From the lower plot, it is clear that the change in $\Delta\mathcal{N}^{(2)}$ due to the change in $R_{\xi 0 \xi 0}$ is greater compared to $R_{n 0 n 0}$.}
    \label{fig:DIE-vs-diff-para-v0-Gauss-Rabcd}
\end{figure*}

In Fig. \ref{fig:DN-vs-diff-para-vne0-Gaussian}, in the left column, we have plotted $\mathcal{N}^{(2)}$ and $\Delta\mathcal{N}^{(2)}$ as functions of the dimensionless detector transition energy for two detectors with different initial velocities respectively. From \eq{eq:exp-f&g} and \eq{eq:gamma-tbp-hj}, one can observe that now, with velocity, many more different curvature terms will contribute to the negativity profile. In the first two figures on the top row of Fig. \ref{fig:DN-vs-diff-para-vne0-Gaussian}, negativity decreases with an increase in velocity, and the same is true if one varies the angle between initial separation and the initial velocity of the two detectors. The change in negativity, as given in the bottom row of Fig. \ref{fig:DN-vs-diff-para-vne0-Gaussian} shows that $\Delta\mathcal{N}^{(2)}$ degrades as initial velocity increases. The change in negativity decreases and then increases when the angle changes from small angles to $\theta=\pi/2$. As the velocity increases, the distance between detectors will increase rapidly and hence the correlation degrades. If the angle is chosen very small, the detectors can come close and entanglement due to communication can be enhanced. Hence choosing larger angles makes sure there is no causal contact between the detectors.

In Figs. \ref{fig:DN-vs-diff-para-v0-Gaussian}, \ref{fig:DN-vs-diff-para-v0-Gaussian-negativeL}, \ref{fig:DN-vs-diff-para-vne0-Gaussian}, \ref{fig:DS-DIe-vs-dE-1}, and \ref{fig:DS-DIe-vs-dE-2} we have plotted either the difference in negativity or the difference in the modulus of the non-local entangling term. In these particular figures, we observed that as the dimensionless detector energy exceeds certain values, in most cases as ${\omega}\,\bar{T}>1$, there is a sudden change in behavior in the entanglement profiles. For instance, in Fig. \ref{fig:DN-vs-diff-para-v0-Gaussian}, we observe that for ${\omega}\,\bar{T}<1$, the curved space entanglement is lower than the flat space, but for ${\omega}\,\bar{T}>1$ the curved space entanglement becomes higher than the flat space entanglement. It is not unusual to believe that there could be some interesting physics hidden behind these phenomena.

Fig. \ref{fig:DN-vperp} portrays the effects due to the magnetic part of the Riemann tensor. The choice of orientation can be used to isolate certain combinations of Riemann tensor components in the expression for negativity. This can be employed as a probe of curvature, especially those components which are otherwise difficult to determine through other experiments.
 
In Fig. \ref{fig:DIE-vs-diff-para-v0-Gauss-Rabcd}, the horizontal axes correspond to different curvature terms. Interestingly in \ref{fig:DIE-vs-diff-para-v0-Gauss-Rabcd}, we observed that the dependence of the negativity profile on different curvature terms is not the same. For instance, with respect to $R_{n\xi n\xi}$ the negativity profiles do not alter at all. While with respect to $R_{n0 n0}$ and $R_{\xi 0 \xi 0}$ the negativity profiles change, but again the change is larger when $R_{\xi 0 \xi 0}$ varies compared to $R_{n0 n0}$. This may provide a means to distinguish between different curved backgrounds in terms of entanglement. 

We would also like to refer to Fig. \ref{fig:Ngen-3D-wrt-v0th} in Appendix \ref{Appn:vel} that depicts qualitative behaviour of the negativity in a general curved spacetime as the initial velocity $v_{0}$ and the angle $\theta$ varies.

\section{Characteristics of $\mathcal{N}^{(2)}$ in de Sitter spacetime}\label{sec:Observations-discussion}

In this section, we will describe several observations about negativity in the de Sitter background, utilizing the expression of Sec. \ref{sec:Negativity-de-Sitter}.
In Sec. \ref{sec:Negativity-de-Sitter}, we have estimated the necessary quantities to investigate the entanglement in a de Sitter background. We will now use numerical plots to extract key features from the resultant expressions.

\begin{figure*}[!h]
\centering
\includegraphics[width=8.0cm]{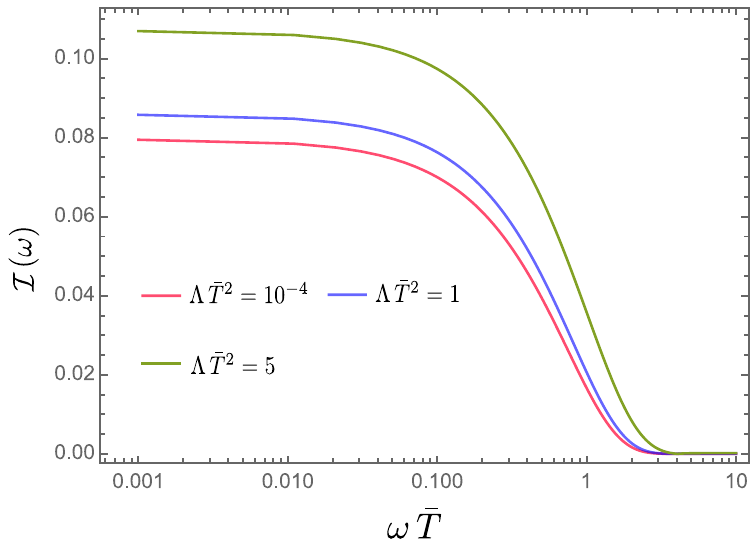}
\hskip 10pt
\includegraphics[width=8.0cm]{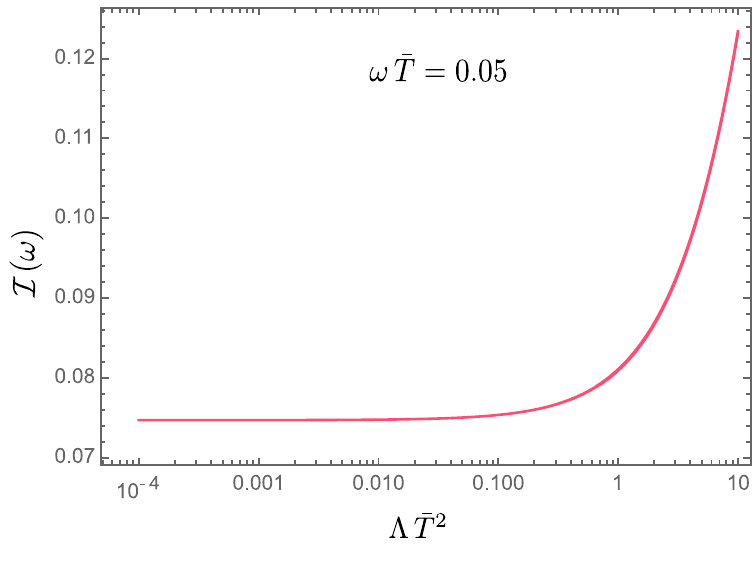}
\caption{The local term $\mathcal{I}$ in entanglement, which denotes individual detector transition probability, is plotted as functions of the detector transition energy and curvature for static detector in de Sitter background. The transition probability depends on the curvature constant. As the curvature increases, the transition probability increases. However, this increment is prominent only in the $\Lambda\,\bar{T}^2>1$ regime, as observed from the right plot.}
    \label{fig:DS-Ij-1}
\end{figure*}

\underline{\textit{Individual detector response}}:

In Fig. \ref{fig:DS-Ij-1}, we have plotted the local terms $\mathcal{I}$ that denote individual detector transition probabilities. One can observe that these probabilities decrease with increasing energy gap and with increasing curvature length scale. However, for any curvature scale $\Lambda\,\bar{T}\ll 1$, it remains almost the same and quantitatively similar to the ones obtained in Minkowski background. In general curved spacetimes, the characteristics of the local terms are expected to be qualitatively similar to the de Sitter ones for large curvature length scales (for geodesic detectors, these will come entirely from the van Vleck determinant).

\begin{figure*}[!h]
\includegraphics[width=8.0cm]{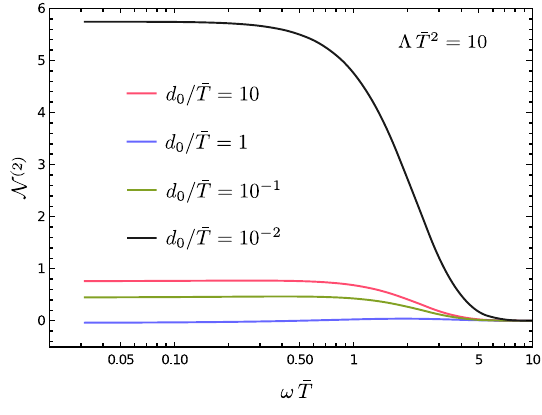}
\hskip 10pt
\includegraphics[width=8.0cm]{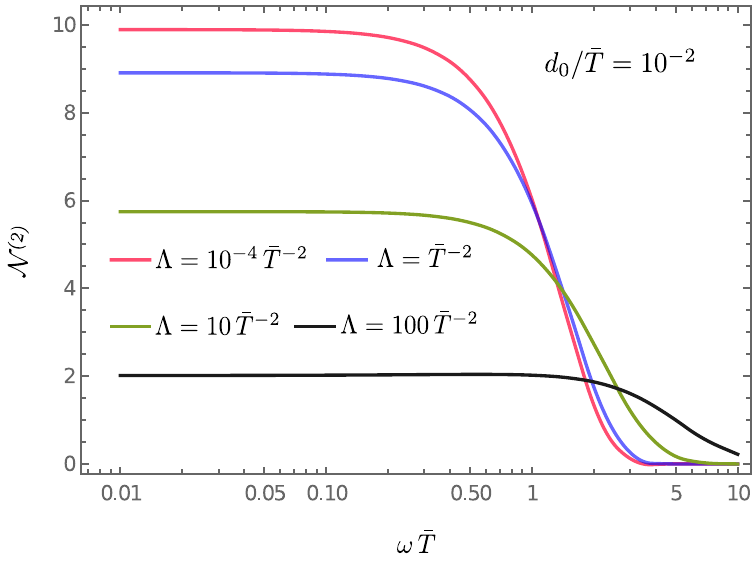}
\caption{In both the above plots we have plotted the negativity $\mathcal{N}^{(2)}$ as a function of the dimensionless detector energy gap ${\omega}\,\bar{T}$ in a de Sitter background. In particular, on the left, we have depicted negativity $\mathcal{N}^{(2)}$ for different initial separations between the detectors. On the right, we have plotted the negativity in the de Sitter background for different curvature length scales. We should mention that the local terms $\mathcal{I}$, along with the non-local terms $\mathcal{I}_{\varepsilon}$, are also affected by high curvature, which results in significant distortion in the negativity profiles. The right plot suggests that one can observe entanglement due to curved backgrounds using detectors with a sufficiently large energy gap.}
    \label{fig:DS-DIe-vs-dE-1}
\end{figure*}

\underline{\textit{Entanglement between the detectors}}:

In Fig. \ref{fig:DS-DIe-vs-dE-1} we have plotted the negativity in a de Sitter background as functions of the dimensionless detector transition energy $({\omega}\,\bar{T})$. The left plot here depicts curves with different values of the initial separation, and the right plot gives curves for a fixed initial separation, but in different curved backgrounds. One can observe that, for a fixed energy gap, as the separation increases, at first the negativity decreases (from black $\to$ green $\to$ blue curve), and then increases (blue $\to$ red curve). Therefore, the behaviour of negativity is \textit{not monotonous} with respect to the initial separation. A similar feature also holds when one looks at negativity as a function of curvature, with fixed initial separation (the plot on the right).
From the right plot, we observe that in the low detector transition energy regime, the negativity decreases monotonically as the background curvature increases. However, there exists a band of energy gap (roughly $1 \lesssim \omega {\bar T} \lesssim 3$ in the given plot), over which negativity is no longer a monotonous function of curvature. We should highlight that, our results are consistent with those in \cite{Kukita:2017etu}, when restricted to the parameter range used in that paper.

\underline{\textit{Imprints of curvature on negativity}}:

We can combine the plots and discussion from previous subsections to explore the possibility of using entanglement as a probe of background curvature, which we now discuss in the context of de Sitter. 
One of the main objectives behind this analysis is to provide qualitative, and to some extent, quantitative, similarities between de Sitter and general curved backgrounds with small curvature. Since our de Sitter results do not assume small $\Lambda$, we can also use them to validate our perturbative technique.

\begin{figure*}[!h]
\centering
\includegraphics[width=8.25cm]{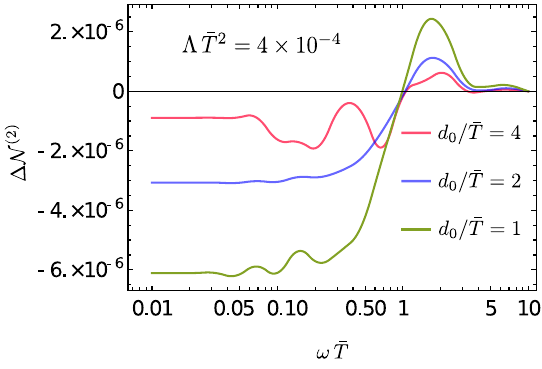}
\hskip 10pt
\includegraphics[width=8.5cm]{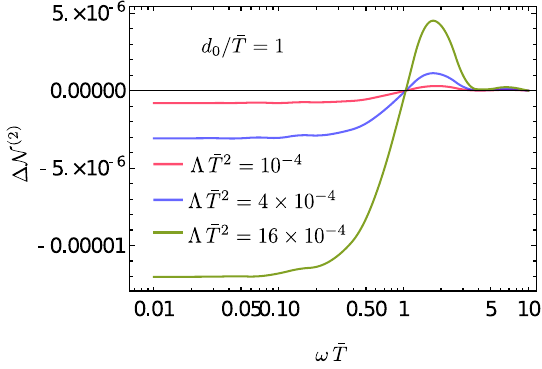}
\caption{The difference in the negativity between the de Sitter and flat spacetimes $\Delta\mathcal{N}^{(2)}$ is plotted as a function of the dimensionless detector transition energy ${\omega}\,\bar{T}$. In both plots, we have considered the curvature length scale to be very large, i.e., the curvature itself is very small so that one can compare these results with the ones from our perturbative approach. Comparing these plots with the perturbative approach of generic curved spacetime, we observe that the plots from the two cases are qualitatively and quantitatively similar.}
    \label{fig:DS-DIe-vs-dE-2}
\end{figure*}

In Fig. \ref{fig:DS-DIe-vs-dE-2} we have plotted the difference in negativity between the de Sitter and Minkowski backgrounds. Unlike in Fig. \ref{fig:DS-DIe-vs-dE-1}, here we have chosen small values for curvature which will facilitate comparison with the perturbative results for general curved backgrounds in the later section. 

\begin{figure*}[!h]
\includegraphics[width=5.5cm]{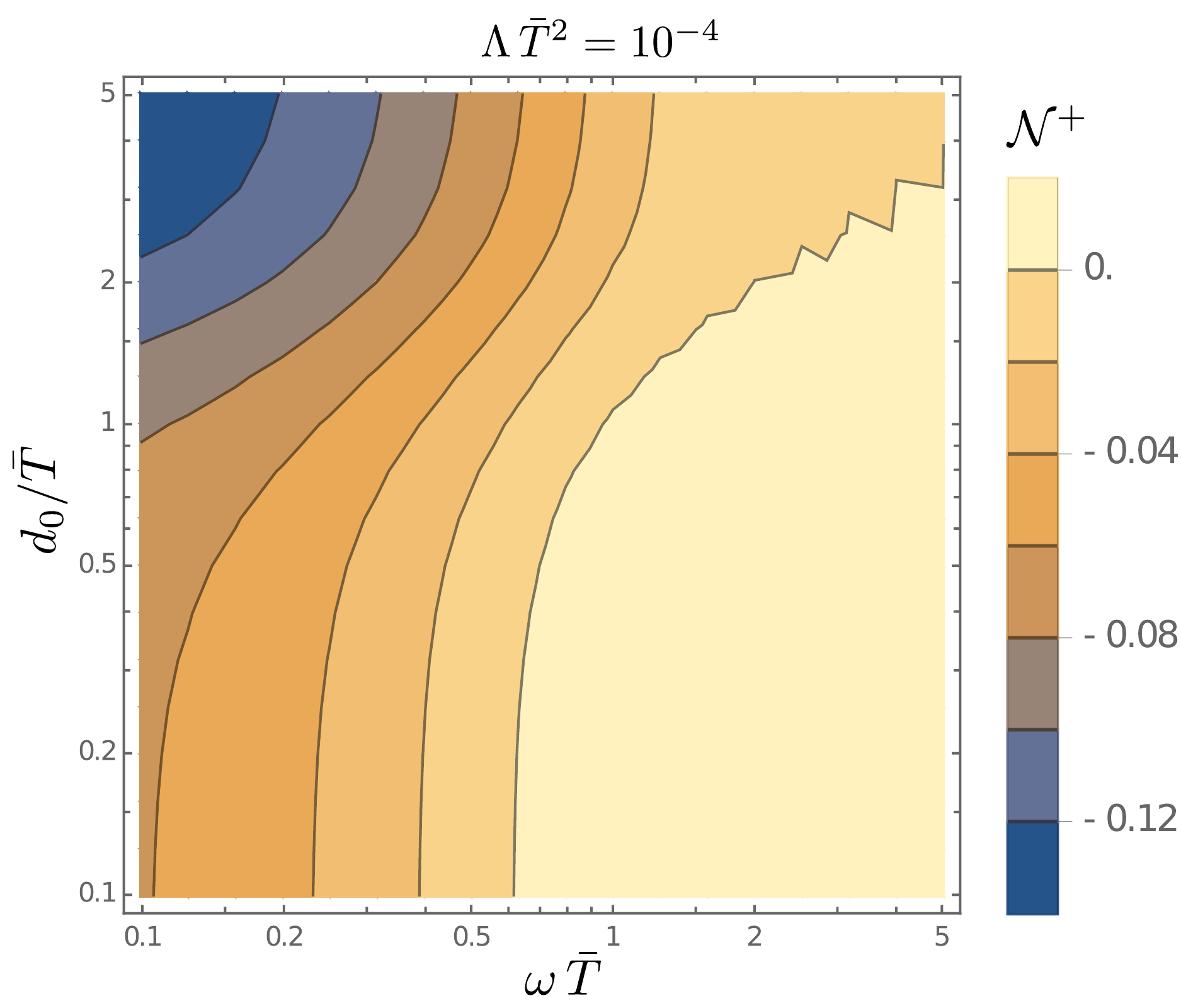}
\hskip 10pt
\includegraphics[width=5.5cm]{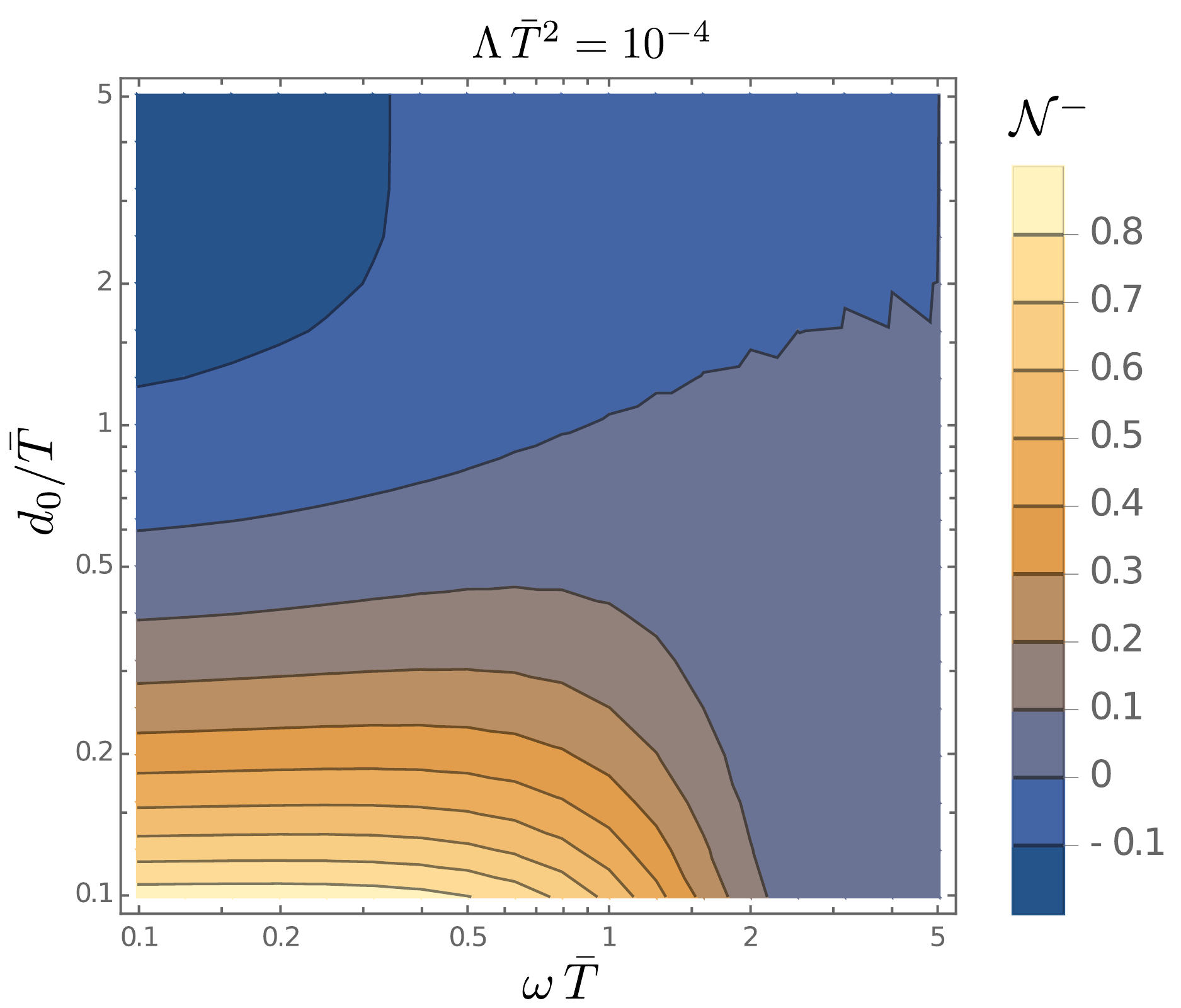}
\hskip 10pt
\includegraphics[width=5.5cm]{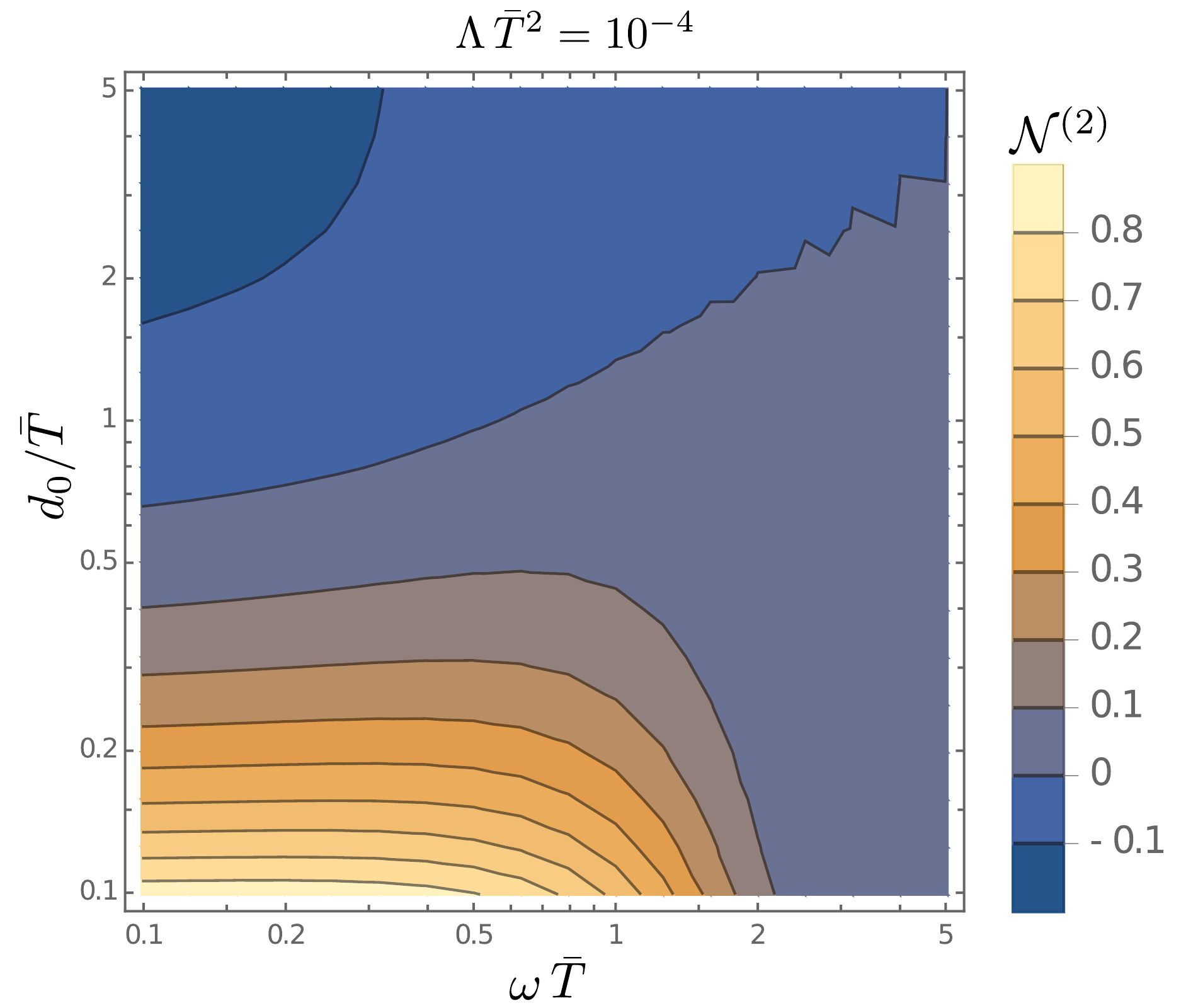}
\vskip 10pt
\includegraphics[width=5.5cm]{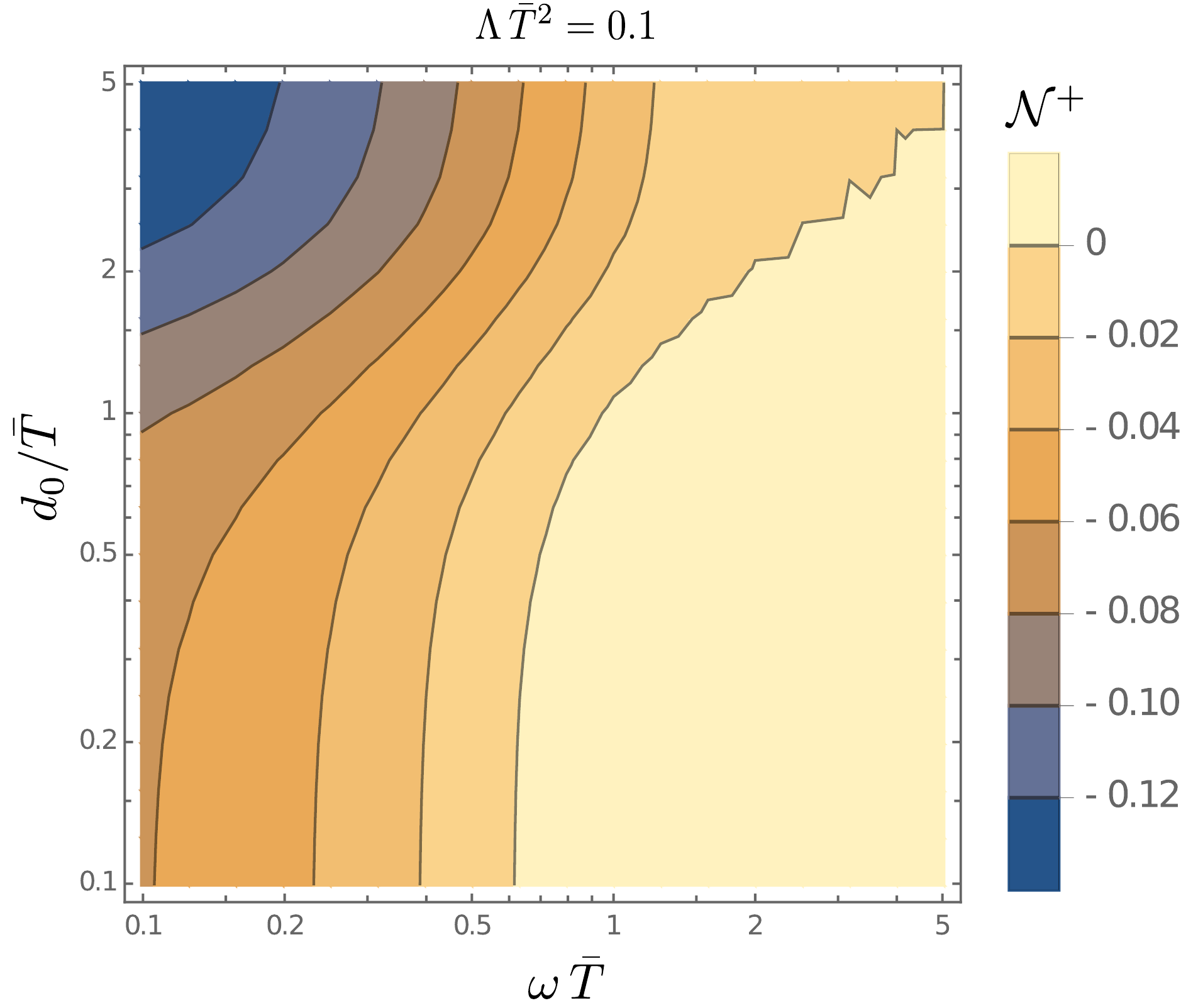}
\hskip 10pt
\includegraphics[width=5.5cm]{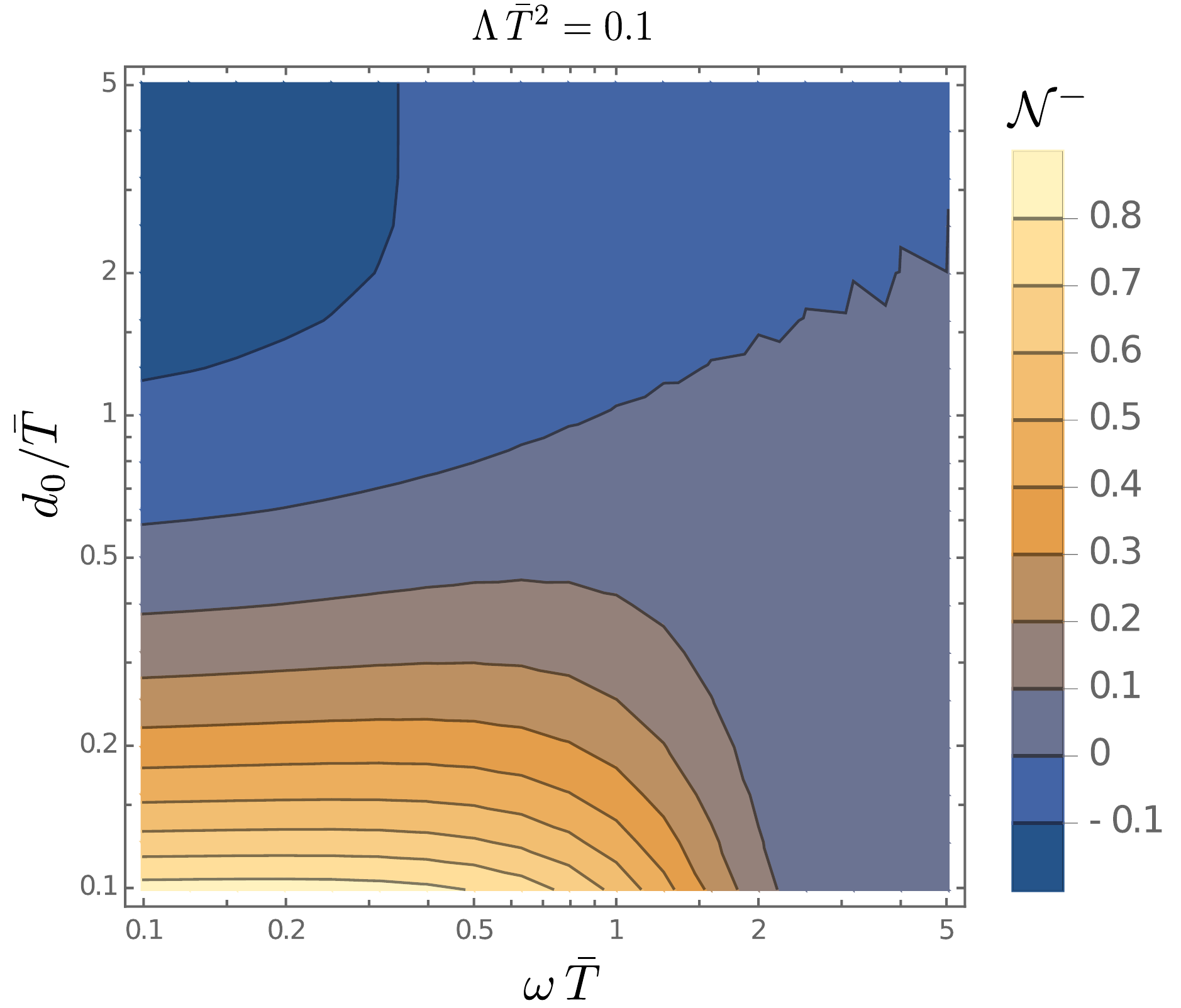}
\hskip 10pt
\includegraphics[width=5.5cm]{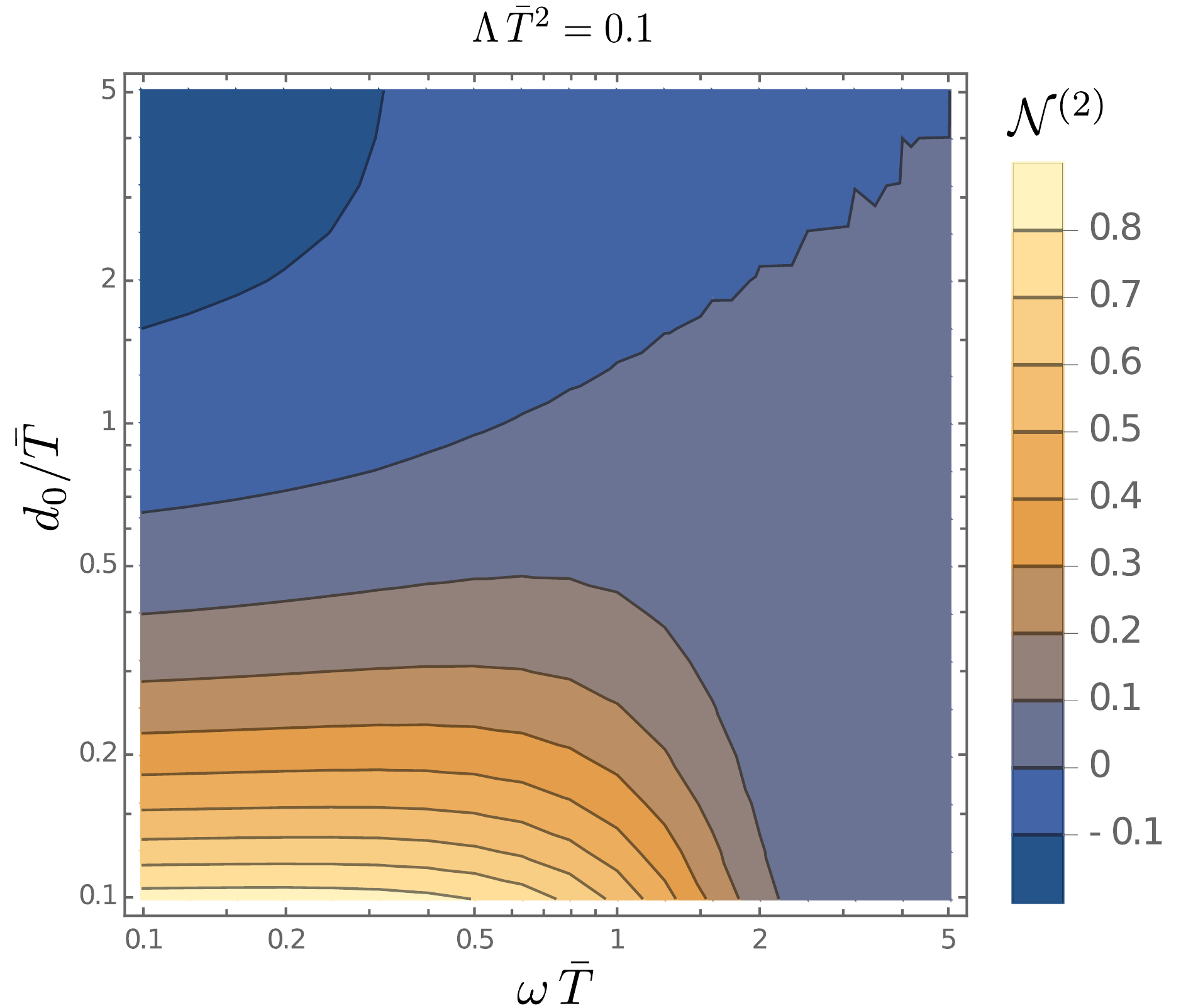}
\caption{The contribution in negativity due to spacelike separation $N^{+}$ (left plot), purely due to the causal communication $N^{-}$ (middle plot), and the entire negativity $N^{(2)}$ (right plot) are plotted for $\Lambda=10^{-4}$ (top row) and $\Lambda=0.1$ (bottom row), see \eq{eq:Npm-general} for the definitions of $N^{\pm}$. These plots correspond to detectors in the de Sitter background with initial separation $d_{0}$. As is evident from the plots, entanglement without causal communication is possible only in the high detector transition energy $\omega\,\bar{T}$ regime. Whereas, entanglement from the causal communication is prominent in the low separation $d_{0}/\bar{T}$ and low transition energy regime ${\omega}\,\bar{T}$.}
    \label{fig:DS-DIe-vs-dE-3}
\end{figure*}

Fig. \ref{fig:DS-DIe-vs-dE-3} shows the contour plot for negativity components that particularly result from spacelike separation $(\mathcal{N}^{+})$, causal communication $(\mathcal{N}^{-})$, and the total negativity $(\mathcal{N}^{(2)})$ in a de Sitter background.  These different negativity components with their significance were mentioned in Sec. \ref{subsec:NpNm-forms}. These contour diagrams are plotted as functions of the dimensionless initial separation and detector energy gap. The two different rows of the figure correspond to two different curvature length scales. One can observe that the negativity due to spacelike separation, often associated with genuine entanglement effects \cite{Martin-Martinez:2015psa, Tjoa:2021roz, Barman:2021kwg}, is positive only in the regime of large energy gaps and initial separations. At the same time, the negativity due to the communication channel is positive in the entire region except for the very large initial. These phenomena are true for both of the considered curvature length scales. 

The physical reason behind the above observations is straightforward: for a finite switching time scale, as the initial separation between the detectors increases, the second detector moves away from the accessible causal lightcone patch of the first detector, see Fig. \ref{fig:Geodesic-trajectory-distance} for illustration. Nevertheless, one observes that there exists a range of parameters that will yield the so-called genuine entanglement effects.

\begin{figure*}[!h]
\includegraphics[width=5.50cm]{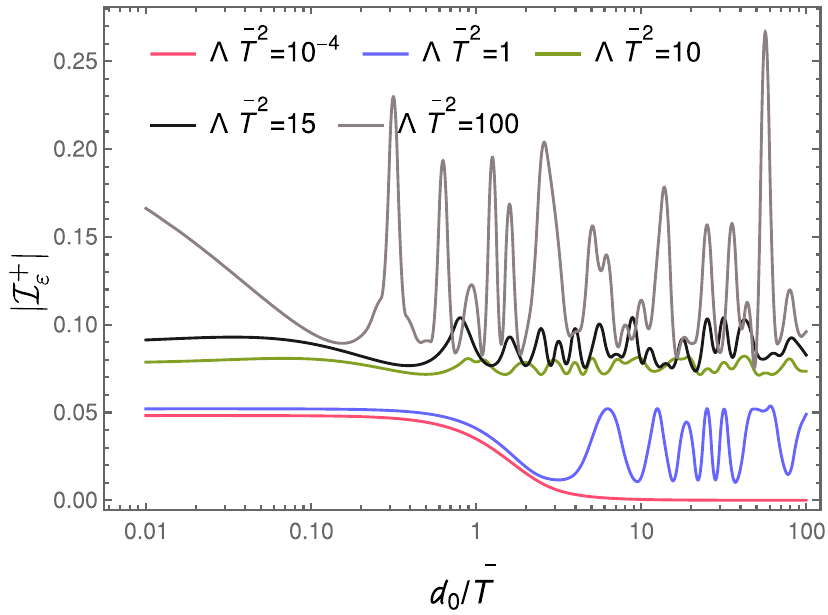}
\hskip 10pt
\includegraphics[width=5.50cm]{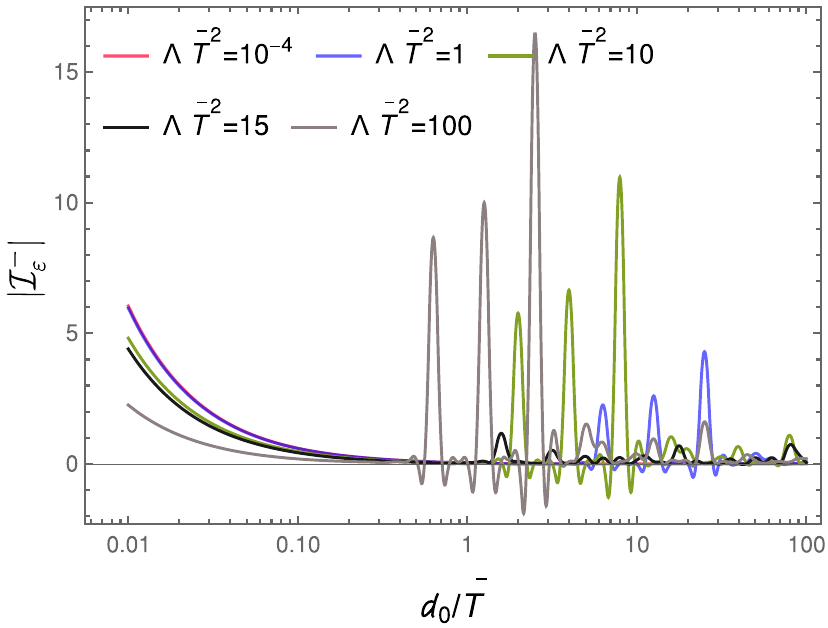}
\hskip 10pt
\includegraphics[width=5.50cm]{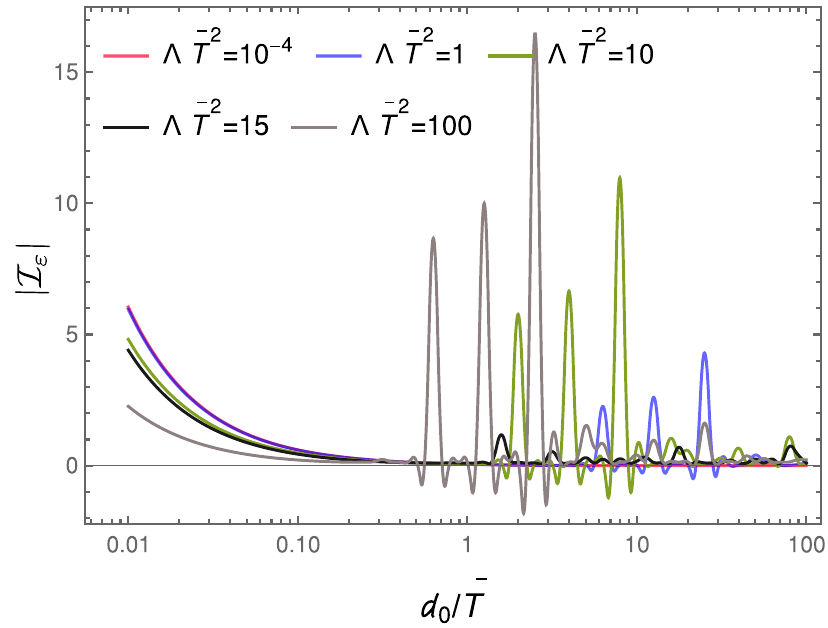}
\vskip 10pt
\includegraphics[width=5.50cm]{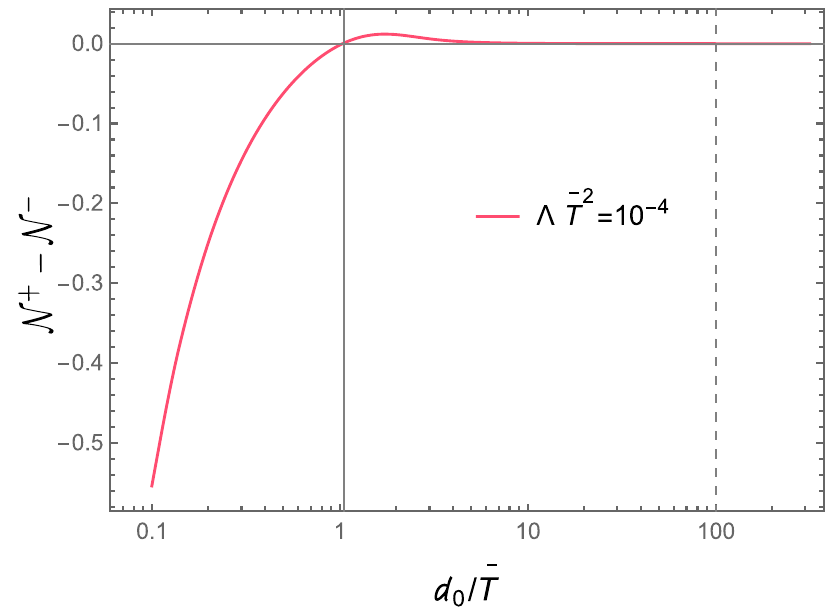}
\hskip 10pt
\includegraphics[width=5.50cm]{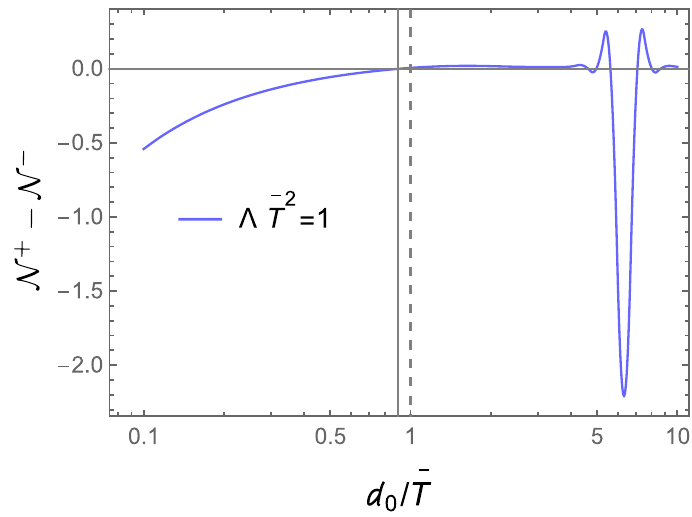}
\hskip 10pt
\includegraphics[width=5.50cm]{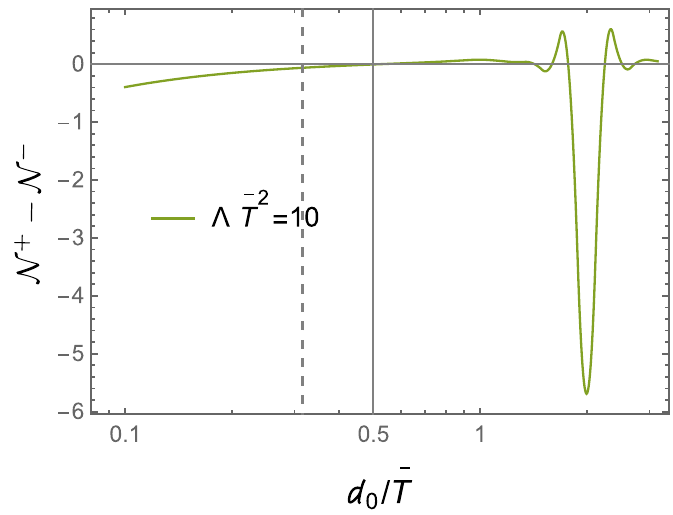}
\hskip 10pt
\includegraphics[width=5.50cm]{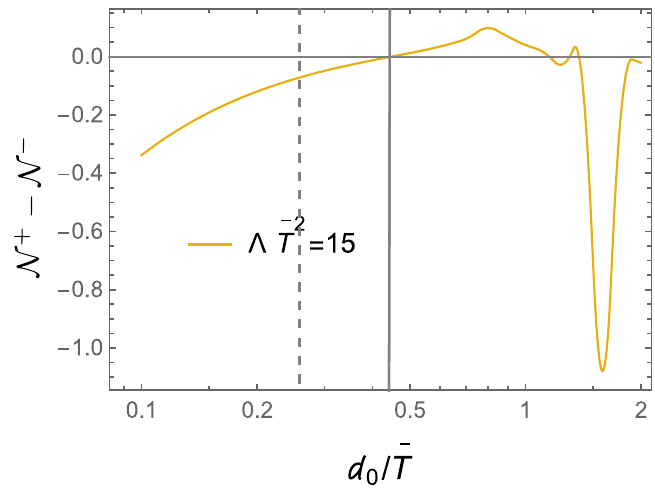}
\hskip 10pt
\includegraphics[width=5.50cm]{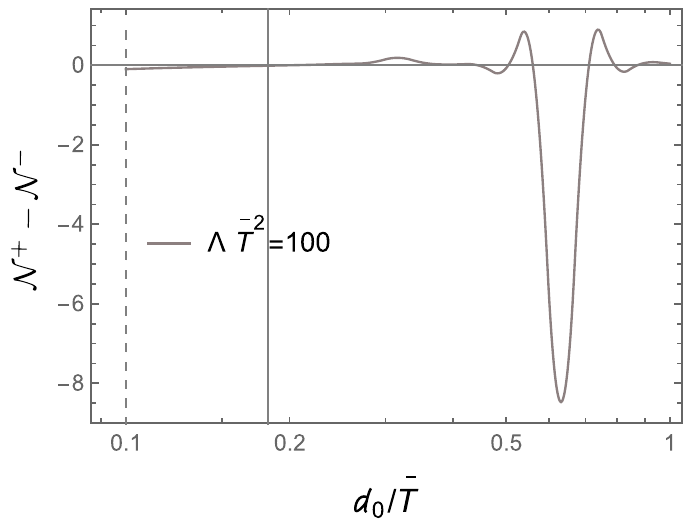}
\caption{
The contribution in the non-local entangling term $\mathcal{I}_{\varepsilon}$ with and without causal communication, which are denoted as $\mathcal{I}_{\varepsilon}^{-}$ and $\mathcal{I}_{\varepsilon}^{+}$, are plotted in the top row of this figure as functions of the dimensionless distance $d_{0}/\bar{T}$ for different curvature scales of the background. We observe that, as the distance decreases, the $|\mathcal{I}_{\varepsilon}^{-}|$ part keeps increasing, while $|\mathcal{I}_{\varepsilon}^{+}|$ seems to reach saturated values. We have also plotted the modulus of the entire entangling term $\mathcal{I}_{\varepsilon}$ that mostly resembles $|\mathcal{I}_{\varepsilon}^{-}|$. In the below two rows of the figure, we have plotted the difference $(\mathcal{N}^{+} -\mathcal{N}^{-})$ as functions of the dimensionless initial separation $d_{0}/\bar{T}$ between the detectors. Each of the plots from these two rows corresponds to a different background curvature. From these plots, we observe that as the separation increases, there is a certain value of $d_{0}/\bar{T}$ when $\mathcal{N}^{+}$ overtakes $\mathcal{N}^{-}$. We further observe that this crossing point in $d_{0}/\bar{T}$ decreases with increasing curvature. In the plots from the lower two rows, the dashed lines correspond to the respective values of $1/(\sqrt{\Lambda}\,\bar{T})$, which indicate the curvature length scales.
}
    \label{fig:DS-DIe-vs-dE-4}
\end{figure*}

Finally, in Fig. \ref{fig:DS-DIe-vs-dE-4}, we have plotted the spacelike negativity component $(\mathcal{N}^{+})$, the causal communication-based negativity component $(\mathcal{N}^{-})$, and their difference as functions of the initial detector separation. For these plots, we have considered a suitable parameter space, taking inspiration from Fig. \ref{fig:DS-DIe-vs-dE-3}, a parameter space where, for large detector separation, the spacelike negativity can become larger compared to the causal negativity. From these plots, one can observe that for small initial separation $d_{0}$, the communication-based negativity $(\mathcal{N}^{-})$ is always greater compared to the spacelike negativity. However, for large initial detector separation, $\mathcal{N}^{+}$ is becoming larger compared to $\mathcal{N}^{-}$. From the difference between these two quantities we observe that, for small background curvature, the transition from $\mathcal{N}^{+}<\mathcal{N}^{-}$ to $\mathcal{N}^{+}>\mathcal{N}^{-}$ happens near $d_{0}/\bar{T}\simeq 1$, and this value keeps on decreasing as the curvature increases. The physical reasoning could be as the curvature increases, the past and the future-directed light cones depart more from each other with increasing separation, see Fig. \ref{fig:Geodesic-trajectory-distance}. Therefore, as the curvature increases, the scope of causal communication for a certain detector separation is also diminished for finite detector switching.

\newpage

\section{Concluding remarks}\label{sec:Concluding-remarks}

Given the already exhaustive discussion and commentary in preceding sections, we will keep the discussions in this concluding section brief, focusing largely on some general observations extracted from the results we have established.

Let us first summarize the key steps of our analysis: We have extracted the role of curvature in the entanglement induced between two Unruh-DeWitt detectors, interacting solely with a background scalar field in an arbitrary background spacetime. Assuming the field state to be Hadamard, the key curvature contribution comes from the relative kinematics of the trajectories. While the curvature will also affect the dynamics of the field, this effect is suppressed compared to that arising from the leading terms in the Hadamard expansion, and this is nicely corroborated by the de Sitter case, where the exact known expressions yield results for entanglement measure that closely matches the one obtained from our perturbative scheme. The key geometric reason for why this works, in terms of the causal structure of spacetime, was already identified in Sec. \ref{sec:GeodesicInterval-GreensFn}.

We now list some general observations based on this work. The list is only representative, to aid in navigating through the bulk of the paper; we hope the reader will be able to extract more insights from the various analytical and graphical results presented here. 
\begin{enumerate}
    \item Contribution to entanglement measure (negativity here) from spacelike correlations, versus that due to timelike correlations, can be identified by decomposing the Feynman propagator into its anti-commutator and commutator parts. For smaller distances, contribution of commutator dominates, while at larger separations, the anti-commutator contribution dominates, which is what one would want to attribute genuine quantum entanglement with. Hence, choosing a suitable $d_0/\bar{T}$ can lead to genuine entanglement (in the suitable range of energy gap), and this ratio is altered by presence of Riemann tensor.
    \item For lower energy gaps, the local term corresponding to the individual detector transition due to vacuum polarisation will dominate over the non-local term for initial separation, $d_0 \ge \bar{T}$. As the energy gap becomes large enough, i.e., $\omega \ge 2\hbar/\bar{T}$, the non-local term starts to dominate over the individual detector transition probabilities.
    \item For parameters chosen such that $d_0 \approx 2 \bar{T}$ and $\omega \approx 2\hbar/\bar{T}$, the genuine entanglement between detectors is present. \textit{Presence of curvature enhances the entanglement for the above parameter range.}
    
    \item When the detectors have an initial relative velocity ${\bm v_0}$, the entanglement depends on this relative velocity and its angle $\theta$ with respective to the initial separation $\bm \xi$. The genuine entanglement decreases with an increase in speed. Overall, the entanglement also increases with increase in $\theta$. 

    The key upshot of the above dependences, and the interplay between them, is that these may allow one to measure spacetime curvature using quantum probes with suitable initial conditions.     
    
    \item Our investigation on the role of curvature in entanglement between the probes indicates that the entanglement is most sensitive to the eigenvalue of tidal part of the Riemann tensor -- $R_{a0c0}=R_{abcd} u^b u^d$ -- along the direction of separation of the detectors $\bm \xi$; that is, $R_{a 0 b 0} \xi^a \xi^b$. This contribution is present even for initially static detectors. Entanglement is also sensitive to $R_{a 0 b 0} n^a n^b$, the eigenvalue of the tidal tensor in the direction $n^a$ of the relative initial velocity of the detectors. This dependence, however, is weaker than the previous one. Interestingly, the entanglement does not depend on the components such as $R_{abcd} n^a \xi^b n^c \xi^d$, which denotes the purely spatial part of the curvature.
    
    The initial relative configuration of the detectors can be used as a probe to determine the curvature of spacetime. An explicit example of one such set-up is demonstrated in Fig. \ref{fig:DN-vperp}, which can be used to extract the contribution of the magnetic part of Riemann tensor. It would be interesting to compare the viability of probing such components of Riemann using entanglement, with classical experiments.
    
    \item There exists in literature a couple of works
    \cite{Cliche:2010fi, Kukita:2017etu} which attempt to study entanglement between detectors in specific backgrounds with non-zero curvature. While helpful, a major drawback of working with specific solutions of Einstein equations is that it hides the explicit role of curvature and hence does not yield physical insights which can be claimed to be generic. Our motivation, set-up, as well overall analysis in this work are much more general and hence have a much wider scope. We can, however, compare our results with the ones in these works in cases when the choice of initial conditions and parameters coincide. In Ref. \cite{Kukita:2017etu}, the authors study entanglement in a de Sitter background and conclude that entanglement at the super horizon scale is possible only with multiple quantum probes. Our present work finds that the parameter values for entanglement extraction are in the same range as those obtained in Ref. \cite{Kukita:2017etu} for the de Sitter background. On the other hand, in Ref. \cite{Cliche:2010fi}, entanglement generation is studied outside of weakly gravitating, spherically symmetric, bodies such as the Earth and the Sun. Their result indicating enhancement of entanglement (compared to Minkowski) for large energy gaps is also observed in our case. 

    \item Our work should be relevant in observationally viable scenarios such as for tests of the violation of Bell's inequality. 
In this context, we would like to specifically mention approaches to understand entanglement between two non-intersecting spatial regions in de Sitter spacetime, which do not suggest the presence of entanglement \cite{Espinosa-Portales:2022yok}. It will be interesting to relate our analysis with these and identify the points of conflicts and similarities between the same. It should not be difficult to construct a set-up similar to \cite{Martin:2021xml, Espinosa-Portales:2022yok, Martin:2021qkg}, using the tools introduced here, to discern entanglement induced in a general curved background. This investigation is currently under progress. 

    \item Finally, we would like to mention a broader context in which this work will serve as the first step. This concerns a known subtlety in application of the \textit{equivalence principle} to deduce the role of gravity in quantum phenomenon by analyzing these phenomenon in non-inertial frames in Minkowski spacetime. While doing so is usually easy since one is working in Minkowski spacetime, it has its pitfalls, and can serve as a hurdle to obtaining insights which a perturbative analysis can never give. For instance, in Ref. \cite{K:2021gns, K:2023sex}, it is shown that even a single accelerated probe can pick up non-trivial effects due to tidal curvature which cannot be seen in a perturbative expansion. Our analysis in this paper, though still perturbative in curvature, develops the necessary mathematical tools that, along with some of the exact results we obtain in de Sitter, have the potential to yield stronger results beyond a perturbative expansion. This is under investigation.
\end{enumerate}


\begin{acknowledgments}

We would like to thank Vincent Vennin, Alessandro Pesci, and L. Sriramkumar for useful discussions and comments on the manuscript. H.K. thanks the Indian Institute of Technology Madras and the Ministry of Human Resources and Development (MHRD), India for financial support. S.B. would like to thank the Science and Engineering Research Board (SERB), Government of India (GoI), for supporting this work through the National Post Doctoral Fellowship (N-PDF, File number: PDF/2022/000428).

\end{acknowledgments}

\appendix

\section{Geodesic interval between events on two timelike trajectories}
\label{app:sigma2}

We start with the following expansion (ref. Fig. (\ref{fig:setup})):
\begin{equation}\label{eq:Synge-WF-1}
\Omega_{\mathcal{G}}=[\Omega_{\mathcal{G}}]+\Omega_{\mathcal{G}_{s}} [t^{i}\nabla_i \Omega_{\mathcal{G}}] + \frac{1}{2} \Omega^2_{B^{\prime}B} \l[t^{i}\nabla_i\l( t^{k} \nabla_{k} \Omega_{\mathcal{G}}\r)\r] + ...
\end{equation}
Here, $\Omega_{xx^{\prime}}:= \Omega(x,x^{\prime})$ represents Synge's world function which is half the square of the geodesic distance between $x$ and $x^{\prime}$ and $[\,]$ represents the limit, $\Omega_{\mathcal{G}s}\rightarrow 0$. The FNC at $B^{\prime}$ is defined by, $x^{i^{\prime}}=\l(\tau_{2}^{\prime}, X^{\alpha^{\prime}}(\tau=\tau_{2}^{\prime})\r)$ and  $X^{\alpha^{\prime}}(\tau=\tau_{2}^{\prime}):=\sigma_{\mathcal{G}s}(\tau_{2}^{\prime}) t^{i^{\prime}}(\tau_{2}^{\prime}) {\rm e}^{\alpha^{\prime}}_{\phantom{\alpha^{\prime}}i^{\prime}}(\tau_{2}^{\prime}) $ where, $\sigma_{\mathcal{G}s}(\tau_{2}^{\prime}) t^{i^{\prime}} = -\sigma^{i^\prime}_{\mathcal{G}s}(\tau_{2}^{\prime}) := -g^{i^\prime k^\prime}\nabla_{k^\prime} \Omega_{\mathcal{G}s}(\tau_{2}^{\prime}) $. Using this definition, the expansion for $\Omega_{\mathcal{G}}$ can be expressed in terms of the Fermi coordinates.
\begin{equation}\label{eq:Synge-WF-2}
\Omega_{\mathcal{G}}=\Omega_{\mathcal{G}^{\prime}}+ [\nabla_{\alpha^{\prime}} \Omega_{\mathcal{G}}] X^{\alpha^{\prime}}(\tau_{2}^{\prime}) + \frac{1}{2}  \l[\nabla_{\alpha^{\prime}}\nabla_{\beta^{\prime}} \Omega_{\mathcal{G}}\r] X^{\alpha^{\prime}}(\tau_{2}^{\prime}) X^{\beta^{\prime}}(\tau_{2}^{\prime})+ ...
\end{equation}
Here, for a reference geodesic curve, $\Omega_{\mathcal{G}^\prime} = -\sigma^2_{\mathcal{G}^\prime}/2 \equiv -1/2 \l(\tau_{2}^{\prime}-\tau\r)^2$. Now, the expansions of $[\,]$ quantities and $X^{\alpha^{\prime}}(\tau_{2}^{\prime})$ should be expressed in terms of data given at $\tau_1=0$. 

The $[\,]$ quantities need to be evaluated carefully. The derivatives of $\Omega_{\mathcal{G}}$ are expanded in double Taylor series of $\Omega_{\mathcal{G}^{\prime}}$ with $\tau$ and $\tau_{2}^{\prime}$ as variables.
\begin{subequations}
\begin{align}\label{eq:Synge-WF-3}
\nabla_{\alpha^{\prime}}\Omega_{\mathcal{G}} &= \nabla_{\alpha^{\prime}}\Omega_{\mathcal{G}^{\prime}} + \tau u^{i} \nabla_{i}\l(\nabla_{\alpha^{\prime}}\Omega_{\mathcal{G}^{\prime}}\r) + \tau_{2}^{\prime} u^{i^{\prime}}\nabla_{i^{\prime}}\l(\nabla_{\alpha^{\prime}}\Omega_{\mathcal{G}^{\prime}}\r) + ... \\
\nabla_{\alpha^{\prime}}\nabla_{\beta^{\prime}}\Omega_{\mathcal{G}} &= \nabla_{\alpha^{\prime}}\nabla_{\beta^{\prime}}\Omega_{\mathcal{G}^{\prime}} + \tau u^i \nabla_i \l(\nabla_{\alpha^{\prime}}\nabla_{\beta^{\prime}}\Omega_{\mathcal{G}}\r) + \tau_{2}^{\prime} u^{i^{\prime}} \nabla_{i^{\prime}} \l( \nabla_{\alpha^{\prime}}\nabla_{\beta^{\prime}}\Omega_{\mathcal{G}} \r) + ...
\label{series-for-omega}
\end{align}
\end{subequations}
Similarly, all higher derivatives are expressed and then the coincidence limit is correctly taken. The calculations are cumbersome and hence we have used \texttt{CADABRA}\cite{peeters2007cadabra,peeters2007introducing} for the calculations.

Note that, $\nabla_{\alpha^{\prime}}\Omega_{\mathcal{G}^{\prime}} \equiv {\rm e}^{i^{\prime}}_{\phantom{i^\prime}{\alpha^{\prime}}}\nabla_{i^{\prime}}\Omega_{\mathcal{G}^{\prime}}$ and note that $u^i\nabla_{i}{\rm e}^{i^\prime}_{\phantom{i^\prime}\alpha^\prime}=0$ since the tetrad is at $x^\prime$ and derivative is taken at $x$. Similarly, $u^{i^{\prime}}\nabla_{i^{\prime}}{\rm e}^{i^\prime}_{\phantom{i^\prime}\alpha^\prime}=0$ since the tetrad is Fermi-Walker transported along a geodesic and will not generally be zero for other curves. Substituting these observations into \eq{series-for-omega}, the RHS is expressed in terms of covariant derivatives of $\Omega_{\mathcal{G}^{\prime}}$ at $x^{\prime}$ and $x$. Then the coincidence limit, $\tau_{2}^{\prime},\tau \rightarrow 0$ is taken.
Some covariant derivatives are as follows:
\begin{align}
    \l[\nabla_{i^{\prime}}\Omega_{\mathcal{G}^\prime}\r]&=0 \nn \\
    \l[\nabla_{j}\nabla_{i^{\prime}}\Omega_{\mathcal{G}^\prime}\r]&=-g_{i j} \nn \\
    \l[\nabla_{j^{\prime}}\nabla_{i^{\prime}}\Omega_{\mathcal{G}^\prime}\r]&=g_{i j} \nn\\
    \l[\nabla_{k^{\prime}}\nabla_{j^{\prime}}\nabla_{i^{\prime}}\Omega_{\mathcal{G}^\prime}\r]=\l[\nabla_{k}\nabla_{j^{\prime}}\nabla_{i^{\prime}}\Omega_{\mathcal{G}^\prime}\r]&=\l[\nabla_{k^{\prime}}\nabla_{j}\nabla_{i^{\prime}}\Omega_{\mathcal{G}^\prime}\r]=\l[\nabla_{k}\nabla_{j}\nabla_{i^{\prime}} \Omega_{\mathcal{G}^\prime}\r] =0\nn \\
    \l[\nabla_{l^{\prime}}\nabla_{k^{\prime}}\nabla_{j^{\prime}}\nabla_{i^{\prime}}\Omega_{\mathcal{G}^\prime}\r] = \l[\nabla_{l}\nabla_{k}\nabla_{j}\nabla_{i}\Omega_{\mathcal{G}^\prime}\r] & = - \l[\nabla_{l}\nabla_{k^{\prime}}\nabla_{j^{\prime}}\nabla_{i^{\prime}}\Omega_{\mathcal{G}^\prime}\r]=\l[\nabla_{l}\nabla_{k}\nabla_{j^{\prime}}\nabla_{i^{\prime}}\Omega_{\mathcal{G}^\prime}\r] = -\frac{1}{3}\l(R_{ikjl}+R_{iljk}\r) \nn \\
    \l[\nabla_{l}\nabla_{k}\nabla_{j}\nabla_{i^\prime}\Omega_{\mathcal{G}^\prime}\r] &= -\frac{1}{3} \l(R_{i jkl}+R_{ikjl}\r) \nn \\
    ..... \nn
\end{align}
Substituting the coincidence limits to the covariant expansion of $\Omega_{\mathcal{G}}$ in \eq{eq:Synge-WF-2} results in the following equation,
\begin{align}
    \Omega_{\mathcal{G}} &= -\frac{1}{2}\l(\tau_{2}^{\prime}-\tau\r)^2 + \frac{1}{2} \l(g_{\alpha \beta}  - \frac{1}{3} \l(\tau_{2}^{\prime}-\tau\r)^2 R_{i j k l} {\rm e}^{i}_{\phantom{i}\alpha} {\rm e}^{k}_{\phantom{k}\beta} u^j u^l\r) X^{\alpha}\l(\tau_{2}^{\prime}\r) X^{\beta}\l(\tau_{2}^{\prime}\r)+\mathcal{O}(\tau^4,X^5,R^2)
    \label{geo-eqn-1}
\end{align}
Notice that, we have assumed that the derivatives of the Riemann tensor are negligible compared to the Riemann tensor itself for the above expression. The FNC needs to be expanded in terms of $\tau_{2}^{\prime}$. For FNC, the expansions are simple Taylor expansion in terms of $\tau_{2}^{\prime}$ given by,
\begin{equation}
X^{\alpha^{\prime}}(\tau_{2}^{\prime})=X^{\alpha^\prime}(0)+ \l. \frac{dX^{\alpha}}{d\tau}\r\vert_{O} \tau_{2}^{\prime}+ \frac{1}{2}  \l.\frac{d^2X^{\alpha}}{d\tau^2}\r\vert_{O} t^2_{B^{\prime}} + ...
\label{taylor-fnc}
\end{equation}
The coefficients of the Taylor expansion can be evaluated using the definition of deviation vector in terms of Synge's world function as given by Ref.\cite{Vines:2014oba}. Let the deviation vector be, $\xi^i=X^{\alpha}{\rm e}^i_{\phantom{i}\alpha}$
\begin{align}
    X^{\alpha} &=-\sigma^{i}_{\mathcal{G}s}(\tau) {\rm e}^{\alpha}_{\phantom{\alpha}i}(\tau) \\
    \frac{\DM X^{\alpha}}{\DM \tau} &= -{\rm e}^{\alpha}_{\phantom{\alpha}i}(\tau)u^{k}\nabla_{k} \sigma^{i}_{\mathcal{G}s}(\tau) -{\rm e}^{\alpha}_{\phantom{\alpha}i}(\tau)v^{\tilde k}\nabla_{\tilde k} \sigma^{i}_{\mathcal{G}s}(s) \\
    ... \nn
    \label{dev-exp-1}
\end{align}
Here, $v^{\tilde k}:=\l({\DM s}/{\DM \tau}\r) u^{\tilde k}_{B} $, where $u^{\tilde k}_{B}$ is the 4-velocity of the detector $B$ along its trajectory. Using the expansion for derivatives of Synge world function gives\cite{Vines:2014oba},
\begin{align}
    \nabla_{k} \sigma^{i}_{\mathcal{G}s}(\tau) & = \delta^{i}_{\phantom{i}k}-\frac{1}{3}R^{i}_{\phantom{i} \alpha k \beta} X^{\alpha}X^{\beta} + \mathcal{O}(R^2,\nabla R,X^3)\\
    \nabla_{\tilde k} \sigma^{i}_{\mathcal{G}s}(s) & = -g^j_{\phantom{j}\tilde k}\l(\delta^i_{\phantom{i} j} + \frac{1}{6} R^i_{\phantom{i}\alpha j \beta}X^{\alpha} X^{\beta} + \mathcal{O}(R^2,\nabla R,X^3)\r)
    \label{dev-exp-2}
\end{align}
Here, $g^j_{\phantom{j}\tilde k}$ is the parallel propagator.
The coefficients for the Taylor expansion of FNC can be evaluated using \eq{dev-exp-1} and \eq{dev-exp-2} and can be obtained as,
\begin{align*}
    \frac{\DM X^{\alpha}}{\DM \tau} &= v^{\alpha}_0 + \frac{1}{3}R^{\alpha}_{\phantom{\alpha}\mu 0 \nu} X^{\mu} X^{\nu} + \frac{1}{6} R^{\alpha}_{\phantom{\alpha}\mu k \nu} X^\mu X^\nu v^k_0 + \mathcal{O}(R^2,\nabla R,X^3) \\
    \frac{\DM^2 X^{\alpha}}{\DM \tau^2} &= - R^{\alpha}_{\phantom{\alpha}0 \beta 0} X^{\beta} + \mathcal{O}(R^2,\nabla R,X^2) \\
    \frac{\DM^3 X^{\alpha}}{\DM \tau^3} &\approx - R^{\alpha}_{\phantom{\alpha}0 \beta 0} v_0^{\beta} + \mathcal{O}(R^2,\nabla R,X^2)
\end{align*}

\begin{equation}
     X^{\alpha}(\tau)=\l(X^{\alpha}+v^{\alpha}_0 \tau_{2}^{\prime} + \frac{1}{3}R^{\alpha}_{\phantom{\alpha}\mu 0 \nu} X^{\mu} X^{\nu} \tau_{2}^{\prime}+ \frac{1}{6} R^{\alpha}_{\phantom{\alpha}\mu k \nu} X^\mu X^\nu v^k_0 \tau_{2}^{\prime} - \frac{1}{2} R^{\alpha}_{\phantom{\alpha}0 \mu 0} X^{\mu} {\tau_{2}^{\prime}}^2 - \frac{1}{6}R^{\alpha}_{\phantom{\alpha}0 \beta 0} v_0^{\beta} {\tau_{2}^{\prime}}^3\r) + \mathcal{O}(R^2,\nabla R)
     \label{eqn:deviation-vector}
\end{equation}
Using the above simplifications in the \eq{geo-eqn-1}, the geodesic distance to $R^2,\nabla R$ order will be,
\begin{align}
    -\sigma^2_{\mathcal{G}} =\Delta \tau_{\mathcal{G}}^2= &\l(\tau_{2}^{\prime}-\tau\r)^2 - \l[ \l( X^{\alpha}(0) + v^{\alpha}_0 \tau_{2}^{\prime} \r)^2 + \frac{2}{3} R_{\alpha \mu 0 \nu}v_0^\alpha X^\mu X^\nu {\tau_{2}^{\prime}}^2 - R_{\alpha 0 \beta 0} X^{\alpha}(0) X^{\beta}(0) {\tau_{2}^{\prime}}^2 - \frac{4}{3} R_{\alpha 0 \beta 0} X^{\alpha}(0) v_0^{\beta} {\tau_{2}^{\prime}}^3  \r. \nn \\
    & \quad \quad \quad \quad \quad \quad \quad  \l. + \frac{1}{3} R_{\alpha \mu \beta \nu} X^\alpha X^\beta v_0^\mu v_0^\nu {\tau_{2}^{\prime}}^2 - \frac{1}{3} R_{\mu 0 \nu 0} v_0^\mu v_0^\nu {\tau_{2}^{\prime}}^4 \r] \nn \\
    & + \frac{1}{3} \l(\tau_{2}^{\prime}-\tau\r)^2 R_{\alpha 0 \beta 0} \l[ X^{\alpha}(0) X^{\beta}(0) + 2 X^{\alpha}(0) v^{\beta}_0 \tau_{2}^{\prime} + v_0^\alpha v_0^\beta {\tau_{2}^{\prime}}^2\r]+\mathcal{O}(\nabla R,R^2)
\end{align}

The initial separation is ${X}^\alpha(0):=d_{0}\, \xi^{\alpha}$ where, $\xi^{\alpha}$ is the directional cosines and $\eta_{\alpha \beta}\xi^{\alpha} \xi^{\beta}=1$. The initial velocity can be decomposed into components along the initial separation and the component along the normal to the initial separation denoted by unit vector $n^\alpha$ that is in the plane orthogonal to $X^\alpha (0)$, $v_0^\alpha = v_{||} \, \xi^{\alpha}+ v_{\perp} \, n^\alpha$. Note that $v_{||}=v_0 \cos\theta$ and $v_{\perp}=v_0 \sin\theta$. Using this decomposition in the above expression, we finally obtain Eq. (\ref{eq:geodesic-dist-gen}).

\textbf{\textit{The mapping between proper times of the detectors}}: Apart from geodesic interval, another quantity we will need to characterize entanglement is the mapping between proper times of the two detectors. The relevant quantitiy here is $\gamma=\l({\DM s}/{\DM \tau}\r)$, which can be easily found using the Fermi metric based on $\mathcal{C}_1$, assumed to bbe geodesic (see, for e.g., \cite{Poisson:2003nc}). 
\begin{eqnarray}
    \DM S^2 &=& -~\DM {\tau_{2}^{\prime}}^2\l(1+R_{0\alpha 0 \beta}X^\alpha X^\beta\r) - 2 \DM X^\mu \DM \tau_{2}^{\prime} \l(\frac{2}{3} R_{0\alpha \mu \beta} X^\alpha X^\beta\r) \nonumber\\
    ~&& ~~~+~ \DM X^{\mu} \DM X^{\nu} \l(\delta_{\mu \nu} - \frac{1}{3} R_{\mu \alpha \nu \beta} X^\alpha X^\beta \r) + \mathcal{O}(X^3,\nabla R, R^2)~.
\end{eqnarray}
Since, $\DM S^2=-\DM s^2$ the ratio is,
\begin{align}
    \gamma(\tau^{\prime}_{2}):= \frac{\DM \tau_{2}^{\prime}}{\DM s} = \l[1-\l(\frac{dX^\alpha}{d\tau_{2}^{\prime}}\r)^2 + R_{0\alpha 0 \beta}X^\alpha X^\beta+\frac{4}{3} R_{0\alpha \mu \beta}X^\alpha X^\beta \frac{dX^\mu}{d\tau_{2}^{\prime}} + \frac{1}{3} R_{\mu \alpha \nu \beta} X^{\alpha} X^{\beta} \frac{dX^{\mu}}{d\tau_{2}^{\prime}}  \frac{dX^{\nu}}{d\tau_{2}^{\prime}} \r]^{-1/2}~.
\end{align}

Using \eq{eqn:deviation-vector} and resolving the components as done for the geodesic distance, in the equation for $\gamma$, to linear order in curvature the equation becomes,
\begin{eqnarray}\label{eq:gamma-tbp}
    \gamma (\tau^{\prime}_{2}) = \frac{\DM \tau_{2}^{\prime}}{\DM s} &=& \l[1- v^2_{0} + \mathpzc{h}_1(v_{0},d_{0}) + 2 d_{0} \mathpzc{h}_2(v_{0},d_{0}) \tau_{2}^{\prime} + \mathpzc{h}_3(v_{0},d_{0}) {\tau_{2}^{\prime}}^2 \r]^{-1/2}   +~ \mathcal{O}(R^2,\nabla R)~.
\end{eqnarray}
Here, $\mathpzc{h}_1(v_{0},d_{0}),\, \mathpzc{h}_2(v_{0},d_{0})$ and $ \mathpzc{h}_3(v_{0},d_{0})$ are auxiliary functions defined by,
\begin{subequations}\label{eq:gamma-tbp-hj}
\begin{eqnarray}
    \mathpzc{h}_1(v_{0},d_{0}) &:=& R_{\xi 0\xi 0} d_{0}^2 + \frac{2}{3} R_{\xi 0 \xi n} d_{0}^2 v_{\perp} \\
    \mathpzc{h}_2(v_{0},d_{0}) &:=& R_{\xi 0\xi 0} v_{||}  + R_{\xi 0n 0} v_{\perp} + \frac{2}{3} v_{\perp}\l( R_{\xi 0 \xi n} v_{||} + R_{n\xi 0 n} v_{\perp} \r)  \\
    \mathpzc{h}_3(v_{0},d_{0}) &:=& R_{\xi 0\xi 0} v_{||}^2 + 2 R_{\xi 0n 0} v_{||} v_{\perp} + R_{n0n0} v_{\perp}^2
\end{eqnarray}
\end{subequations}
The proper time of $B$ can be expressed as, $s=\int d\tau_{2}^{\prime}/\gamma(\tau_{2}^{\prime})$. (When the coordinate velocity of second detector is zero, $\gamma=\sqrt{1+R_{\xi 0\xi 0}d_{0}^2}$.)

\section{\textbf{Evaluation of $\mathcal{I}_{\varepsilon}$ in general curved spacetime: complete perturbative analysis}}\label{Appn:perturbative-Ie}

We observed that the individual detector transition probabilities $\mathcal{I}$ and non-local entangling terms $\mathcal{I}_{\varepsilon}$ depend on the geodesic distance that in a general curved spacetime was estimated in a perturbative manner in (\ref{eq:geodesic-dist-gen}). We have already analytically estimated the individual local terms $\mathcal{I}$ for small curvatures in \ref{sec:Negativity-GenCurved}. Now we shall evaluate the non-local entangling term $\mathcal{I}_{\varepsilon}$ by expressing it perturbatively around the Minkowski scenario, i.e., by assuming the corrections due to the curvature to be very small compared to other system parameters specifically the initial separation between the detectors. Let us proceed this way and express $\mathcal{I}_{\varepsilon}$ for two initially static and general geodesic detectors.\vspace{0.15cm}

\subsubsection{Evaluation of $\mathcal{I}_{\varepsilon}$ for two initially static detectors}

Let us first consider the case of two static detectors. In this scenario, the only quantity that depends on curvature terms and which is present in the geodesic distance between the two detectors is $R_{\alpha 0 \beta 0}X^\alpha(0)X^\beta(0):=k_1d_{0}^2$. Let us term this quantity $c_{1}$, i.e., $c_{1} = k_1d_{0}^2$. We take the geodesic distance between the detectors $\sigma_{\mathcal{G}}^2(\tau^{\prime}_{2}-\tau,\tau^{\prime}_{2})$ from Eq. (\ref{eq:geodesic-dist-gen}). We consider a change of variables $\tau_{2}^{\prime}-\tau=\Delta \tau$ and $\tau_{2}^{\prime}=v$, similar to the one done in the case for evaluating $\mathcal{I}_{j}$. First, the Feynman propagator in the non-local entangling term can be expanded considering $c_{1}\ll 1$. Second, the non-local term contains other factors involving the switching functions and the exponential with the detector transition energy. These factors will also contain the curvature term $c_{1}$ due to their representation in terms of Fermi-normal coordinates with respect to a certain geodesic, see \eq{eq:gamma-bar-tbp}. Finally, the non-local entangling term can be expressed as 
\begin{equation}\label{eq:Appn-Ie-Full-perturbative}
\mathcal{I}_{\varepsilon} = \mathcal{I}^{M}_{\varepsilon} + \mathcal{I}^{C}_{\varepsilon}~,
\end{equation}
where with the Gaussian switching function we have
\begin{subequations}
\begin{eqnarray}\label{eq:Appn-IeM-IeC}
    \mathcal{I}^{M}_{\varepsilon} ({\omega}) &=& -\frac{1}{4 \pi ^2 }\int_{-\infty}^{\infty}dv\,\int_{-\infty}^{\infty}d\Delta\tau \,\,\frac{e^{-\frac{(v -\Delta \tau )^2+v ^2}{\Bar{T}^2}+i {\omega} (2 v -\Delta \tau )}}{ \left(\Delta \tau ^2-d_{0}^2-i \epsilon \right)}~,\\
    \mathcal{I}^{C}_{\varepsilon} ({\omega}) &=& -\frac{1}{4 \pi ^2 }\int_{-\infty}^{\infty}dv\,\int_{-\infty}^{\infty}d\Delta\tau\,\,\frac{e^{-\frac{(v -\Delta \tau )^2+v ^2}{\Bar{T}^2}+i {\omega} (2 v -\Delta \tau )}}{ \left(\Delta \tau ^2-d_{0}^2-i \epsilon \right)}\Bigg[\frac{(\Delta\tau^2/3)-v^2}{\left(\Delta \tau ^2-d_{0}^2-i \epsilon \right)}-\left\{\frac{i \,v\, {\omega}+1}{2}-\frac{v^2}{\Bar{T}^2}\right\}\Bigg]\,c_{1}\, \nonumber\\
    ~&&~~~~~~~~~~~\hspace{10cm} +~\mathcal{O}(c_{1}^2)~.
\end{eqnarray}
\end{subequations}
Let us first evaluate the Minkowski part $\mathcal{I}^{M}_{\varepsilon} ({\omega})$. Here the integration over $\tau$ can be readily carried out as the denominators of $G^{M,C}_{F} (\Delta \tau, v)$ are independent of $v$. Using the Gaussian integration formula $\int_{-\infty}^{\infty}dv\,e^{-\alpha(v-\beta)^2}=\sqrt{\pi/\alpha}$ one can express 
\begin{eqnarray}\label{eq:Appn-IeM-1}
    \mathcal{I}^{M}_{\varepsilon} ({\omega}) = -\frac{\Bar{T} e^{-\frac{{\omega}^2 \Bar{T}^2}{2}}}{4 \sqrt{2} \pi ^{3/2} } \int_{-\infty}^{\infty}d\Delta\tau\,\frac{ e^{-\frac{\Delta \tau ^2}{2 \Bar{T}^2}}}{ \left(\Delta \tau ^2-d_{0}^2-i \epsilon \right)}~.
\end{eqnarray}
To evaluate this integral, we express the involved Gaussian factor as $e^{-\frac{\Delta \tau ^2}{2 \Bar{T}^2}} = (\Bar{T}/\sqrt{2\pi})\int_{-\infty}^{\infty}d\kappa\,e^{-(\zeta ^2 \Bar{T}^2/2)+i \Delta \tau \, \kappa }$, i.e., in terms of its Fourier transform. The integral of the previous equation has poles at $\Delta\tau = \pm \sqrt{d_{0}^2+i \epsilon }$, which are in the upper and the lower half complex planes respectively, depending on the `+ve' and the `-ve' signs. Therefore, we must consider the contour in the upper or lower semicircle to dampen the necessary parts of the contour integral. After finding the residues, the integration limit of $\kappa$ will also be restricted. For example, for the upper pole, we have $\kappa\in [0,\infty)$, and for the lower pole, we have $\kappa\in (-\infty,0]$. Then, carrying out the integration over $\kappa$, we have the final expression
\begin{eqnarray}\label{eq:Appn-IeM-2}
    \mathcal{I}^{M}_{\varepsilon} ({\omega}) = -\frac{i\,\Bar{T} \left(\erf\left(\frac{i\,d_{0}}{\sqrt{2}\, \Bar{T}}\right)+1\right)}{4\, \sqrt{2 \pi }\, d_{0}}\, e^{-\frac{d_{0}^2}{2\Bar{T}^2}-\frac{{\omega}^2 \Bar{T}^2}{2}}\,.
\end{eqnarray}

Let us now evaluate the integral $\mathcal{I}^{C}_{\varepsilon} ({\omega})$ that corresponds to the contribution entirely from the curved spacetime due to the terms up to $\mathcal{O}(c_{1})$. It is to be noted that several numerators in the term $\mathcal{I}^{C}_{\varepsilon} ({\omega})$ also depend on $\tau$. If we integrate over this variable $\tau$, we shall get 
\begin{eqnarray}\label{eq:Appn-IeC-1}
    \mathcal{I}^{C}_{\varepsilon} ({\omega}) &=& \frac{c_{1}e^{-\frac{{\omega}^2 \Bar{T}^2}{2 }}}{48 \sqrt{2} \pi ^{3/2} \Bar{T}}\int_{-\infty}^{\infty} \frac{d\Delta\tau\,e^{-\frac{\Delta \tau ^2}{2 \Bar{T}^2}}}{\left(\Delta \tau ^2-d_{0}^2-i \epsilon \right)^2}\,\Bigg[ \bigg\{3 \Delta \tau ^2 \left(\Delta \tau ^2-i \epsilon \right)-3 d_{0}^2 \left(\Delta \tau ^2+\Bar{T}^2 (-1+i {\omega} \Delta \tau )\right)-3 {\omega}^2 \Bar{T}^6~\nonumber\\
    ~&&~~\hspace{4cm} +~\Bar{T}^4 (3+6 i {\omega} \Delta \tau )+\Bar{T}^2 \left(\Delta \tau ^2 (4+3 i {\omega} \Delta \tau )+3 \epsilon  ({\omega} \Delta \tau +i)\right)\bigg\}\Bigg]~.
\end{eqnarray}
Here also the poles are at the same places $\Delta\tau = \pm \sqrt{d_{0}^2+i \epsilon }$ but with order two, and we proceed in a similar fashion to integrate over the variable $\Delta\tau$. The final expression for $\mathcal{I}^{C}_{\varepsilon} ({\omega})$ will be given by 
\begin{eqnarray}\label{eq:Appn-IeC-2}
    \mathcal{I}^{C}_{\varepsilon} ({\omega}) &=& \frac{c_{1}}{192\, \pi\,  d_{0}^3\, \Bar{T}} \, e^{-\frac{{\omega}^2 \Bar{T}^2}{2}-\frac{d_{0}^2}{2 \Bar{T}^2}} \bigg[\sqrt{2 \pi }\, \erf\text{i}\left(\frac{d_{0}}{\sqrt{2} \Bar{T}}\right) \left\{d_{0}^4+d_{0}^2 \left(2 \Bar{T}^2-3 {\omega}\,^2 \Bar{T}^4\right)-3 {\omega}\,^2 \Bar{T}^6+3 \Bar{T}^4\right\}-i\, \sqrt{2 \pi }\, d_{0}^4\nn\\
    ~&& +~i \sqrt{2 \pi }\, d_{0}^2\, \Bar{T}^2 \left(3 {\omega}\,^2 \Bar{T}^2-2\right)+6 d_{0} \Bar{T}^3 e^{\frac{d_{0}^2}{2 \Bar{T}^2}} \left({\omega}\,^2 \Bar{T}^2-1\right)-14 d_{0}^3 \Bar{T} e^{\frac{d_{0}^2}{2 \Bar{T}^2}}+3 i \sqrt{2 \pi } \Bar{T}^4 \left({\omega}\,^2 \Bar{T}^2-1\right)\bigg].
\end{eqnarray}
Therefore, we have the analytical expression for both the $\mathcal{I}^{M}_{\varepsilon} ({\omega})$ and $\mathcal{I}^{C}_{\varepsilon} ({\omega})$, which will lead one to the entire expression of $\mathcal{I}_{\varepsilon} ({\omega})$ from \eq{eq:Appn-Ie-Full-perturbative}. We have verified that the plots of this non-local entangling term are similar in nature compared to the ones from Figs. \ref{fig:Ij-Ie-v0e0}, \ref{fig:DN-vs-diff-para-v0-Gaussian}, and \ref{fig:DN-vs-diff-para-v0-Gaussian-negativeL}. It further validates our previous procedure.\vspace{0.15cm}

\subsubsection{Evaluation of $\mathcal{I}_{\varepsilon}$ for two geodesic detectors}

For two detectors in curved spacetime with an initial separation and relative velocity between them, one can express the non-local entangling term like the previous case as $\mathcal{I}_{\varepsilon} = \mathcal{I}^{M}_{\varepsilon} + \mathcal{I}^{C}_{\varepsilon}$. Here $\mathcal{I}^{M}_{\varepsilon}$ corresponds to the contribution from the Minkowski spacetime and is given by
\begin{eqnarray}\label{eq:Appn-IeM-vne0-1}
    \mathcal{I}^{M}_{\varepsilon} = -\frac{1}{4 \pi ^2 \gamma_{0}}\int_{-\infty}^{\infty}dv\,\int_{-\infty}^{\infty}d\Delta\tau\,\frac{e^{-\frac{(v -\Delta \tau )^2}{\Bar{T}^2}-\frac{v ^2}{\Bar{T}^2\gamma_{0}^2}+i {\omega} (v(1+1/\gamma_{0}) -\Delta \tau )}}{ \left\{ \Delta \tau ^2-(d_{0}^2+ 2 d_{0}\, v_{0}\, \cos{(\theta)}\, v +v_{0}^2 v^2)-i \epsilon \right\}}~.
\end{eqnarray}
Whereas $\mathcal{I}^{C}_{\varepsilon}$ corresponds to the term solely dependent on the presence of curvature. This quantity is given by 
\begin{eqnarray}\label{eq:Appn-IeC-vne0-1}
    \mathcal{I}^{C}_{\varepsilon} &=& \frac{c_{1}}{4 \pi ^2 \gamma_{0}}\int_{-\infty}^{\infty}dv\,\int_{-\infty}^{\infty}d\Delta\tau\,\frac{e^{-\frac{(v -\Delta \tau )^2}{\Bar{T}^2}-\frac{v ^2}{\Bar{T}^2\gamma_{0}^2}+i {\omega} (v(1+1/\gamma_{0}) -\Delta \tau )}}{ \left\{ \Delta \tau ^2-(d_{0}^2+ 2 d_{0}\, v_{0}\, \cos{(\theta)}\, v +v_{0}^2 v^2)-i \epsilon \right\}}\,\nn\\
    ~&\times& \Bigg[-\frac{ \bar{\mathpzc{f}}(v)\, \Delta \tau ^2+\bar{\mathpzc{g}}(v)}{\left\{ \Delta \tau ^2-(d_{0}^2+ 2 d_{0}\, v_{0}\, \cos{(\theta)}\, v +v_{0}^2 v^2)-i \epsilon \right\}}+\bigg\{\frac{\mathcal{H}_{1}(v)\,\gamma_{0}^2+iv\,\gamma_{0}{\omega}\,\mathcal{H}_{2}(v)}{2}-\frac{v^2\,\mathcal{H}_{2}(v)}{\bar{T}^2}\bigg\}\Bigg]\,.
\end{eqnarray}
Here $\bar{\mathpzc{f}}(v)=\mathpzc{f}(v)-1$, and $\bar{\mathpzc{g}}(v)=\mathpzc{g}(v)-\l( d_{0}^2 + v_0^2 v^2+ 2d_{0} v_{||} v \r)$ can be obtained from \eq{eq:exp-f&g}. Whereas, $\mathcal{H}_{1}(v)= \left\{\mathpzc{h}_1(v_{0},d_{0}) + 2d_{0}\mathpzc{h}_2(v_{0},d_{0}) + \mathpzc{h}_3(v_{0},d_{0})v^2\right\}$, and $\mathcal{H}_{2}(v)= \left\{\mathpzc{h}_1(v_{0},d_{0}) + d_{0}\mathpzc{h}_2(v_{0},d_{0}) + \mathpzc{h}_3(v_{0},d_{0})v^2/3\right\}$ are obtained utilizing the expressions from \eq{eq:gamma-tbp-hj}. Therefore, one can easily understand that the first part inside the brackets in \eq{eq:Appn-IeC-vne0-1} appears from the curvature corrections to the geodesic interval. On the other hand, the second part inside the brackets in \eq{eq:Appn-IeC-vne0-1} appears from the transformation from $s$ to $\tau_{2}^{\prime}$.
The above integrals can be easily performed using the Fourier transform $e^{-\frac{(\Delta \tau-v)^2}{\Bar{T}^2}} = (\Bar{T}/\sqrt{2\pi})\int_{-\infty}^{\infty}d\kappa\,e^{i\, \kappa(\Delta \tau-v) - \kappa^2 \Bar{T}^2/4}$. One can first perform the integral over $\Delta \tau$ through the help of contour integration and then over $\kappa$. Both of these integrations can be performed analytically. Finally, one is left with the integration over $v$, for which we could not find an analytical result. We have sought the help of numerical integration to perform this final integration. We have compared the results obtained from here with the ones from our previous calculation from Fig. \ref{fig:DN-vs-diff-para-vne0-Gaussian}, and confirmed similar outcomes.

\section{\textbf{3D plots - dependence of negativity on velocity and angle}}\label{Appn:vel}
\begin{figure*}[h!]
\centering
\includegraphics[width=8.25cm]{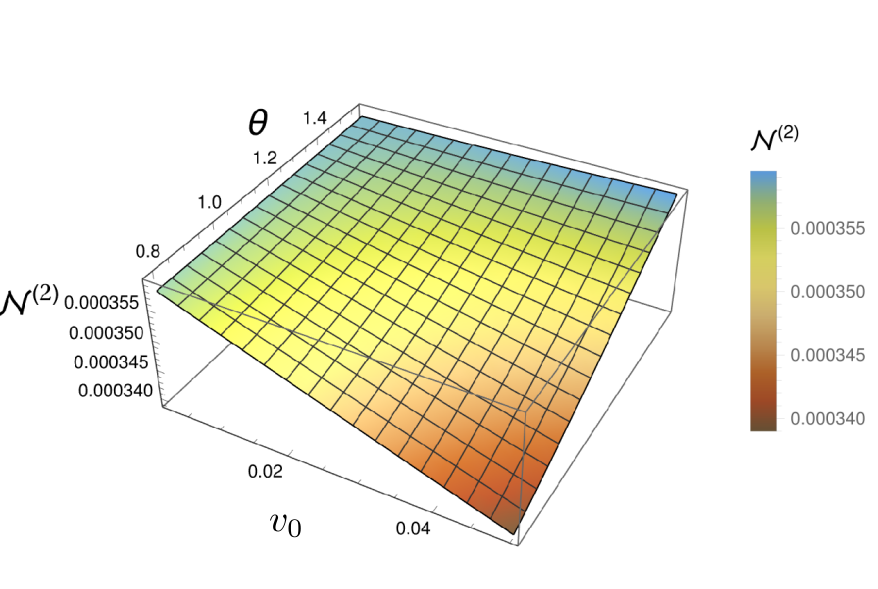}
\hskip 10pt
\includegraphics[width=8.5cm]{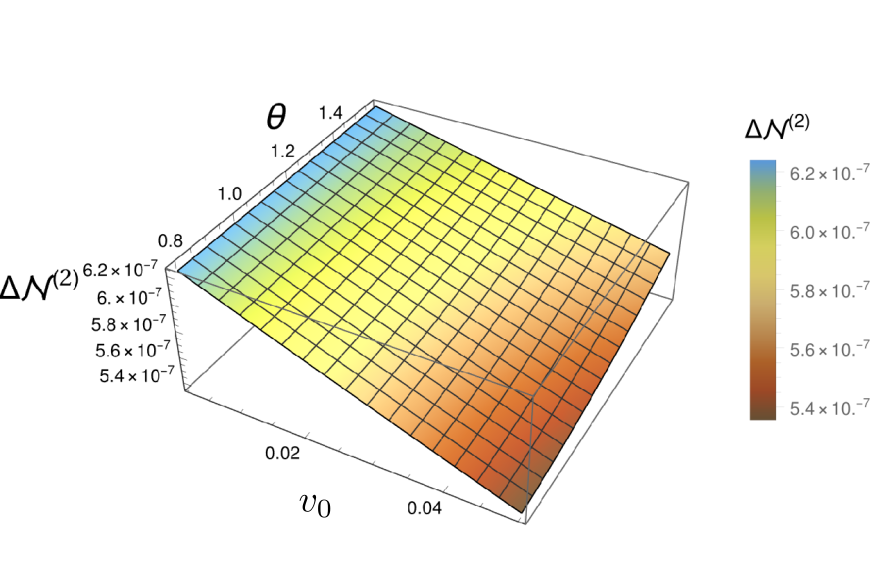}
    \caption{The negativity as a function of both initial velocity and angle between initial separation and initial velocity is shown in the left. The difference in negativity between the curved and flat spacetimes $\Delta\mathcal{N}^{(2)}$ is plotted as functions of the initial velocity $v$ of the detector $B$ and the angle $\theta$ between the initial velocity and the separation on the right. In both of the plots, the curvature length scales are of the order of $\ell_R=50\bar{T}$, the dimensionless detector energy gap is chosen as, $\omega\,\bar{T} = 2.5$ and the initial separation distance is $d_0/T=2$.}
    \label{fig:Ngen-3D-wrt-v0th}
\end{figure*}


\end{document}